\documentclass{aa}  

\usepackage{bm}
\usepackage{subfig}
\usepackage{rotating}
\usepackage{amssymb,amsmath}
\usepackage{graphicx}
\usepackage{txfonts}
\begin{document} 

        \titlerunning{Improving the RSM map algorithm for exoplanet imaging}
   \title{Improving the RSM map exoplanet detection algorithm}
   \subtitle{PSF forward modelling and optimal selection of PSF subtraction techniques}

   \author{C.-H. Dahlqvist\inst{1} \and O. Absil\inst{1}}

   \institute{STAR Institute, Universit\'{e} de Li\`{e}ge, All\'{e}e du Six Ao\^{u}t 19c, 4000 Li\`{e}ge, Belgium\\
              \email{carl-henrik.dahlqvist@uliege.be} }
   \date{}

\abstract
   { High-contrast imaging (HCI) is one of the most challenging techniques for exoplanet detection. It relies on sophisticated data processing to reach high contrasts at small angular separations. Most data processing techniques of this type are based on the angular differential imaging (ADI) observing strategy to perform the subtraction of a reference point spread function (PSF). In addition, such techniques generally make use of signal-to-noise (S/N) maps to infer the existence of planetary signals via thresholding.}
   {An alternative method for generating the final detection map was recently proposed with the regime-switching model (RSM) map, which uses a regime-switching framework to generate a probability map based on cubes of residuals generated by different PSF subtraction techniques. In this paper, we present several improvements to the original RSM map, focusing on novel PSF subtraction techniques and their optimal combinations, as well as a new procedure for estimating the probabilities involved. }
   {We started by implementing two forward-model versions of the RSM map algorithm based on the LOCI and KLIP PSF subtraction techniques. We then addressed the question of optimally selecting the PSF subtraction techniques to optimise the overall performance of the RSM map. A new forward-backward approach was also implemented to take into account both past and future observations to compute the RSM map probabilities, leading to improved precision in terms of astrometry and lowering the background speckle noise.}
   {We tested the ability of these various  improvements to increase the performance of the RSM map based on data sets obtained with three different instruments: VLT/NACO, VLT/SPHERE, and LBT/LMIRCam via a computation of receiver operating characteristic (ROC) curves. These results demonstrate the benefits of these proposed improvements. Finally, we present a new framework to generate contrast curves based on probability maps. The contrast curves highlight the higher performance of the RSM map compared to a standard S/N map at small angular separations.}
   {}


   \maketitle
%

\section{Introduction}
\label{sec:intro}

High-contrast imaging (HCI) is a method aimed at disentangling the faint signal of an exoplanet from the much brighter signal of its host star.It relies on large telescopes, adaptive optics, coronagraphs, and sophisticated data processing. 
While state-of-the-art HCI instruments provide near-infrared images with a high level of sensitivity and angular resolution, they are not immune to residual aberrations. These aberrations originate from time-averaged uncorrected atmospheric turbulence or from the optical train of the telescope and instrument. The resulting high and low spatial frequency structures can be problematic when searching for exoplanetary candidates since some of them, which are known as quasi-static speckles, share the same properties. Quasi-static speckles are often correlated across the exposures of a typical observation \citep{Bloemhof_2001}, which leads to their variation either being not slow enough to be subtracted through instruments calibration nor fast enough to be time-averaged.

Data processing represents a key ingredient for reducing the intensity of these speckles and further increasing the sensitivity of HCI. Different post-processing techniques, coupled with observing strategies, have been proposed in recent decades. Spectral differential imaging \citep[SDI,][]{Marois00,Sparks02} and angular differential imaging \citep[ADI,][]{Marois06} are the most commonly used observing strategies in dealing with quasi-static speckles. 
Most ADI-based methods rely on the subtraction of a reference point spread function (PSF), which models the speckle field based on the original set of images. The resulting residual images are then co-aligned and combined, cancelling part of the residual noise while increasing the intensity of the potential exoplanet signal. 

The most common subtraction methods include: ADI median-subtraction, which subtracts the median of a set of reference
frames from the original set of images, the locally optimised combination of images\citep[LOCI,][]{Lafreniere07}, which uses a least-squares approach to optimally estimate the speckle field, and the principal component analysis \citep[PCA/KLIP,][]{Soummer12,Amara12}, which computes the high-order singular modes of the speckle field. More recently, other speckle subtraction methods, such as non-negative matrix factorisation \citep[NMF,][]{Ren18} and the local low-rank plus sparse plus Gaussian decomposition \citep[LLSG,][]{Gonzalez16} have been proposed. These methods generally rely on signal-to-noise (S/N) maps to detect planetary candidates via the definition of a S/N threshold \citep[e.g.,][]{Mawet14,Bottom17,Pairet19}. Other methods such as ANDROMEDA \citep{Cantalloube15} or KLIP-FMMF \citep{Pueyo16,Ruffio17} use a forward model of the point source to identify the planetary signal in the residual images via the maximum likelihood. These approaches aim to take into account the distortion of the point source caused by the speckle field subtraction.

A new approach, known as a regime-switching model (RSM) map \citep{Dahlqvist20}, was recently proposed to better take advantage of the numerous PSF subtraction techniques that have been developed in the past decade. The RSM map can be seen as a multi-ADI alternative to the estimation of S/N map as it allows for the creation of a probability map based on several cubes of residuals obtained with different ADI-based PSF subtraction techniques. In contrast with most S/N map based techniques, the RSM map does not make the assumption of a Gaussian and white distribution for the speckle residuals after subtraction. It was indeed demonstrated that the speckle residual distribution is often closer to a Laplacian distribution, especially at small separations \citep{Pairet19,Dahlqvist20}. The RSM map takes into account the radial evolution of the speckle noise distribution and its dependence on the instrument, switching between Gaussian and Laplacian distributions. \cite{Dahlqvist20}  successfully tested this approach on a subset of PSF subtraction techniques (annular PCA, NMF, and LLSG).

The goal of this paper is to further develop the RSM approach by considering a larger set of PSF subtraction techniques that includes the forward modelling of the point source. Indeed, as for ANDROMEDA and KLIP-FMMF, the RSM map relies on a matched filter to infer the existence of planetary candidates in residuals images. However, the initial version of the algorithm, summarised in Sect.~2, uses solely an off-axis PSF for the detection. Forward modelling could significantly improve  the sensitivity of the algorithm to faint companions by taking into account the distortion generated by the speckle field subtraction. We propose a method that would rely on the KLIP forward model (KLIP-FM) developed by \cite{Pueyo16} as well as a forward model version of LOCI for investigating the added value of forward-modelled point sources. Section~3 is devoted to the development of the two forward-model versions of the RSM algorithm, while Sect.~4 provides a performance assessment of these versions.

As mentioned above, the RSM map can accommodate several PSF subtraction techniques to generate a final probability map. That raises questions regarding the selection of the optimal set of techniques to reach the highest sensitivity as well as its dependence on the HCI instrument and on the radial distance. We compare, in Sect.~5, the performance of several set of techniques via ROC curves and investigate the impact of the considered instruments on this selection, considering three state-of-the-art HCI instruments: NACO, SPHERE, and LMIRCam. 

Finally, in Sect.~6, we propose an improved method for the probability estimation, relying on a forward-backward approach that allows the use of both past and future observations within the cube of residuals to generate the RSM map. The original RSM map uses a simple forward approach, which considers only past observations to build up the probabilities. We compare the performance of both approaches with standard S/N maps through the use of ROC curves and contrast curves. In Sect.~7, we present a new framework developed to compute these contrast curves, as it is not possible when dealing with probability maps to rely on the standard procedure used for S/N maps.


\section{The RSM map}
\label{sec:model}

The RSM approach \citep{Dahlqvist20} relies on a two-state Markov chain to model the pixel intensity evolution inside the de-rotated cube of residuals generated by an ADI-based PSF subtraction technique. The regime-switching model is applied annulus-wise to account for the radial evolution of the residual noise properties. A residuals time series, $\bm{x}_{i_a}$, is built for each annulus, $a,$ by vectorising the selected set of patches along the time axis and then the spatial axis. The index $i_a \in \{1,\dots, T \times L_a \}$ provides the position of the considered patches within the cube of residuals, where $T$ and $L_a$ are respectively the number of frames in the cube and the number of pixels in a given annulus of radius, $a$. The time series is then described by a set of two equations accounting for the two considered regimes: the first regime, $S_{i_a}=0$, where $\bm{x}_{i_a}$ is described by a residual noise following the statistics of the quasi-static speckle residuals; and a second regime, $S_{i_a}=1$, where $\bm{x}_{i_a}$ is described by both the residual noise and the planetary signal model\footnote{The off-axis PSF in the original paper.}.
 Thus,\begin{eqnarray}
\label{maineq}
\bm{x}_{i_a} = \mu+ \beta F_{i_a} \bm{m}+ \bm{\varepsilon_{s,i_a}} =
  \begin{cases}
    \mu+  \bm{\varepsilon_{0,i_a}}  & \quad \text{if } S_{i_a}=0\\
    \mu+ \beta \bm{m}+ \bm{\varepsilon_{1,i_a}}      & \quad \text{if } S_{i_a}=1\\
  \end{cases}
,\end{eqnarray}
where $\beta$ and $\bm{m}$ provide, respectively, the flux and a model of the planetary signal, $\mu$ is the mean of the quasi-static speckle residuals, and $\bm{\varepsilon_{s,i_a}}$ is their time and space varying part that is characterised by the quasi-static speckle residuals statistics (see Appendix A for a summary of all the variables used for the RSM map computation). The parameter $F_{i_a}$ is a realisation of a two-state Markov chain that provides a short-term memory function to the model. The parameter $F_{i_a}$ being the realisation of a stochastic process, the time series $\bm{x}_{i_a}$ is described via a probability-weighted sum of the values generated by Eq.~\ref{maineq}.

The key feature of the regime-switching model is the fact that it provides for every pixel in the selected annulus, $a$, a probability for their being in each regime. The RSM map is constructed based on the probabilities of being in the planetary regime, that is, $S_{i_a}=1$. Relying on a two-state Markov chain, the probability, $\xi_{s,i_a}$, of being in regime $s$ at index $i_a$ will depend on the probability at the previous step, $i_a-1$, on the likelihood of being currently in a given regime, $\eta_{s,i_a}$, and on the transition probability between the regimes given by the matrix, $p_{q,s}$. The resulting probability of being in the regime, $s$, at index, $i_a$, is given by:

\begin{eqnarray}
\label{proba}
\xi_{s,i_a}=\sum^{1}_{q=0} \frac{\eta_{s,i_a} p_{q,s} \; \xi_{1,i_a-1} }{\sum^{1}_{q=0} \sum^1_{s=0} \eta_{s,i_a} p_{q,s} \; \xi_{q,i_a-1}} ,
\end{eqnarray}
where $\sum^{1}_{q=0} \sum^1_{s=0} \eta_{s,i_a} p_{q,s} \; \xi_{q,i_a-1}$ is a normalisation factor and $\eta_{s,i_a}$ is the likelihood associated with the regime, $s$, which is given for each patch, $i_a$, in the Gaussian case by
\begin{eqnarray}
\label{like}
\eta_{s,i_a}= \sum^{\theta^2}_n \frac{1}{\theta^2} \frac{1}{\sqrt{2 \pi}\sigma} \exp\left[- \frac{(\bm{x}^n_{i_a} - F_{i_a} \beta \bm{m}^{n} -\mu)^2}{2\sigma^2}\right],
\end{eqnarray}
where $\theta$ gives the size in pixels of the planet model, $\bm{m}$, and $n$ is the pixel index within the patch.

Due to the dependence on the previous step, the RSM approach relies on an iterative algorithm to estimate the probability for every pixel contained in the selected annulus, $a$. Once all the probabilities of being in the planetary regime, $\xi_{1,i_a}$, have been computed, those probabilities are averaged along the time axis to eventually provide the final RSM probability map. These different steps summarise the RSM approach, which was first proposed in \cite{Dahlqvist20} and where a more detailed description of the algorithm can be found.

\section{Using forward models in RSM}
\label{sec:fm}

The original RSM map relies on an off-axis PSF to model the planetary signal. A promising development of the current method would be to take into account, via a forward model, the effects of the PSF subtraction techniques on the planetary signal. Indeed, most PSF subtraction techniques lead to distortions of the planetary signal, such as over-subtraction and self-subtraction \citep{Pueyo16}. Over-subtraction is attributable to quasi-static speckles inside the set of reference frames, while self-subtraction is due to the presence of the planetary signal itself inside the same set of reference frames. The signature of self-subtraction is specific to planetary candidates, as quasi-static speckles coming from the optical train do not rotate with the field. Because of field rotation, the evolution of the reference frames composition leads to the appearance of a negative wing travelling in time from one side of the planet to the other in the azimuthal direction. The temporal motion of this negative wing should therefore help disentangling a planetary candidate from a bright speckle. 

We investigate, in this section, two forward model versions of the RSM map relying on the KLIP and LOCI algorithms. Both algorithms can accommodate an analytical estimation of the forward-modelled PSF, which is not the case for other ADI-based techniques such as NMF and LLSG. This avoid the complex task of choosing the fake companion intensity when constructing a forward-modelled PSF empirically by comparing an initial cube of residuals and one in which a fake companion has been injected.

\subsection{KLIP-based forward modelling}

Karhunen-Loève image processing (KLIP) is a popular speckle subtraction technique first proposed by \cite{Soummer12} and further improved by \cite{Pueyo16} who developed its forward model version. Similarly to PCA, KLIP\ estimates the reference PSF via a low-rank approximation of a reference library built to limit the impact of potential planetary signal on the speckle field estimation. For each frame of an ADI sequence, the KLIP algorithm computes the directions of maximal variance from the reference library. It keeps the principal components up to a rank, $K,  $ that is smaller than the dimension of the reference library, discarding the higher order modes that should contain more of the planetary signal. The principal components are found via a decomposition of the covariance matrix of the mean-subtracted reference frames $R$ via the Karhunen-Loève transform. They are given by:
\begin{eqnarray}
\bm{Z}_K= \sqrt{\bm{\Lambda}^{-1}}\bm{V}_K \bm{R},
,\end{eqnarray}
with $\bm{\Lambda}=diag(\mu_1,\mu_2,...,\mu_k)^{\top}$ as the diagonal matrix with the eigenvalues of the image-to-image sample covariance matrix $\bm{R}\bm{R}^{\top}$ with $\bm{R}$  the mean-subtracted reference library matrix and $\bm{V}_K=\left[ \bm{v}_1,\bm{v}_2,...,\bm{v}_k \right] $ as the respective eigenvectors up to the order $K\leq N_R$, $N_R$ being the number of images in the reference library used to compute the reference PSF (see Appendix A for a summary of all the variables used in the KLIP-based forward modelling). The reference PSF is then found by projecting the initial science image, $\bm{i}$, or a subsection of this science image, onto the selected principal components. The reference PSF is subtracted from the science image yielding the residual image $\bm{x}$ as follows:
\begin{eqnarray}
\label{klip}
\bm{x}= \bm{i}-\bm{Z}^{\top}_K\bm{Z}_K \bm{i}.
\end{eqnarray}

The selection of the reference library is done via the definition of a minimal field of view (FOV) rotation between the science image, $\bm{i},$ and the set of selected reference images, $\bm{R}$. The minimal FOV rotation should be large enough to limit the distortion due to the planetary signal contained in the library \citep[see][]{Marois10} but not too large so that it is possible to keep a sufficient correlation between the speckle field contained in the science image and the reference library. \cite{Pueyo16} proposed to model the distortion via an analytical expansion of the principal components to account for the presence of planetary signal inside the reference library. In the case of self-subtraction, the planetary signal appears in the principal components estimation via the covariance matrix and therefore the distortion is non linear as a result of the projection of $\bm{m}$ onto the perturbed components $\Delta \bm{Z}_K$. In contrast, the distortion due to over-subtraction is linear in $\bm{m}$ as it is defined as the projection of the planetary signal, $\bm{m},$ on the  unperturbed components, $\bm{Z}_K$. The forward model of the planetary signal considers both type of subtractions as follows:
\begin{eqnarray}
\label{fmklip}
\bm{p}= \bm{m} - \bm{Z}^{\top}_K \bm{Z}_K \bm{m} - \left( \bm{Z}^{\top}_K \Delta \bm{Z}_K + (\bm{Z}^{\top}_K \Delta \bm{Z}_K)^{\top}\right)  \frac{\bm{i}}{\beta},
\end{eqnarray}
where $\bm{m}$ represents the normalised planetary signal before reference PSF subtraction, typically the instrument off-axis PSF, and $\bm{p}$ is the forward model of the planet after subtraction. The second term on the right provides the over-subtraction of the point source while the third term gives the self-subtraction due to rotation via  $\Delta \bm{Z}_K$ \citep[see][for the detailed derivation of this expression]{Pueyo16}.

Having documented the estimation of the forward-modelled planetary signal and of the cube of residuals, we now consider how to include these elements in the RSM map framework. We rely on an annulus-wise estimation but in contrast with annular PCA or LLSG \citep{Gonzalez17}, we do not estimate the speckle field for consecutive non-overlapping annuli. We estimate instead a specific speckle field for every radial distance, $a$. The self-subtraction wings appearing azimuthally, the brightest part of the planetary signal is contained in an annulus segment of width equal to one full width at half maximum (FWHM). The selected annulus with a width of one FWHM is centred on $a$ and shifted by one pixel between each radial distance $a$ instead of being shifted by one FWHM in the case of annular PCA or LLSG. This approach simplifies the forward model PSF estimation, provides a more accurate estimation of the speckle field, and avoids any non-linearities due to transitions between annuli. 

        \begin{figure}[t]
\begin{center}
\includegraphics[width=200pt]{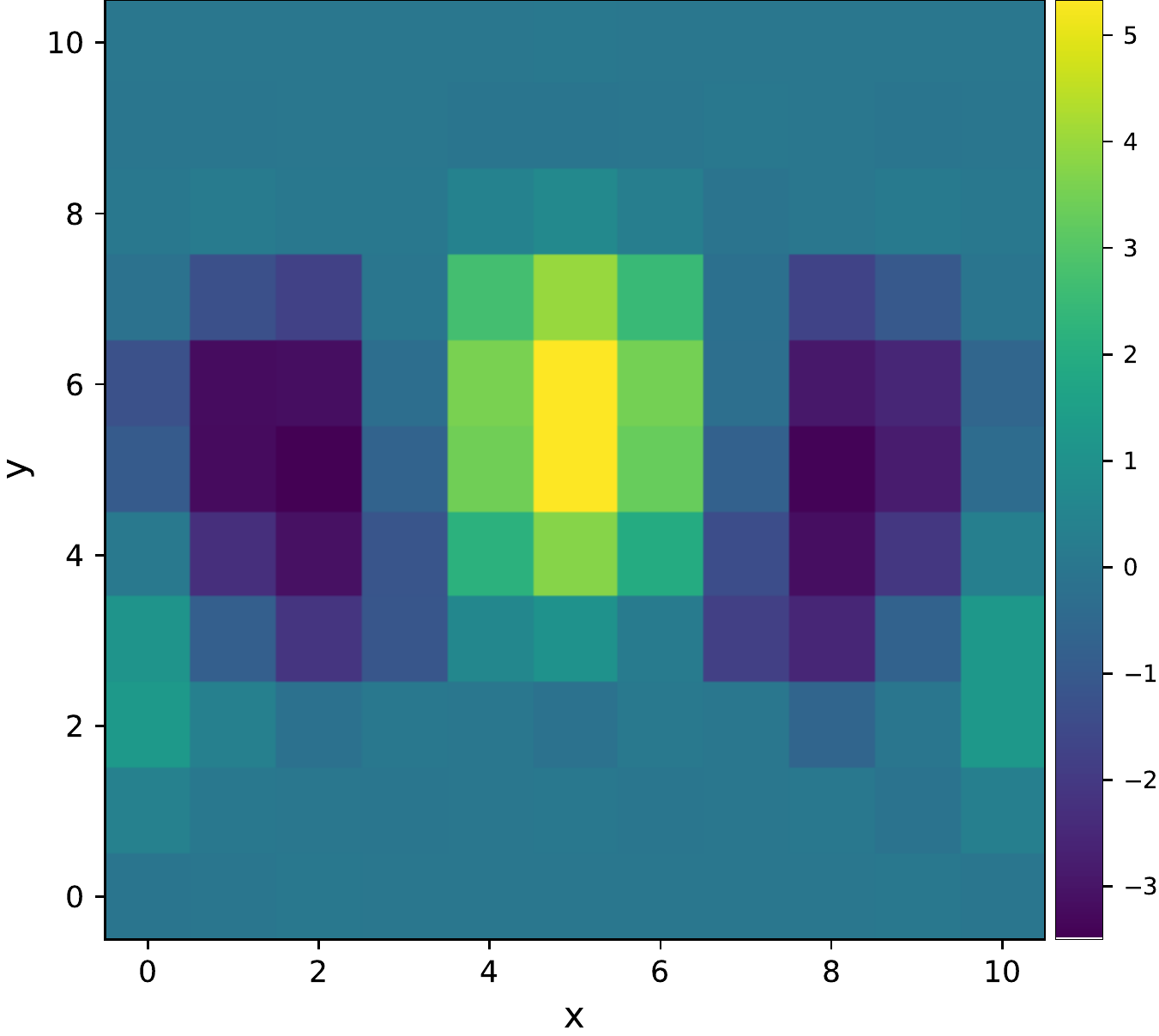} 
                        \caption{\label{fig1} KLIP forward-modelled PSF for a NACO $\beta$ Pictoris ADI sequence taken at a distance of $2\lambda/D$, cropped at two FWHM and summed along the time axis.}
                        \end{center}
                        
                          \end{figure}

The reference PSF and the forward-modelled PSF computation is done via  Eq.~\ref{klip} and Eq.~\ref{fmklip}, respectively. The resulting forward-modelled PSF is then derotated and cropped to form a set of patches, $\bm{p}_{i_a}$, where all the elements outside the selected annulus segment is set to zero as can be seen in Fig.~\ref{fig1}. Several crop sizes, from one to two FWHM, are tested in the next section. Once injected in the expression of the likelihood given in Eq.~\ref{like}, this allows us to focus the model only on the region where strong intensity variations occur. The expression of the likelihood of being in the state, $s,$ for every patch, $i_a$, becomes (in the Gaussian case):\ \begin{eqnarray}
\label{like2}
\eta_{s,i_a}= \sum^{\theta^2}_n \frac{1}{\theta^2} \frac{1}{\sqrt{2 \pi}\sigma} \exp\left[- \frac{(\bm{x}^n_{i_a} - F_{i_a} \beta \bm{p}^{n}_{i_a} -\mu)^2}{2\sigma^2}\right] ,
\end{eqnarray}
where $\bm{x}_{i_a}$ are the derotated and cropped patches obtained from the residual image, $\bm{x}$.

When relying on a forward-modelled PSF, the intensity parameter $\beta$ directly provides an estimation of the planet luminosity, which may be helpful for characterizing the planetary candidate beyond its detection. Two methods are considered for estimating this intensity parameter. The first method is similar to the one proposed in \cite{Dahlqvist20}, with $\beta$ defined as a multiple of the estimated standard deviation of the pixel intensity in the annulus, $\beta=\delta \sigma$. In this case, the standard deviation $\sigma$ is estimated empirically by considering all the frames and an annulus with a width of one FWHM centred on the annulus of interest, $a$. The $\delta$ parameter is defined via the maximisation of the total likelihood of the annulus. The second method relies on the definition of the intensity via a Gaussian maximum likelihood \citep[see][for more details]{Cantalloube15,Ruffio17} before the computation of the RSM map itself, which allows the use of the following analytical form for the flux parameter $\beta:$
\begin{eqnarray}
\label{gausslike}
\tilde{\beta}= \frac{ \sum^{T}_j \bm{i}^{\top}_j  \bm{p}_j /\sigma_j}{\sum^{T}_j \bm{p}^{\top}_j  \bm{p}_j /\sigma_j},
\end{eqnarray}
with the standard deviation $\sigma_j$ computed separately for each frame by considering an annulus with a width equal to one FWHM centred on $a$.

The main advantage of this second method is the simplicity of the intensity computation. It provides also a specific intensity for each pixel, which may help the algorithm to differentiate bright speckles from planetary candidates. The only drawback of this approach is that it makes the assumption of a Gaussian residuals distribution, which is not always the case especially near the host star. As for the parametrisation of the RSM likelihood function in this second case, the standard deviation used for the flux estimation is taken and the computation of the mean is also done frame-wise using the same procedure. Both approaches are investigated in Sect.~\ref{resfm}.

\subsection{LOCI-based forward modelling}

The second PSF subtraction technique that we propose for investigating the RSM framework is a forward model version of LOCI, the locally optimised combination of images \citep{Lafreniere07}. Specifically, LOCI relies on a linear combination of reference images to model the speckle field in a given science image, $\bm{i}$. As for the KLIP algorithm, the definition of the reference library is based on the definition of a minimal FOV rotation between the frames composing the reference library and the selected science image, that is, a minimal distance by which a potential point source in the science image would be displaced in the frames composing the reference library. 

Besides this angular distance, LOCI relies on the definition of two different subsections within the ADI sequence. A first section, $\bm{O}_K,$ is used for the computation of the linear combination factors, while a second smaller subsection $\bm{R}_K$ is selected for the speckle field subtraction\footnote{We consider an annulus of three FWHM for $\bm{O}_K$ and one FWHM for $\bm{R}_K$.}. The use of a larger section $\bm{O}_K$ aims to reduce the weight of the potential planetary candidate in the estimation of the linear combination. Once the reference library is defined, the computation of the linear combination factors is simply done via the minimisation of the sum of squared residuals \citep{Lafreniere07}:
\begin{eqnarray}
\epsilon^2= \sum^{N_p}_{j=0} \left[ \bm{o}^j_i-\sum^{N_R}_k c_k\bm{o}^j_k \right]^2 ,
\end{eqnarray}
with $\bm{o}_i$ the section of the frame for which a model of the speckle field is computed via the factors $c_k$, and $\bm{o}_k$ the section of the reference frame $k$ (see Appendix A for a summary of all the variables used in the LOCI-based forward modelling).
The minimum of this last expression has an analytical form obtained by setting all the partial derivatives with respect to $c_k$ equal to zero,
which is equivalent to solving a simple system of linear equations of the form Ax = b given by:
\begin{eqnarray}
\sum^{N_R}_k c_k \left[ \sum^{N_p}_{j=0} \bm{o}^j_l \bm{o}^j_k \right] = \left[ \sum^{N_p}_{j=0}  \bm{o}^j_l \bm{o}^j_i \right],
\end{eqnarray}
which holds $\forall l \in K$.

Once obtained, the factors, $c_k$, are multiplied by the subsection of the reference frames, $\bm{r}_k$, and subtracted from the subsection of the science image, $\bm{i},$ to get the residual, $\bm{x},$ as follows:
\begin{eqnarray}
\bm{x}=\bm{i} - \sum^{N_R}_k c_k \bm{r}_k \, ,
\end{eqnarray}
$\bm{i}$  defined here as the subsection of the science image corresponding to the reference frames, $\bm{r}_k$.

The forward model of the planetary signal is easily computed using the same factors, $c_k$, and the planetary signal, $\bm{m}$:
\begin{eqnarray}
\bm{p}=\bm{m}_i - \sum^{N_R}_k c_k \bm{m}_k \, .
\end{eqnarray}

As in the case of the KLIP forward model, for both the residual image and the forward model PSF, an annulus with a width of one FWHM is used to focus on the region where the planetary signal is the most visible. The forward model PSF, $\bm{p},$ and the residuals images, $\bm{x},$ are again derotated and cropped to from the time series ,$\bm{x}_{i_a}$ and  $\bm{p}_{i_a}$. The two methods used to estimate the flux parameter $\beta$ are again considered.

\subsection{Forward model RSM map: summary}

We briefly summarise the main steps for the RSM map estimation when relying on the LOCI and KLIP forward-model approach as follows:

\begin{enumerate}

\item Compute the residuals for an annulus centred on $a$ for each frame using the KLIP or LOCI procedure.
\item Compute the PSF forward model for every frame and every position within the annulus, $a.$
\item Derotate the resulting annuli and crop the forward-modelled PSF and science image to form the times series, $\bm{x}_{i_a}$ and $\bm{p}_{i_a.}$
\item Estimate the mean and variance of the residuals for every frame, considering the annulus of width equal to one FWHM centred on $a.$
\item Using the iterative procedure described in Sect.~\ref{sec:model}, estimate, $\xi_{1,i_a}$ for each index, $i_a$, using the forward model version of the likelihood (see Eq.~\ref{gausslike}).
\item Repeat steps 1 to 5 for every annulus.
\item Average the resulting probability matrix along the time axis to obtain the final RSM detection map.
\end{enumerate}

\section{Performance assessment of a forward-modelled RSM map}
\label{resfm}

\subsection{Data sets}

We propose a reliance on data sets provided by three different instruments to assess the performance of the two forward model versions of the RSM map. The two first ADI sequences were acquired with two instruments of the Very Large Telescope (VLT), NACO, and SPHERE, while the third sequence was acquired with the LMIRCam instrument of the Large Binocular Telescope (LBT). This choice of data sets aims to investigate the behaviour of the algorithm when facing different noise profiles generated by a variety of instruments.

The first data set is an ADI sequence on $\beta$ Pictoris and its planetary companion $\beta$ Pictoris b obtained in L band in January 2013 with NACO in its AGPM coronagraphic mode \citep{Absil13}. The ADI sequence is composed of 612 individual frames obtained by integrating 40 successive individual exposures of 200 ms. Every third frame was selected here to reduce the computation time, resulting in a final cube of 204 frames. The parallactic angle ranges from -15$^{\circ}$ to +68$^{\circ}$.

The second ADI sequence focuses on 51 Eridani. It was obtained in K1 band in September 2015 with the SPHERE-IRDIS instrument, using an apodised pupil Lyot coronagraph \citep{Samland17}. The data set regroups 194 frames pre-processed using the SPHERE
Data Center pipeline \citep[for more details about the reduction see][]{Delorme17,Maire19}. The integration time is 16 s and the parallactic angle ranges from 297$^{\circ}$ to 339$^{\circ}$. 

The last data set is an ADI sequence on HD183324 produced by the LMIRCam instrument of the LBT. The images were obtained in October 2018 in L' band without coronograph using a single telescope.  The pre-processed data set\footnote{Courtesy of Arianna Musso-Barcucci.}contained 1394 frames with integration time of 109 ms. They were binned over 10 successive individual exposures to reduce the computation time, leading to 139 frames with an integration time of 1.09 s. The parallactic angles range from -13$^{\circ}$ to -39$^{\circ}$. A region with a radius of one arcsecond is considered for all three data sets, which corresponds to around 16 $\lambda/D$ for the SPHERE data set and 8 $\lambda/D$ for the NACO and LMIRCam data sets.

\subsection{Results}

The performance assessment of the two forward model versions of the RSM map is done via the estimation of ROC curves. In contrast to the ROC curve usually used for assessing the performance of binary classifiers, the false positive rate (FPR) is replaced by the number of false positive (FP) for the entire frame, averaged over the number of test data sets used for the ROC curve computation (see \cite{Dahlqvist20} and \cite{Gonzalez18}). Synthetic data sets are generated based on the three selected ADI sequences by injecting fake companions at two different angular separations to account for the radial evolution of the noise profile. The known companions and some bright disk structures for the $\beta$ Pictoris data set were removed via the negative fake companion technique \citep{Lagrange10} prior to generating the synthetic data sets. The fake companions, which are simply  defined as the normalised off-axis PSF, are injected at 16 different position angles with five different flux values for a given angular separation.  This allows us to test the sensitivity of the forward model RSM map to different contrasts and mitigates the impact of local speckles on the estimation of the ROC curves. The contrasts used for the three ADI sequences are given in Table \ref{fluxval}. The relatively low contrasts used for the LMIRCam data set arise from the short integration time, the low number of frames after binning as well as the small angular rotation, all of which affect the performance of the PSF subtraction techniques. This provides an interesting way of exploring the algorithm performance in different HCI regimes.

\begin{table}[t]
                        \caption{Injected companions contrasts range for the two considered separations and the three ADI sequences. }
                        \label{fluxval}
\centering

                        \begin{tabular}{llll}
                        \hline
                        \hline
&\textbf{NACO}  & \textbf{SPHERE}  & \textbf{LMIRCam} \\                        
 Separation   & Contrast   & Contrast & Contrast  \\    
 \hline
$2\lambda/D$ & 3.3-8.2 $\times 10^{-4}$ & 1.0-2.6 $\times 10^{-4}$ & 3.4-8.6 $\times 10^{-3}$\\
$8\lambda/D$  & 1.3-3.3 $\times 10^{-5}$ & 2.1-5.2  $\times 10^{-6}$ & 3.4-8.6 $\times 10^{-4}$\\
\hline
                        \end{tabular}
                                \end{table}

We consider a true positive (TP) for a given threshold to be a peak value above the threshold in a circle with a diameter of one FWHM centred on the position of the injected fake companion. A value above the selected threshold at any other location is considered as a FP. In order to avoid double counting, we impose the condition that peak values outside the fake companion region should be separated by a minimal distance of one FWHM.

\subsubsection*{KLIP-FM RSM map}

The forward model version of the KLIP algorithm was developed along a Gaussian matched filter to detect potential planetary candidates in the cube of residuals using the PSF forward model. We propose therefore to compare the performance of the forward model RSM map with the performance of the KLIP forward model matched filter (KLIP-FMMF) developed by \cite{Ruffio17}. We include additionally the original RSM map applied on the cube of residuals generated by the KLIP PSF subtraction techniques and the S/N map obtained with KLIP.  In the last case, the S/N map is generated annulus-wise using the procedure of \citet{Mawet14}, which estimates the S/N by comparing the flux inside a one FWHM aperture centred on the considered pixel with the flux of all the other apertures included in the annulus. This procedure implements a small-sample statistics correction, relying on a student t-test to determine the S/N. Increasing S/N or probability thresholds are applied to generate the different ROC curves for all the considered methods.

\begin{figure*}[h!]
  \centering
  \subfloat[NACO at $2\lambda/D$]{\includegraphics[width=160pt]{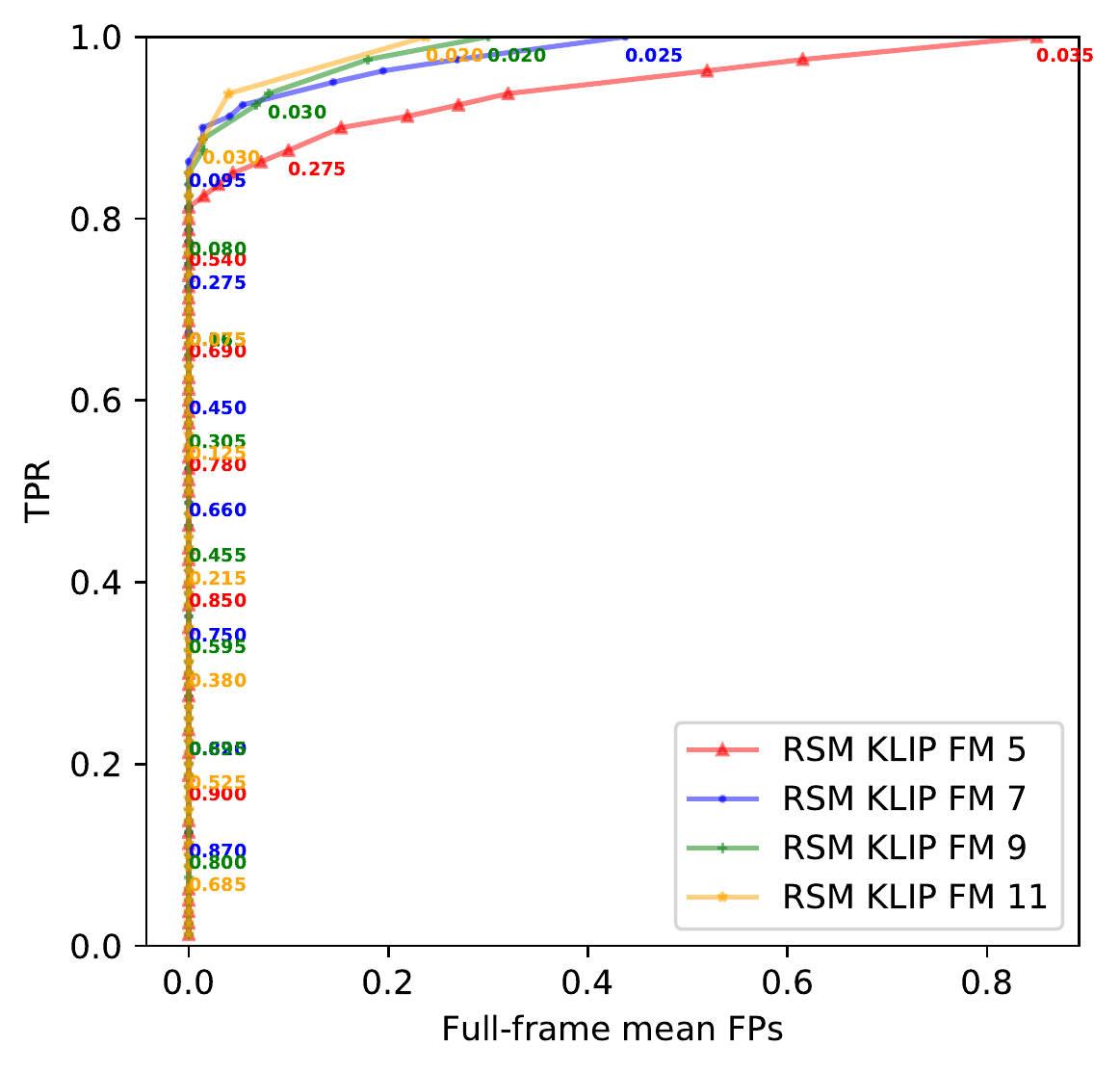}}
  \subfloat[SPHERE at $2\lambda/D$]{\includegraphics[width=160pt]{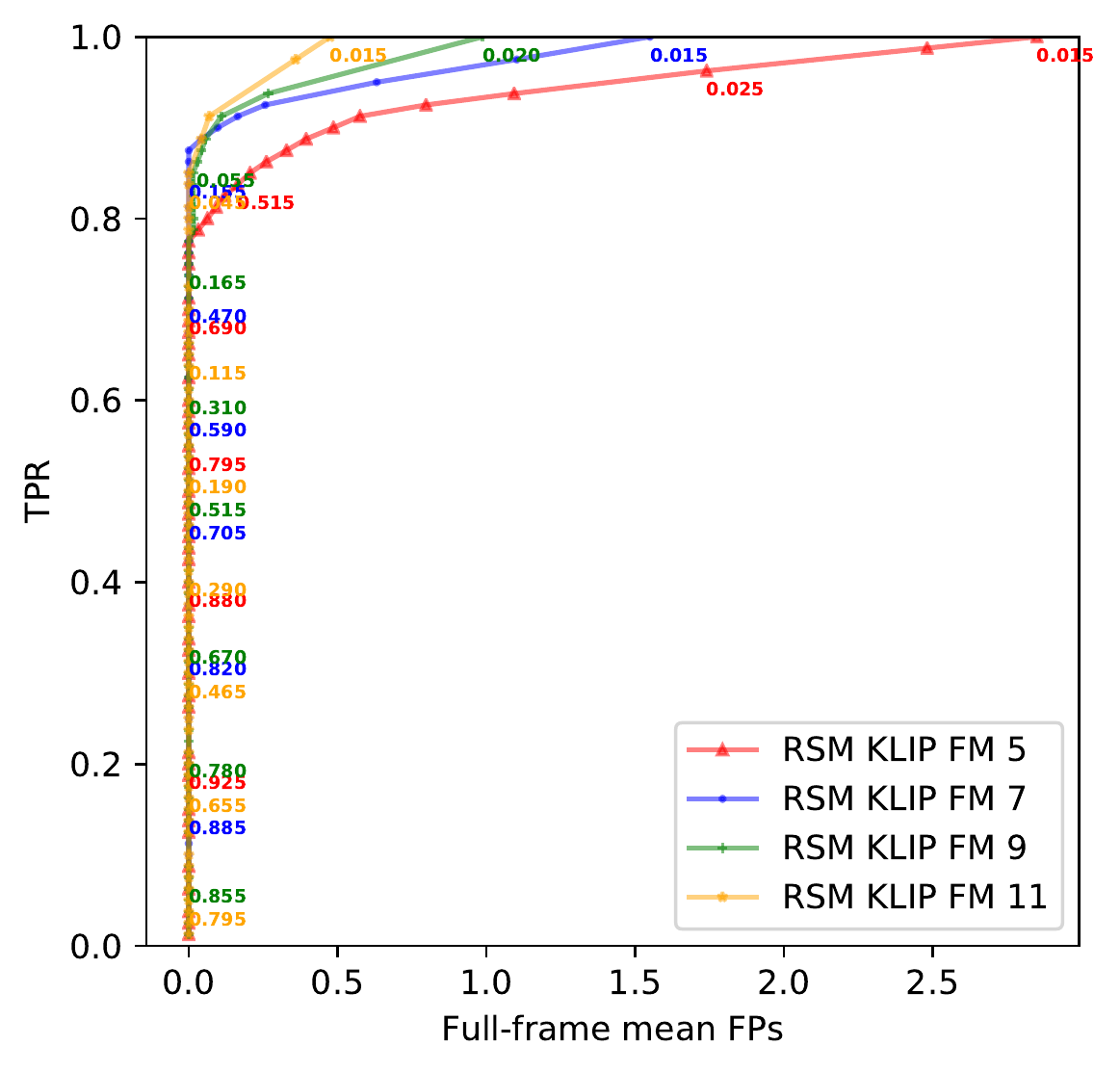}}
  \subfloat[LMIRCam at $2\lambda/D$]{\includegraphics[width=160pt]{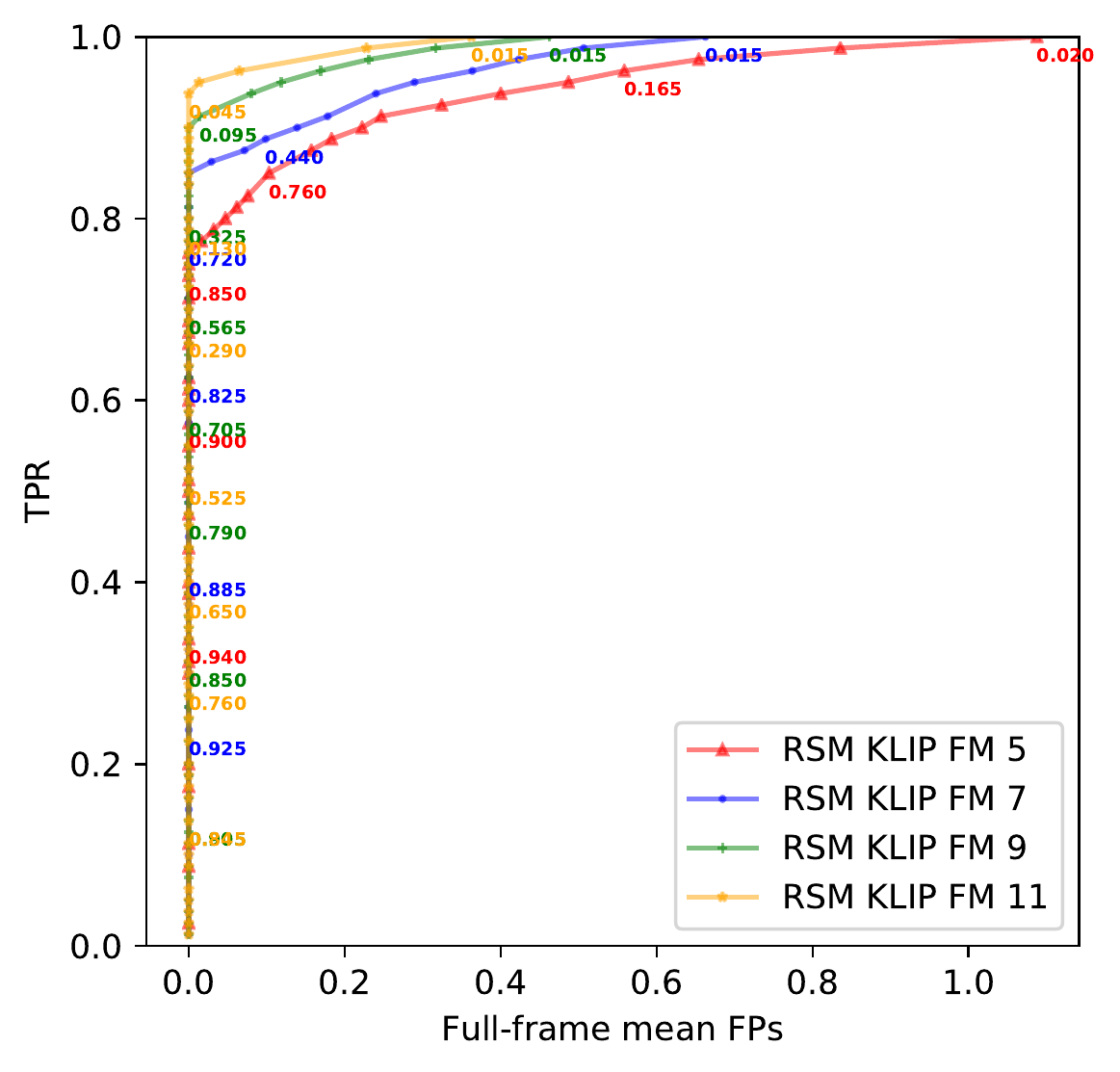}}\\
    \subfloat[NACO at $8\lambda/D$]{\includegraphics[width=160pt]{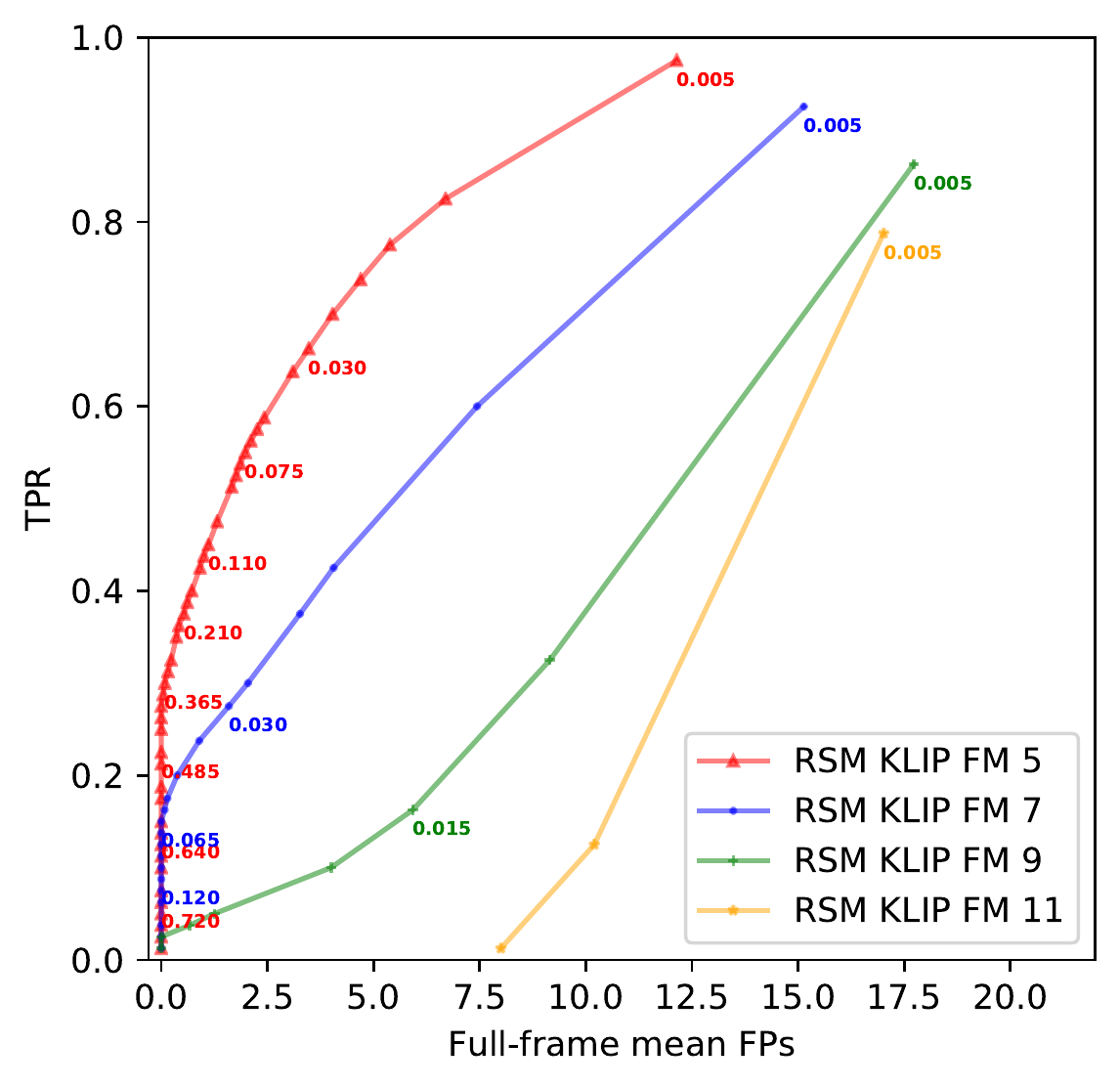}}
  \subfloat[SPHERE at $8\lambda/D$]{\includegraphics[width=160pt]{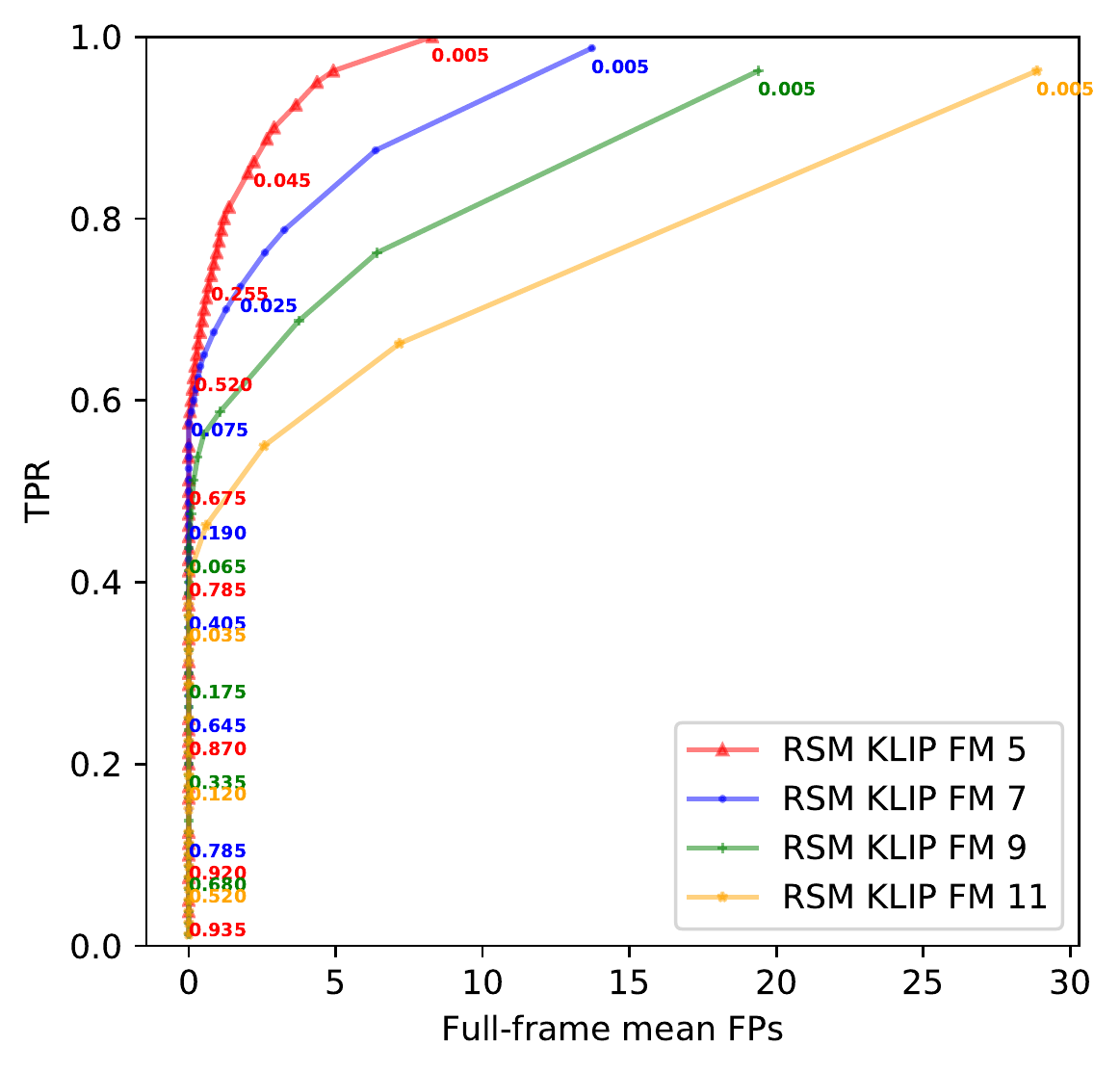}}
  \subfloat[LMIRCam at $8\lambda/D$]{\includegraphics[width=160pt]{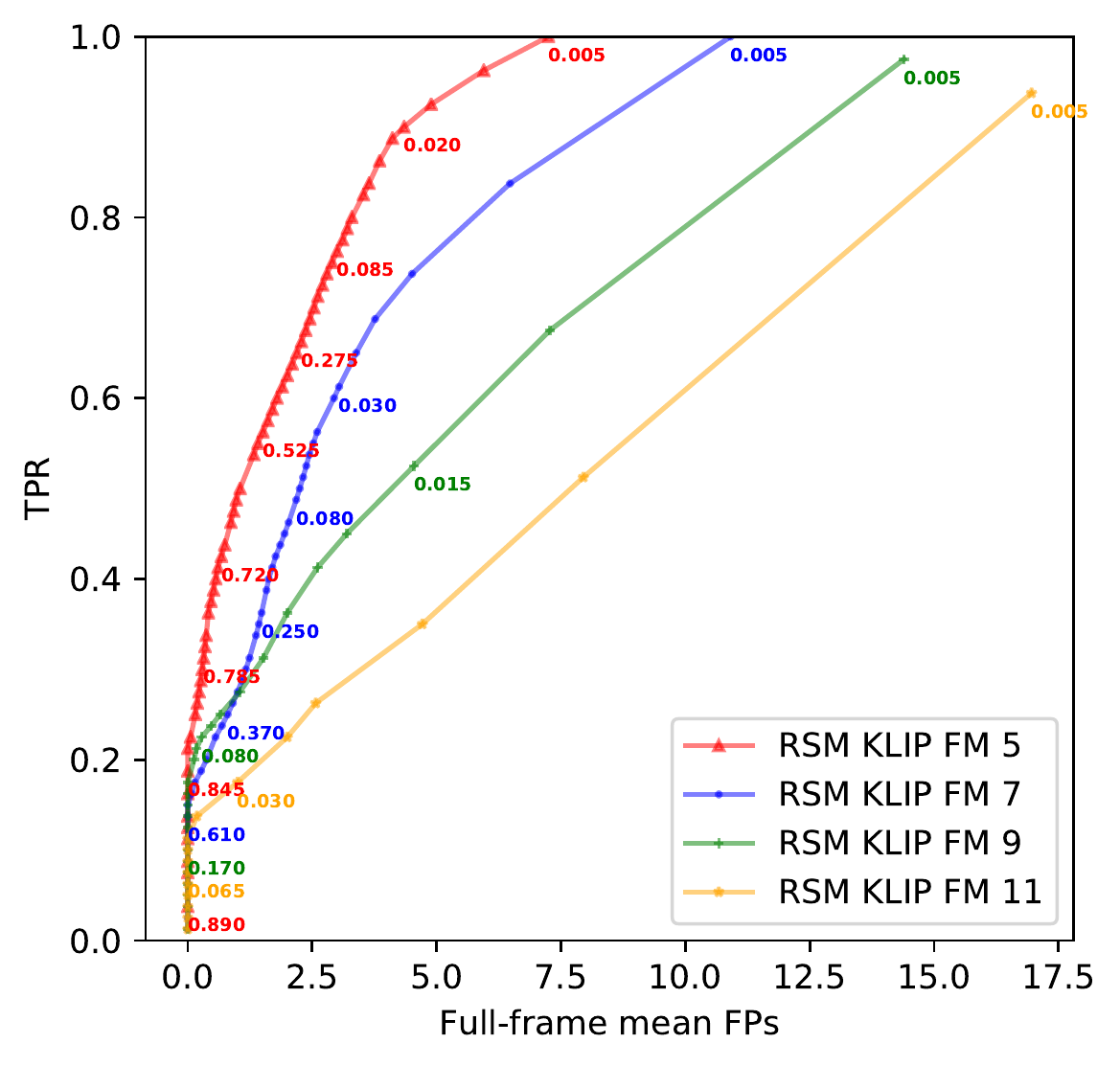}}
  
  \caption{\label{WSKLIP} ROC curves for the NACO, SPHERE, and LMIRCam data sets, with the KLIP-FM RSM map using respectively a crop size for the froward modelled PSF of 5 (red), 7 (blue), 9 (green), 11 (orange) pixels (FWHM$\approx 5$ pixels for all three data sets).}
\end{figure*}

The parameters of the KLIP algorithm, namely, the number of principal components and minimum FOV rotation, were selected to optimise the ROC curves\footnote{We mean by optimizing the ROC curve, maximizing the true positive rate (TPR) while minimizing the number of FPs for the set of considered thresholds.} for the two considered angular separations. A single set of parameters was defined for each data set. The number of principal components was set to 20 for the SPHERE and NACO data sets while a value of 18 principal components was chosen for the LMIRCam data set. The FOV rotations expressed in terms of FWHM are respectively 0.5, 0.3, and 0.3. 

We start by considering for all three data sets the impact of the selected crop size for the forward-modelled PSF used in the KLIP-FM RSM map. As can be seen from Fig.~\ref{WSKLIP}, the larger crop sizes seem to outperform the crop size of one FWHM for the small angular separation while the reverse is true for the large angular separation. However we observe a much larger gap for the largest separation, especially in the case of the NACO data set. This may be explained by the reduced self-subtraction observed at large angular separations, the movement of potential astrophysical signals increasing linearly with the angular separation. This implies that the brightness of the negative lobes appearing on the sides of the main peak reduces with the angular separation. This makes the larger crop sizes unnecessary and more prone to speckle noise.

\begin{figure}[t]
  \centering
  \subfloat[First frame]{\includegraphics[trim=12 10 0 10, clip,width=130pt]{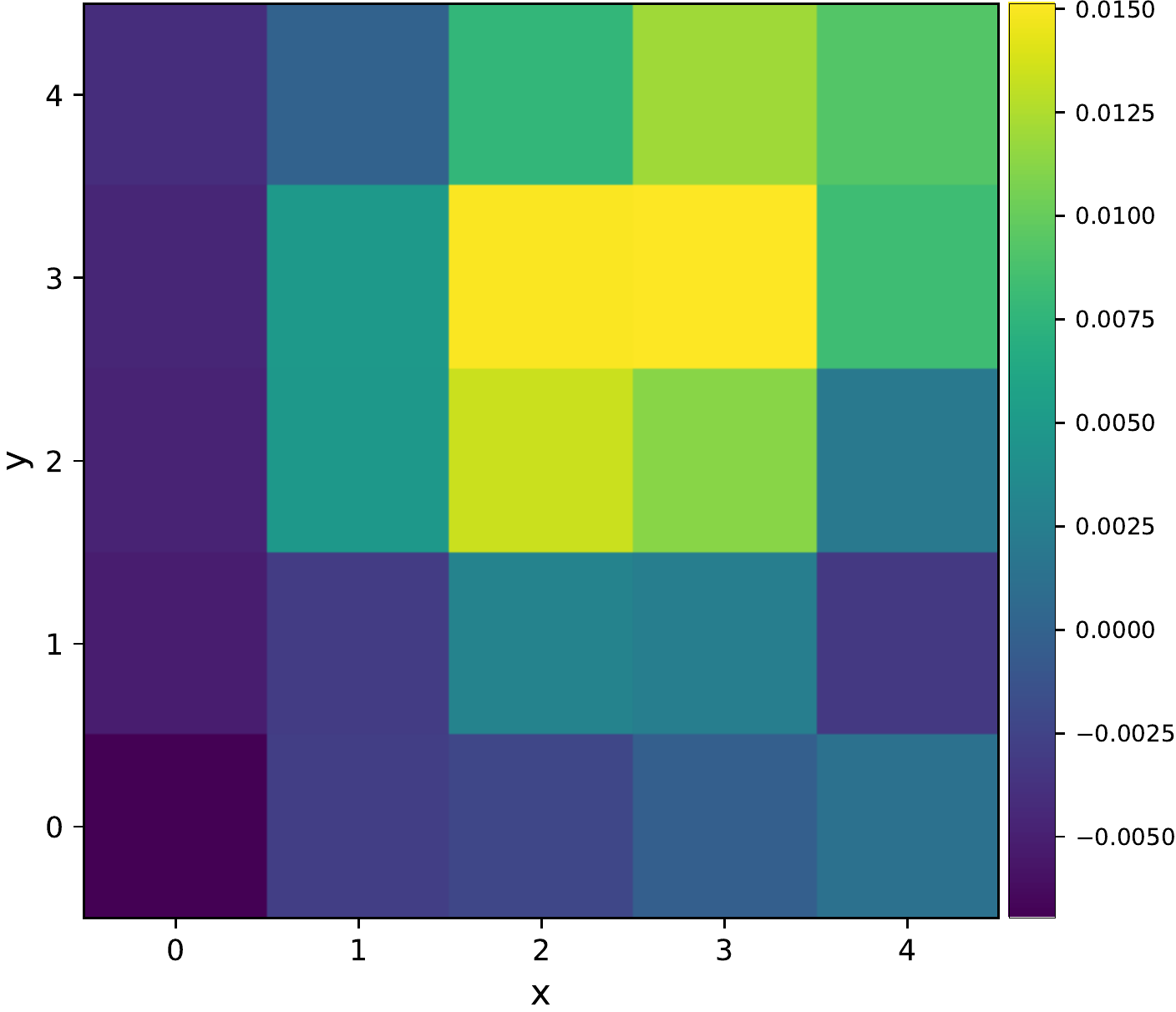}}
  \subfloat[Last Frame]{\includegraphics[trim=12 10 0 10, clip,width=130pt]{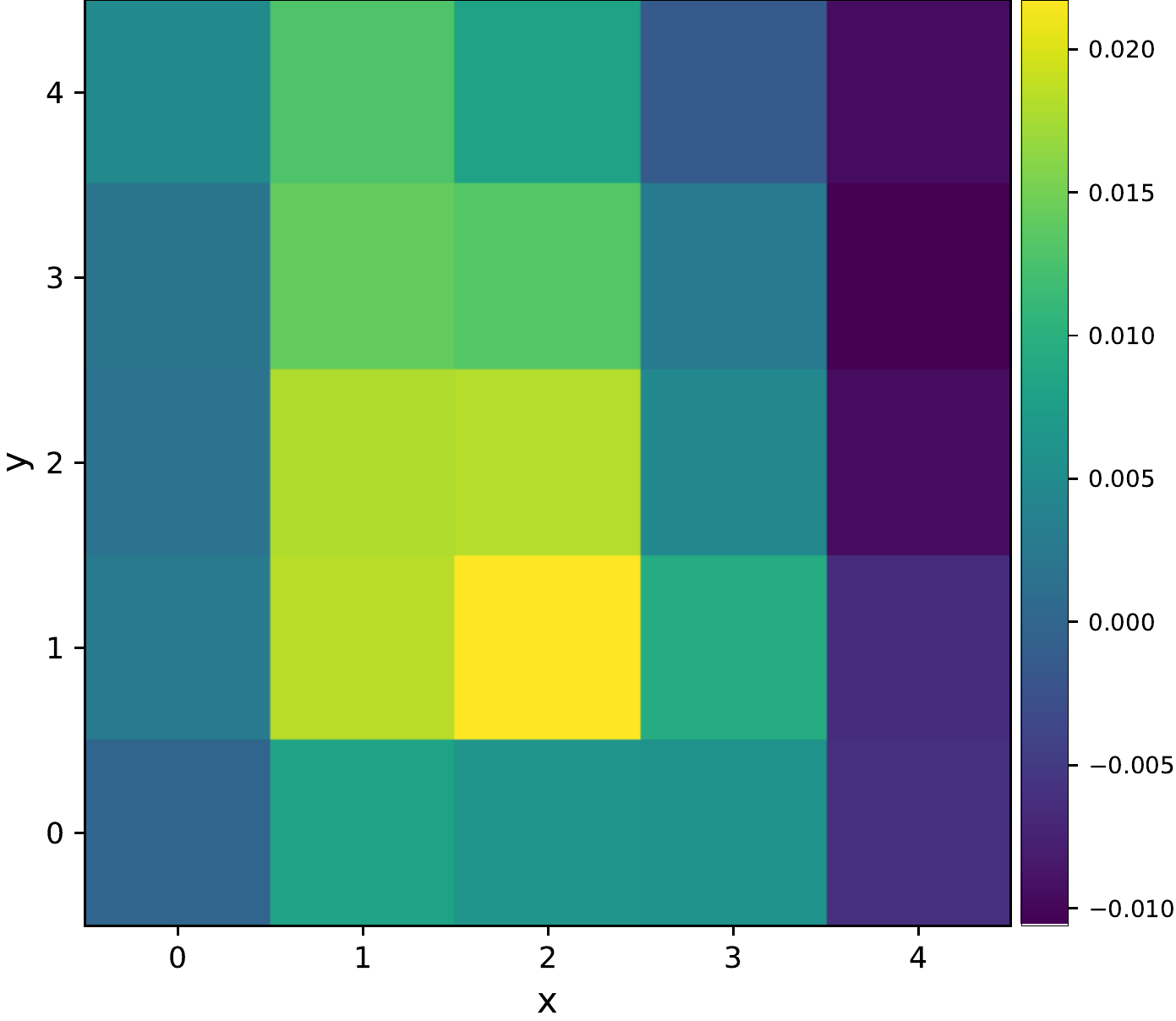}}
  
  \caption{\label{FMPSF} KLIP forward-modelled PSF for the NACO $\beta$ Pictoris ADI sequence taken at a distance of $2\lambda/D$ with the same azimuthal orientation as in Fig.~\ref{fig1} and cropped at one FWHM. The two images correspond respectively to the first and the last frame.}
\end{figure}

\begin{figure*}[h!]
  \centering
  \subfloat[NACO at $2\lambda/D$]{\includegraphics[width=160pt]{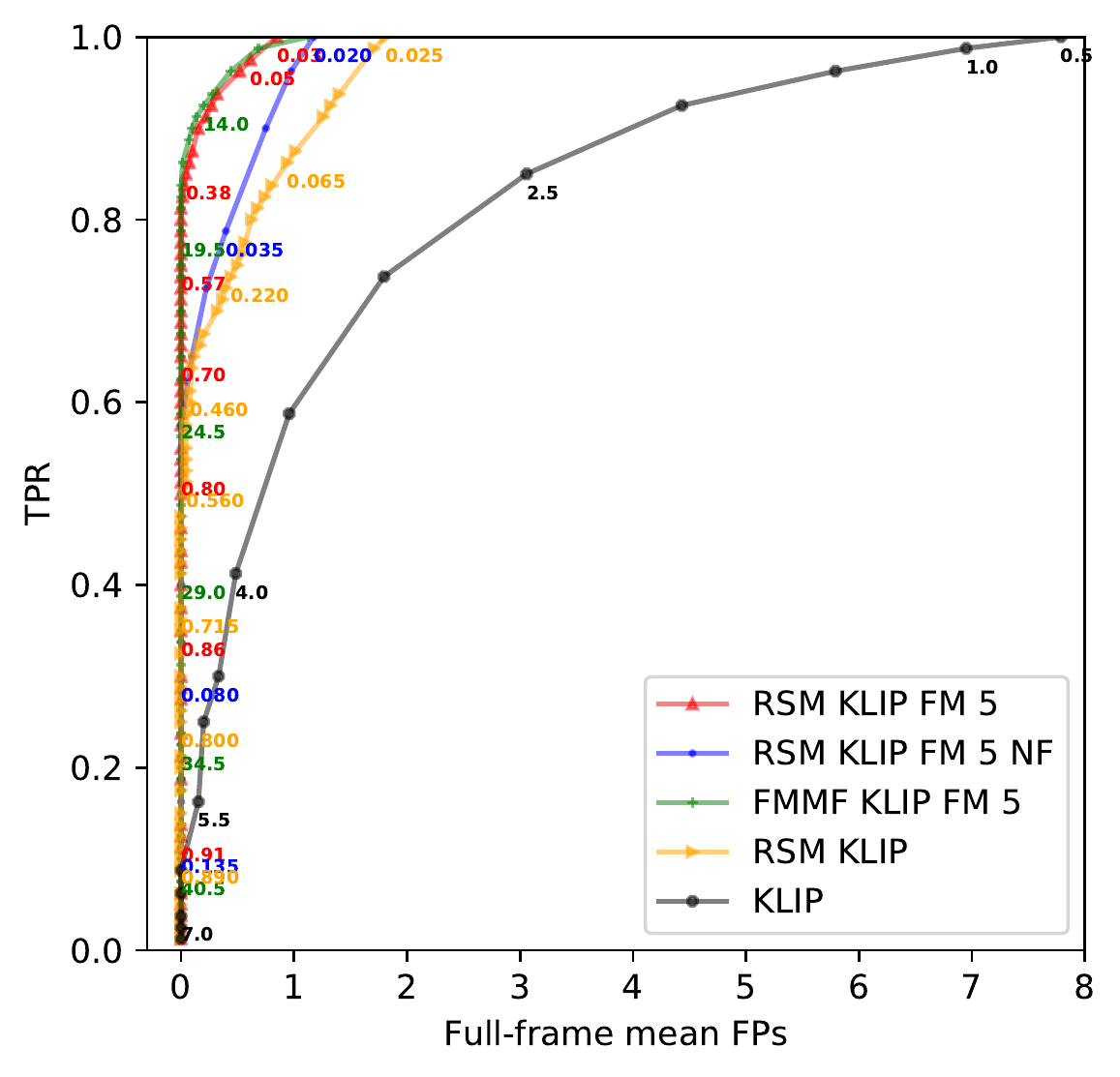}}
  \subfloat[SPHERE at $2\lambda/D$]{\includegraphics[width=160pt]{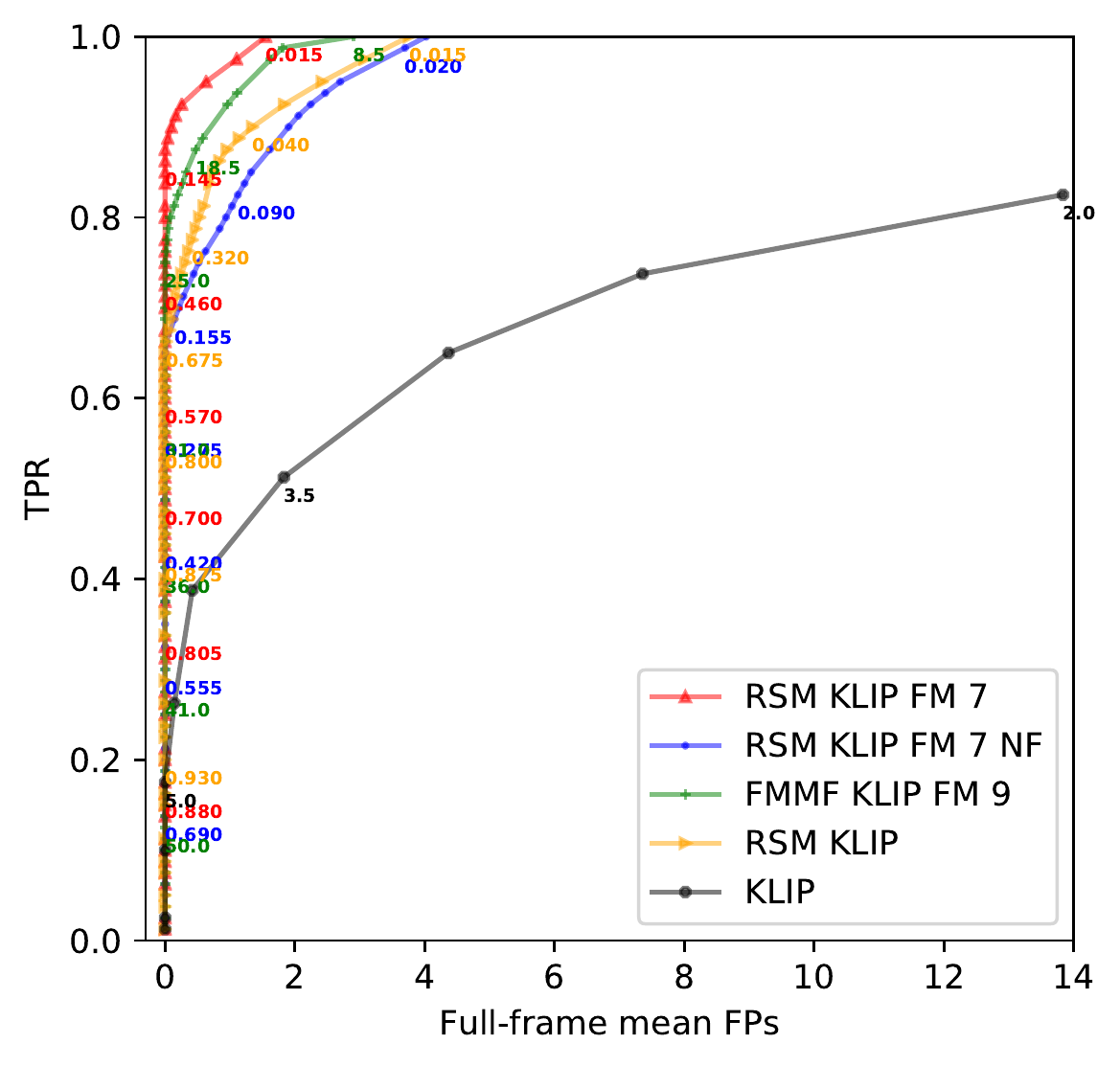}}
  \subfloat[LMIRCam at $2\lambda/D$]{\includegraphics[width=160pt]{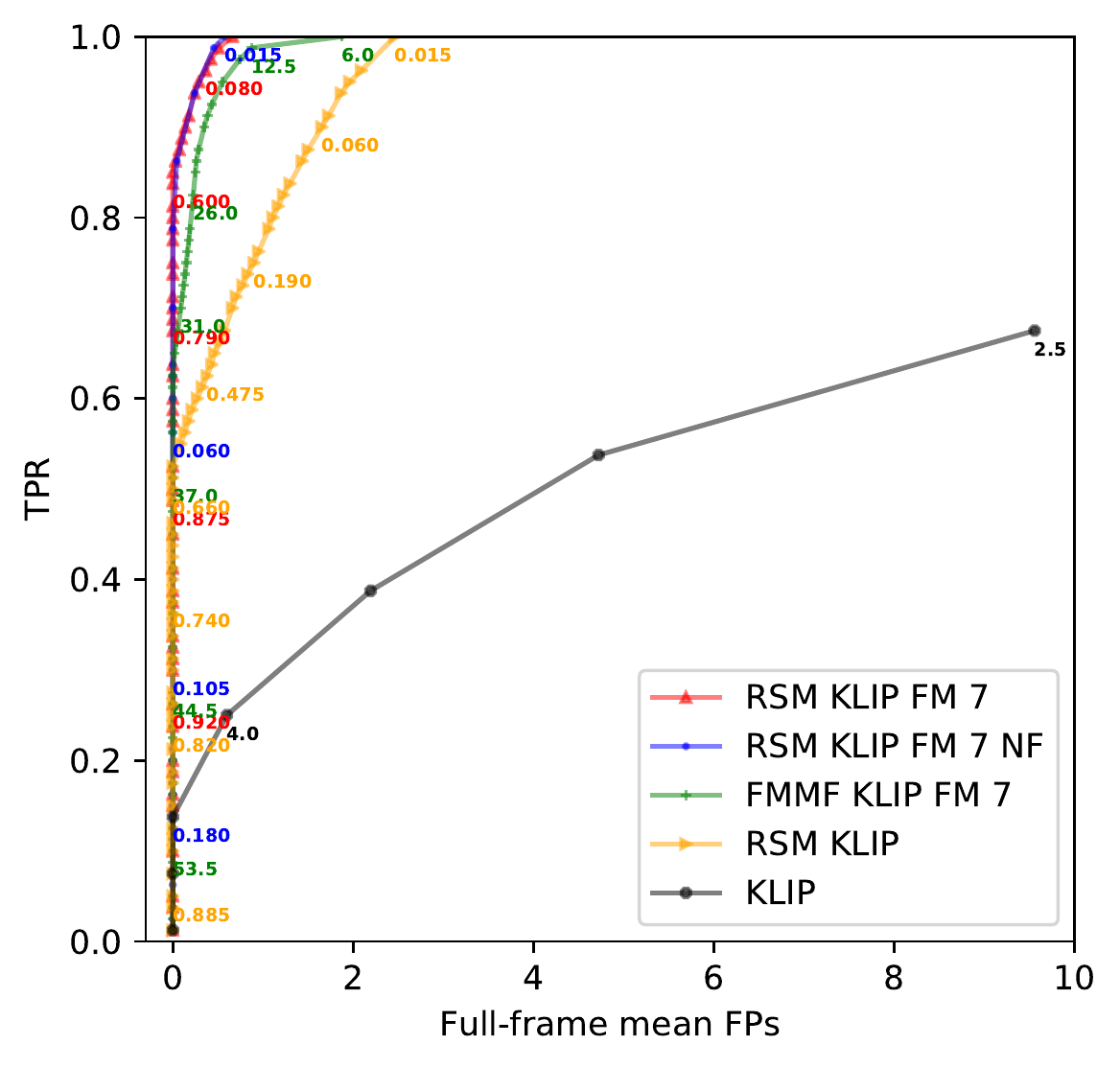}}\\
    \subfloat[NACO at $8\lambda/D$]{\includegraphics[width=160pt]{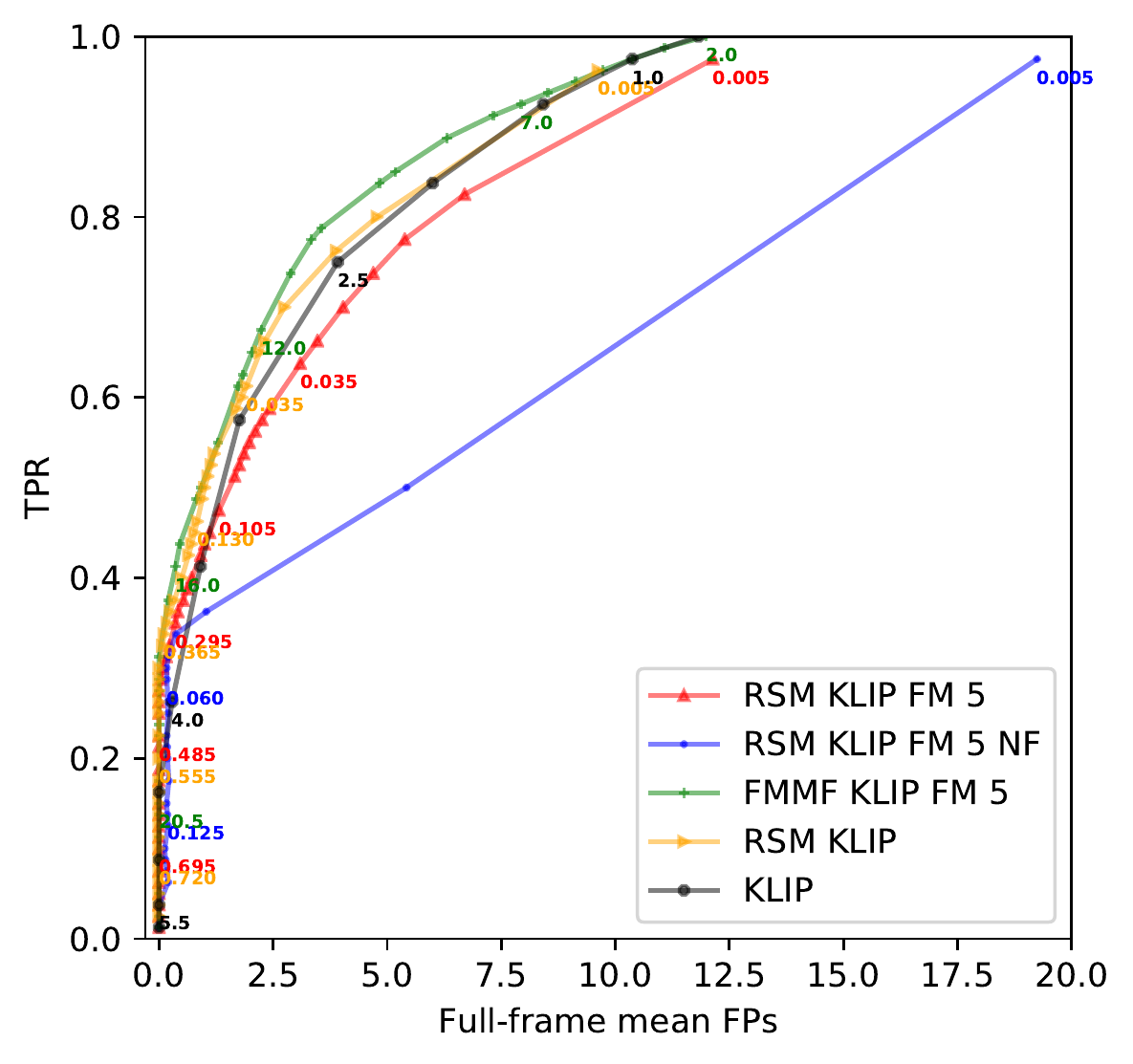}}
  \subfloat[SPHERE at $8\lambda/D$]{\includegraphics[width=160pt]{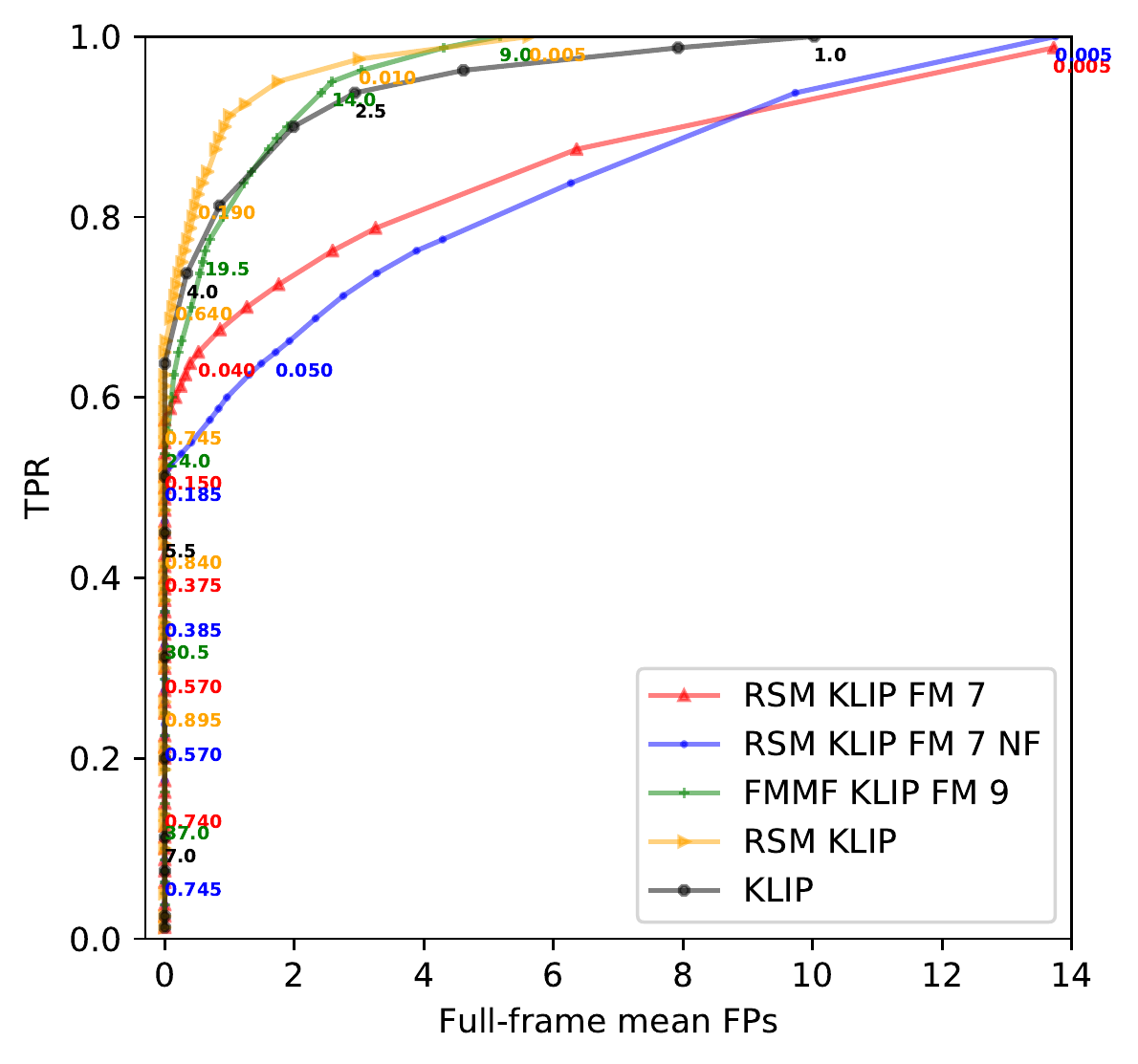}}
  \subfloat[LMIRCam at $8\lambda/D$]{\includegraphics[width=160pt]{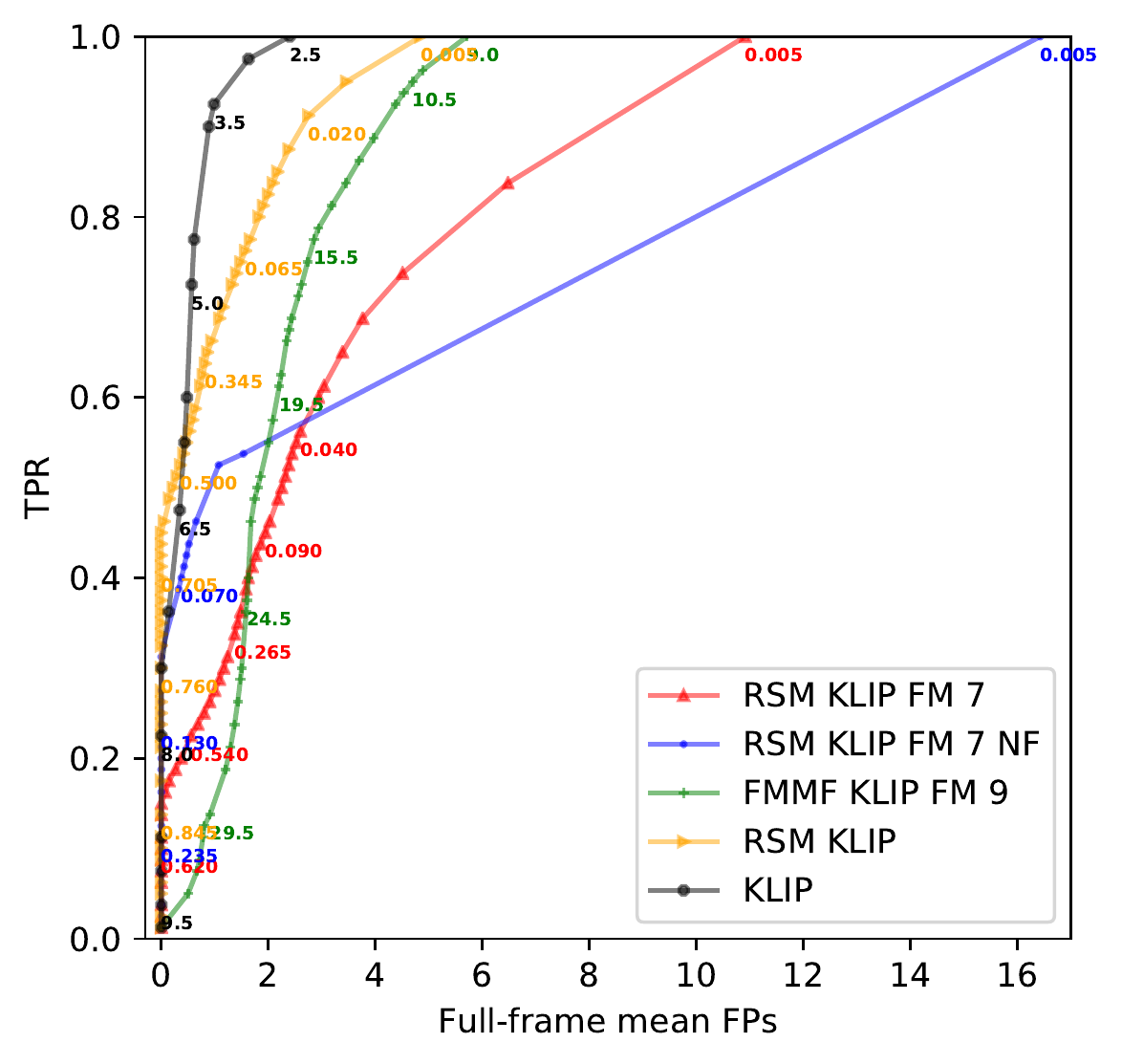}}
  
  \caption{\label{FMKLIP} ROC curves for the NACO, SPHERE and LMIRCam data sets, with respectively, the KLIP-FM RSM map using the Gausssian maximum likelihood for the pre-optimisation of the flux parameter $\beta$ (red), the KLIP-FM RSM map with no flux pre-optimisation (NF), which relies on the standard maximum likelihood used in the original RSM map for the estimation of flux parameter $\beta$ (blue), the forward model matched filter KLIP-FMMF (green), the RSM map using KLIP (orange) and KLIP using the standard S/N map (black).}
\end{figure*}

Considering the previous results, we select for the NACO data set a crop size of five pixels (one FWHM) and a crop size of seven pixels for the other two data sets. This provides a good performance trade-off between small and large angular separations.  As can be seen from Fig.~\ref{FMPSF}, a crop size of one FWHM still captures part of the negative wing azimuthal translation. The results for the two versions of KLIP-FM RSM map, as well as KLIP-FMMF, KLIP RSM map  and the KLIP using S/N map are given in Fig.~\ref{FMKLIP}. We see from these plots that, at small separation, the KLIP-FM RSM map seems to slightly outperform KLIP-FMMF, while the reverse is true at large separation. The KLIP  approach using S/N map has a higher ability to detect faint companions at large radial distances but it is no match to the other methods at small separations. The KLIP RSM map provides surprisingly good results, being often the closest to the KLIP S/N map for large separations and being relatively close to KLIP-FM RSM and KLIP-FMMF at $2\lambda/D$ from the host star. A combination of both KLIP RSM and KLIP-FM RSM could be interesting to keep the high sensitivity of KLIP-FM RSM at close separations while improving the sensitivity at larger radial distances. It seems also clear from Fig.~\ref{FMKLIP} that the version of the KLIP-FM RSM map using the Gaussian approximation for estimating the flux parameter $\beta$ (Eq.~\ref{gausslike}) outperforms the one relying on the maximum likelihood approach used in \cite{Dahlqvist20}, providing in all cases equivalent or better results. We also tested the maximum likelihood based approach and the Gaussian approximation with the KLIP RSM map with similar results (see Appendix B for a comparison between the two approaches in the case of KLIP RSM), demonstrating the efficiency of this new way of estimating $\beta$ on top of its faster estimation.

\subsubsection*{LOCI-FM RSM map}

\begin{figure*}[h!]
  \centering
  \subfloat[NACO at $2\lambda/D$]{\includegraphics[width=160pt]{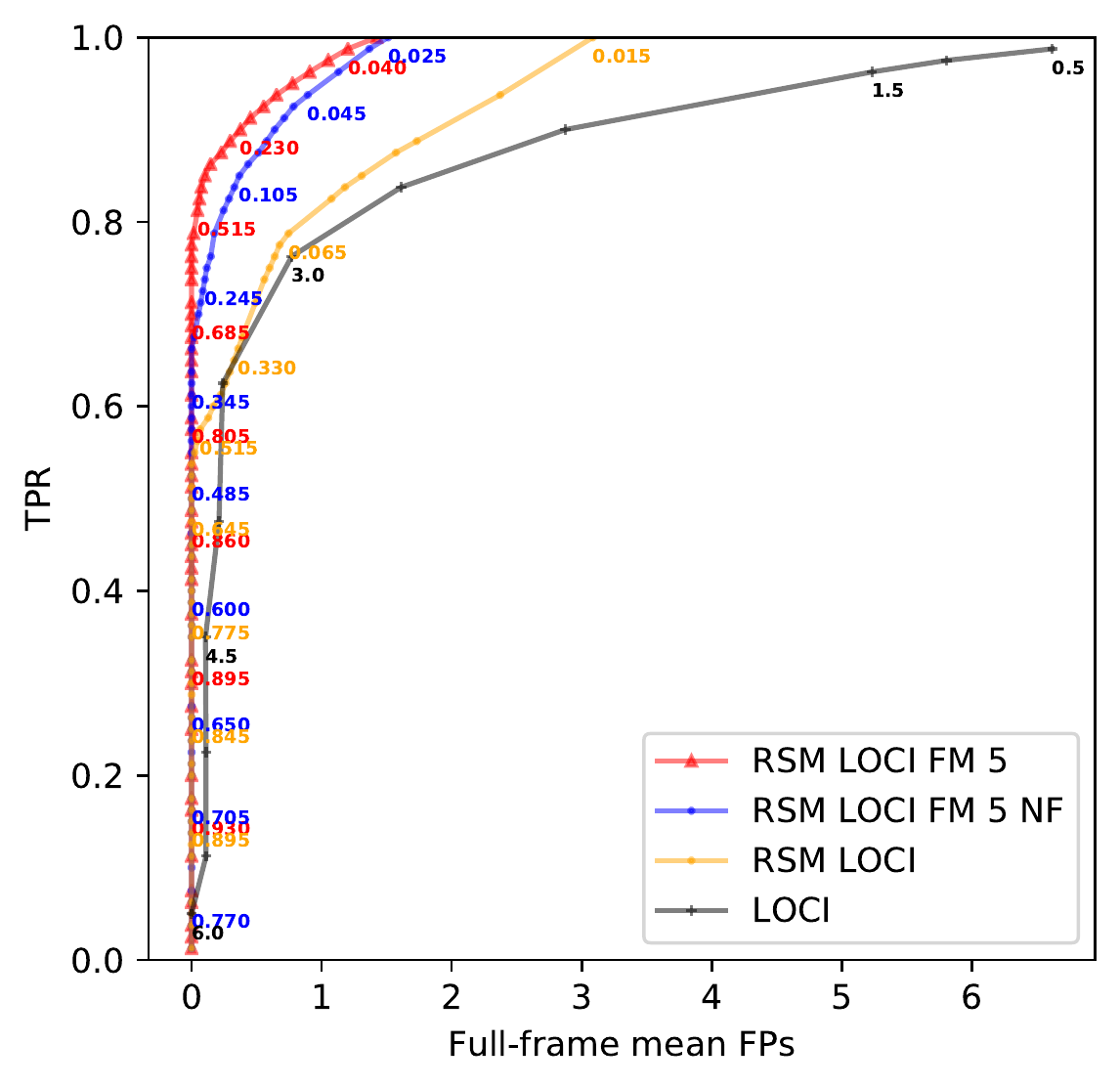}}
  \subfloat[SPHERE at $2\lambda/D$]{\includegraphics[width=160pt]{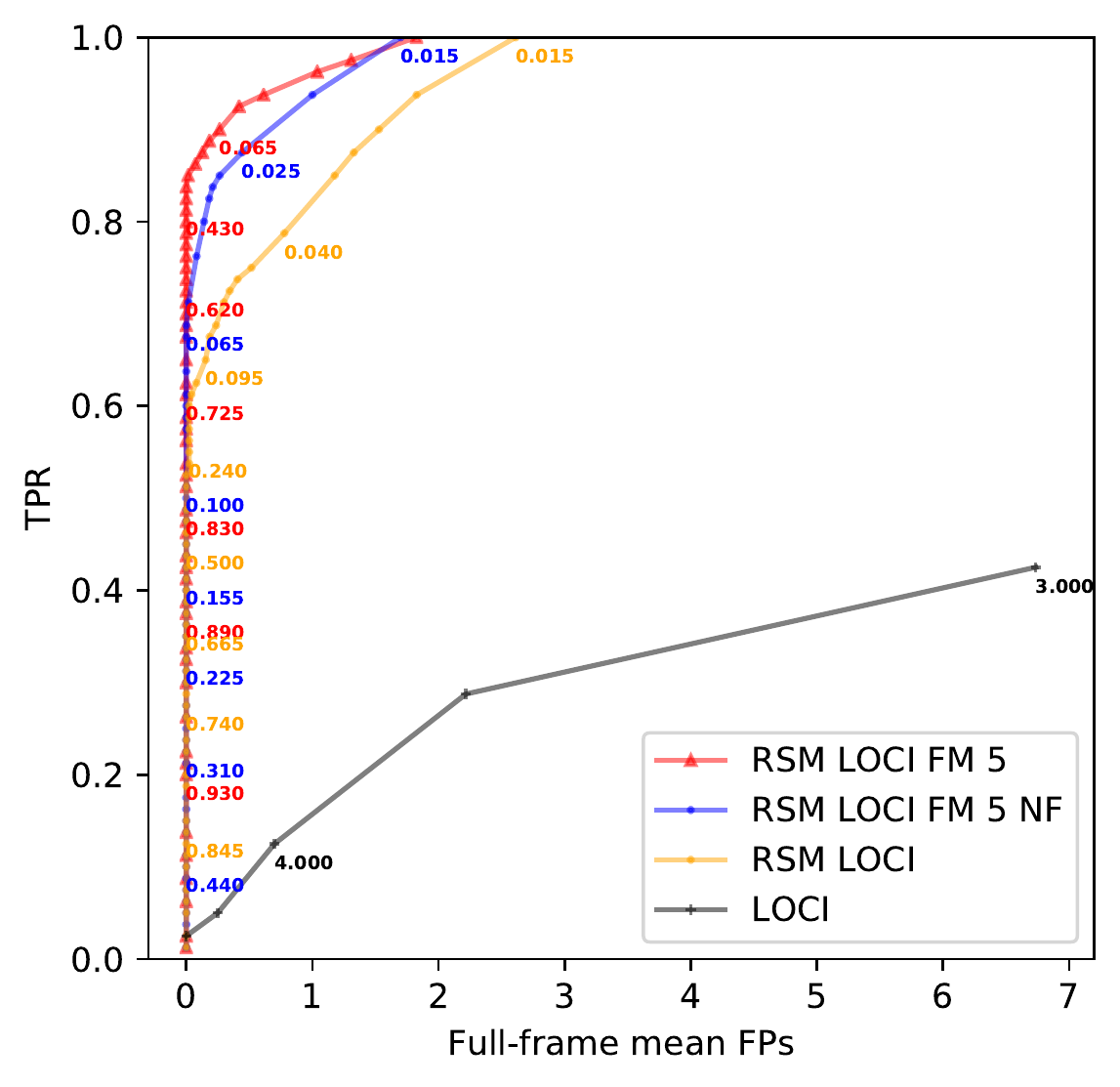}}
  \subfloat[LMIRCam at $2\lambda/D$]{\includegraphics[width=160pt]{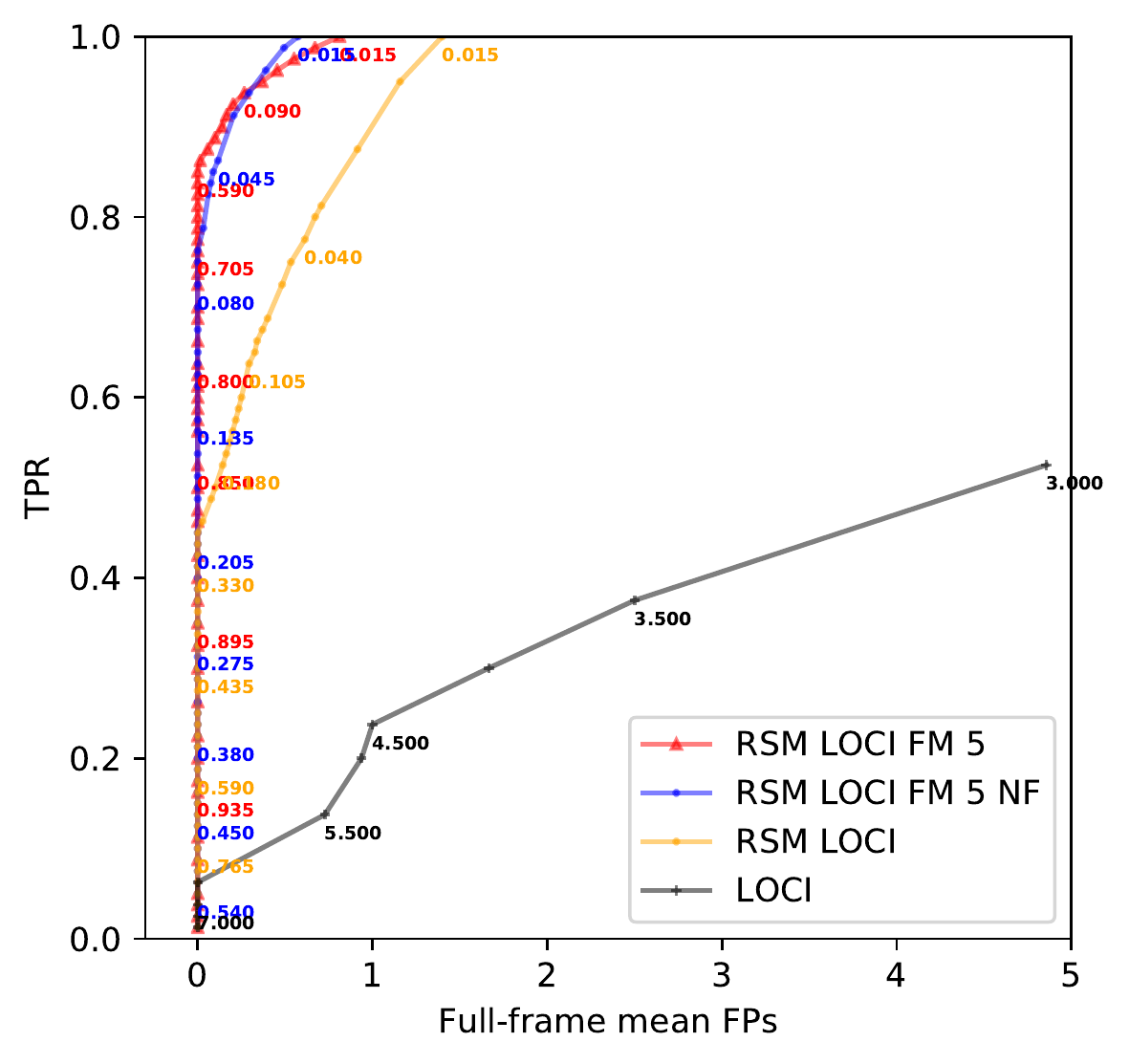}}\\
    \subfloat[NACO at $8\lambda/D$]{\includegraphics[width=160pt]{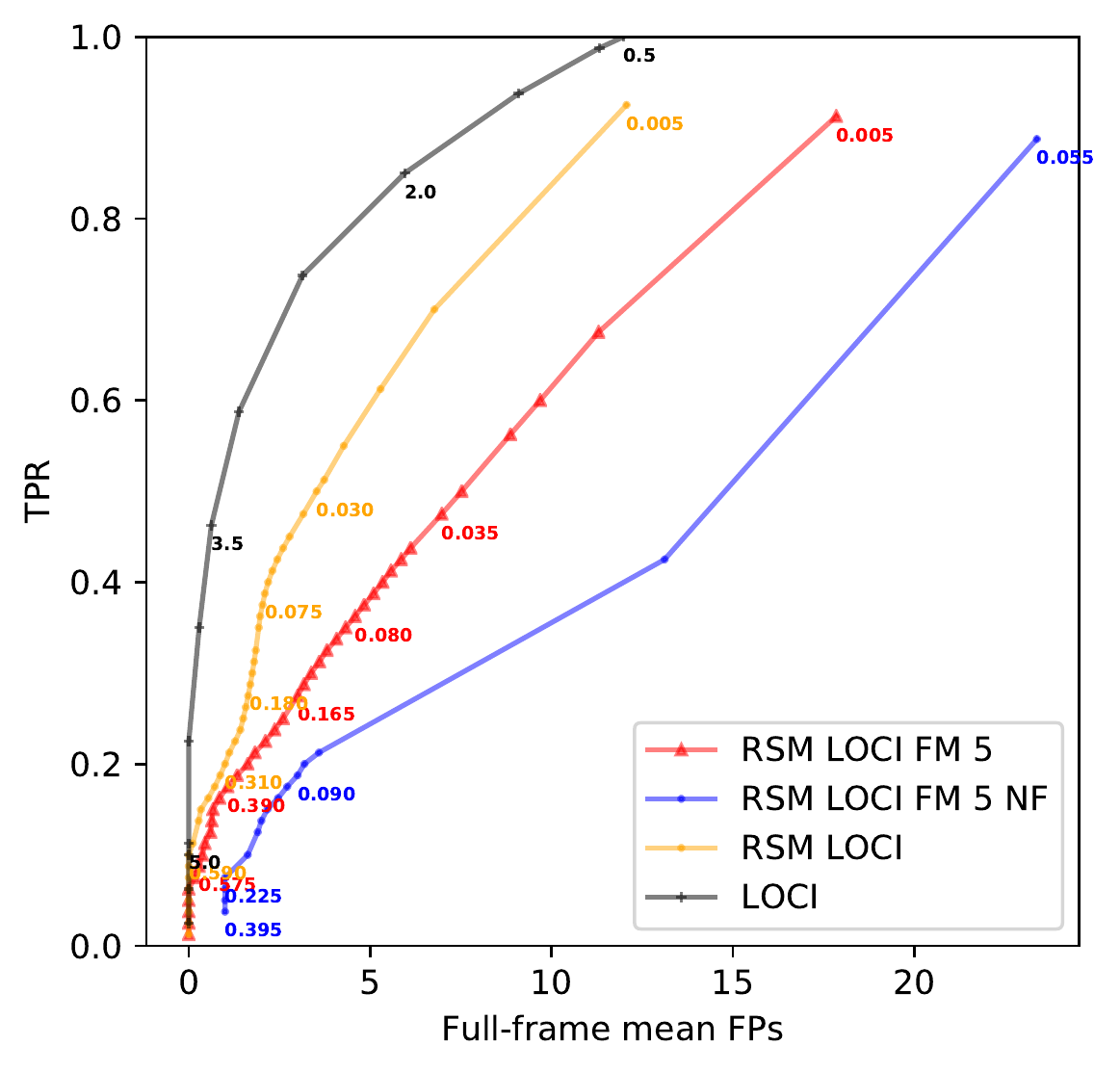}}
  \subfloat[SPHERE at $8\lambda/D$]{\includegraphics[width=160pt]{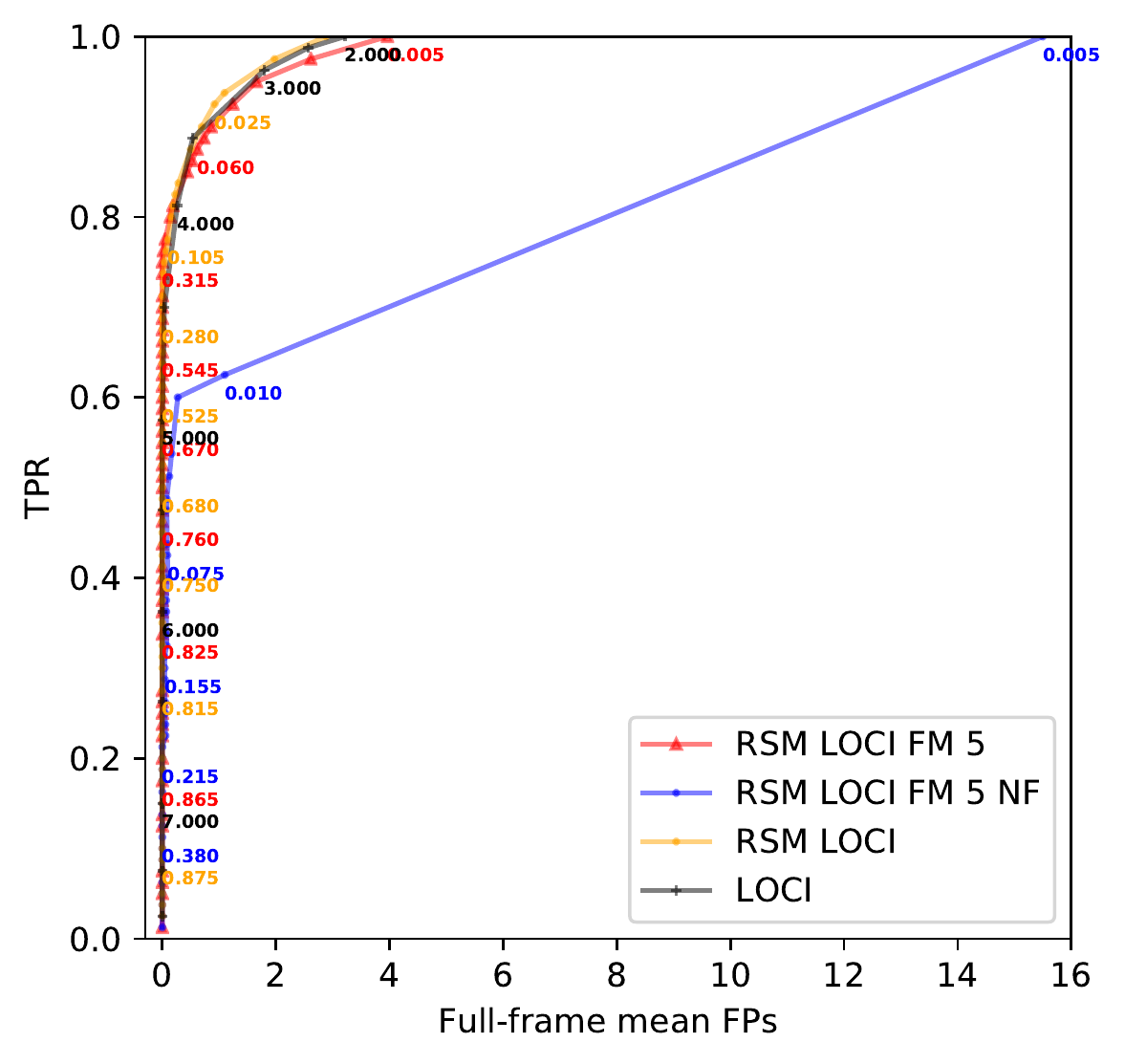}}
  \subfloat[LMIRCam at $8\lambda/D$]{\includegraphics[width=160pt]{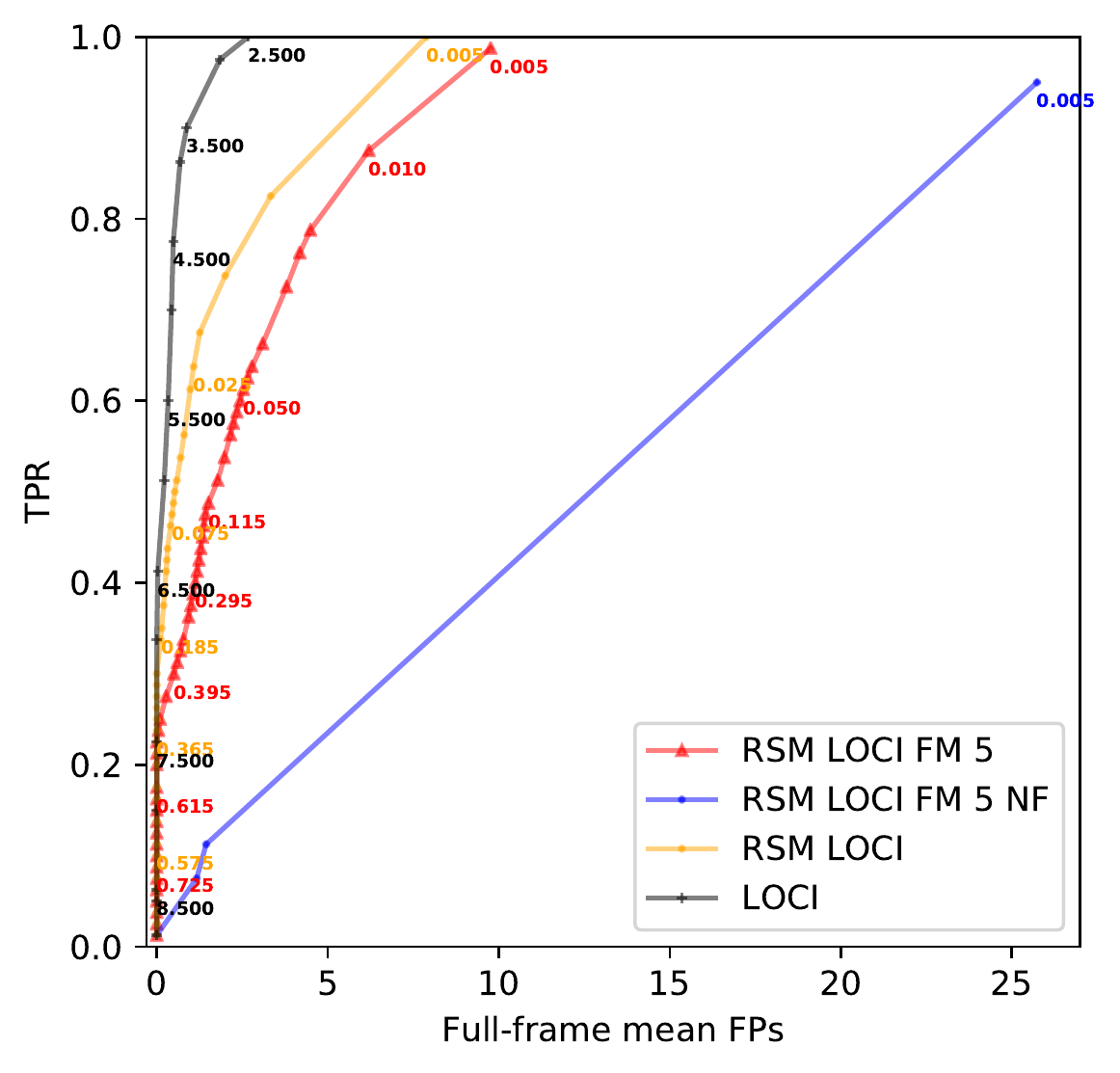}}
  
  \caption{\label{FMLOCI} ROC curves for the NACO, SPHERE and LMIRCam data sets, with, respectively, the LOCI-FM RSM using the Gausssian maximum likelihood for the pre-optimisation of the flux parameter $\beta$ (red), the LOCI-FM RSM map with no flux pre-optimisation (NF), which relies on the standard maximum likelihood used in the original RSM map for the estimation of flux parameter $\beta$ (blue), the RSM map using LOCI (orange) and LOCI using the standard S/N map (black).}
\end{figure*}

Turning to the RSM LOCI FM map, the tolerance level for the square-residuals minimisation and the minimum FOV rotation were also selected to provide the best overall performance. A tolerance of 9 $\times 10^{-3}$ was chosen for NACO and SPHERE and a tolerance of 1 $\times 10^{-2}$ for LMIRCam. The minimum FOV rotations are respectively, $0.6$, $0.2$ and $0.2$ FWHM. The analysis of the ROC curves obtained with different crop sizes leads to similar conclusions to the case of KLIP-FM RSM. The crop size of one FWHM performs better, in a global sense, even though larger crop sizes lead to slightly better results at small angular separations. The ROC curves corresponding to the crop sizes performance comparison are presented in Appendix C. Regarding the performance of LOCI-FM RSM, the results in Fig.~\ref{FMLOCI} demonstrate again the interest of the Gaussian maximum likelihood to define the flux parameter $\beta$ mainly for the largest separation. Both LOCI-FM RSM and LOCI RSM outperform clearly the LOCI S/N map for the $2\lambda/D$ angular separation, while the reverse is true for the $8\lambda/D$ angular separation. The ordering is similar to that of the KLIP case with LOCI-FM RSM leading at $2\lambda/D$ and LOCI RSM being closer to the LOCI S/N map at $8\lambda/D,$ which again seems to favour a combination of both the LOCI and LOCI-FM to benefit from their respective strength. The search for an optimal mix between the different PSF subtraction techniques is investigated in the next section.

\section{Optimal PSF subtraction techniques selection}
\label{optimix}

Having demonstrated the added value of the forward model versions of the RSM map, at least at small angular separations, we are now left with five different PSF subtraction techniques (annular PCA, KLIP, NMF, LLSG, LOCI) plus two forward model versions to generate the RSM maps. Given that the Annular PCA and KLIP are relatively close in their definition, we decided to focus solely on KLIP, as preliminary results demonstrated their similarities in terms of performance and their non-complementarity. We address in this section the difficult question of optimally selecting these PSF subtraction techniques to optimise the overall performance of the resulting RSM maps. In particular, we investigate the dependence of  the optimal combination on the instrument and radial distance.  We rely again on ROC curves to assess the performance of the various combinations we considered. 

In order to speed up the multiple RSM map estimations, we slightly modified the original RSM map procedure as presented in \cite{Dahlqvist20}, with, however, no impact on the final outcome of the algorithm. We divided the procedure into two separate steps, the first one being the estimation of the likelihood provided in Eq.~\ref{like} and the second one the estimation of the probability of being in the planetary regime given by Eq.~\ref{proba}. A separate likelihood cube is estimated for every considered PSF subtraction technique for the entire set of annuli. Some of these likelihood cubes are then stacked along the time axis depending on the selected combination. The probabilities are eventually estimated annulus-wise for every pixel of every frame and averaged along the time axis to generate the final probability map\footnote{This architecture is implemented in the PyRSM python package, which includes all the developments presented here, and is available on GitHub: \url{https://github.com/chdahlqvist/RSMmap}.}. This allows us to estimate only once the likelihood cubes for the different PSF subtraction techniques, with the second step, which is also the fastest, as the only one to be repeated for each combination. The parametrisation of the underlying PSF subtraction techniques were selected to maximise the overall performance for each data set.

\begin{figure*}[h!]
  \centering
  \subfloat[NACO at $2\lambda/D$]{\includegraphics[width=160pt]{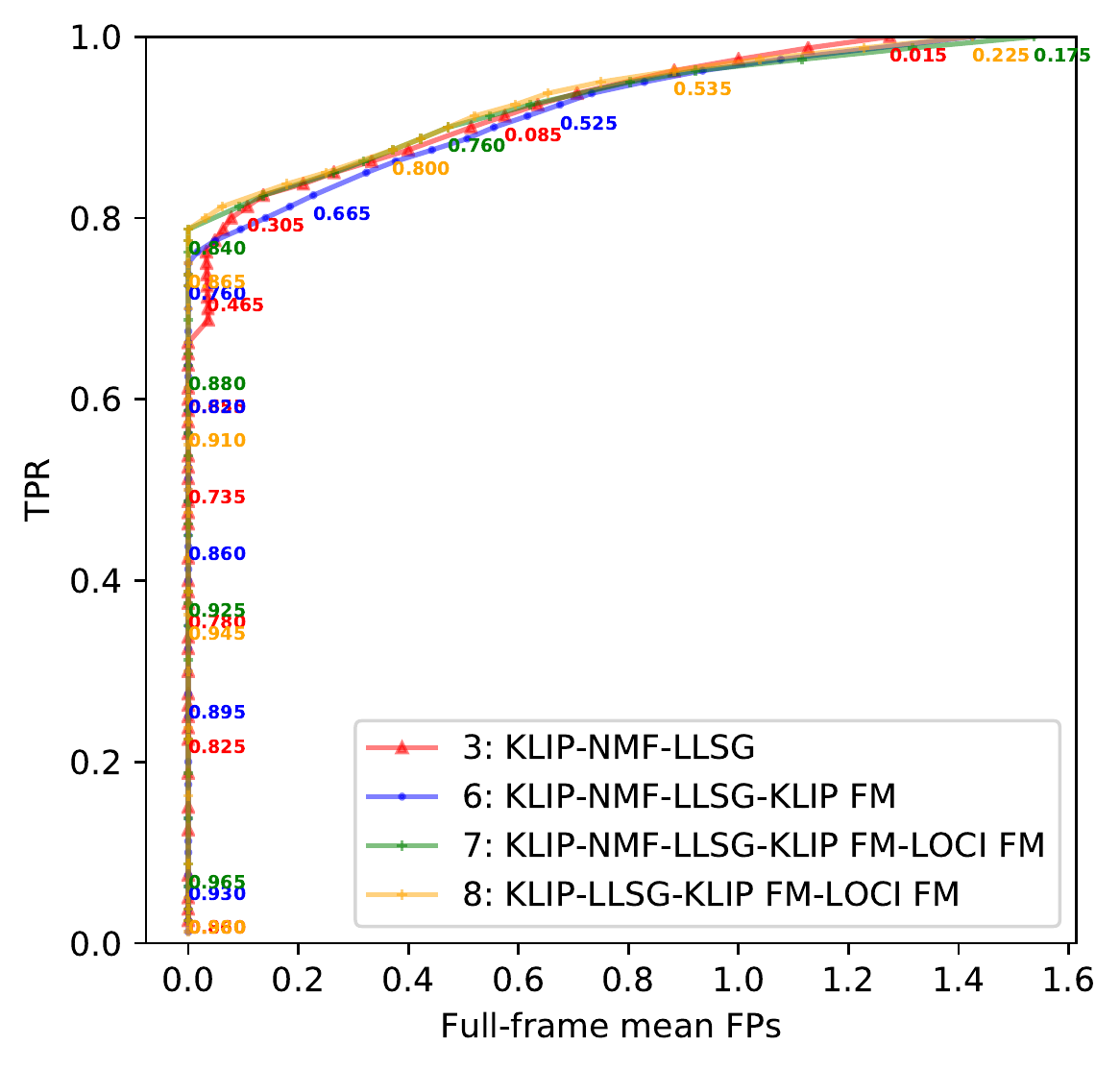}}
  \subfloat[SPHERE at $2\lambda/D$]{\includegraphics[width=160pt]{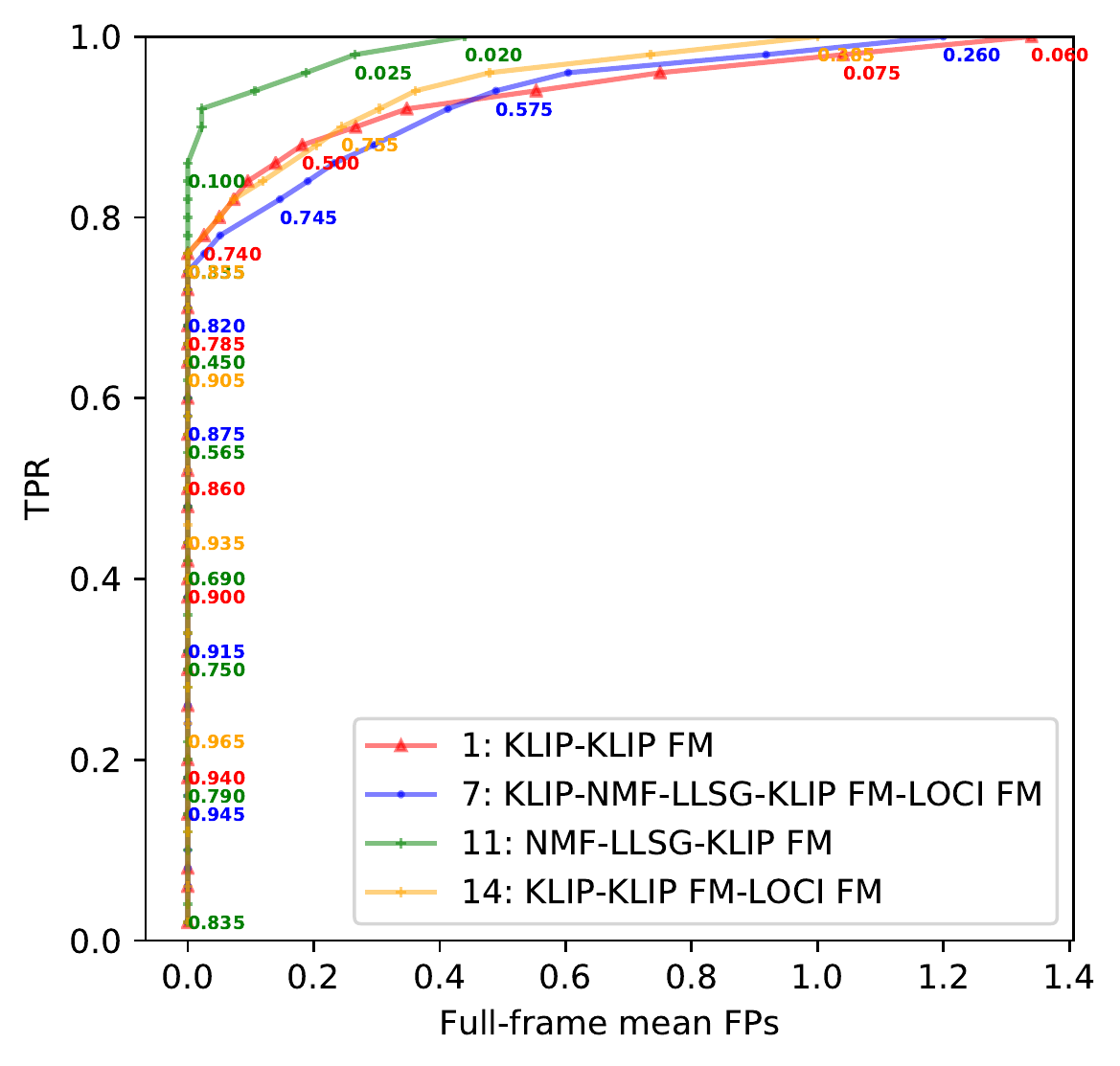}}
  \subfloat[LMIRCam at $2\lambda/D$]{\includegraphics[width=160pt]{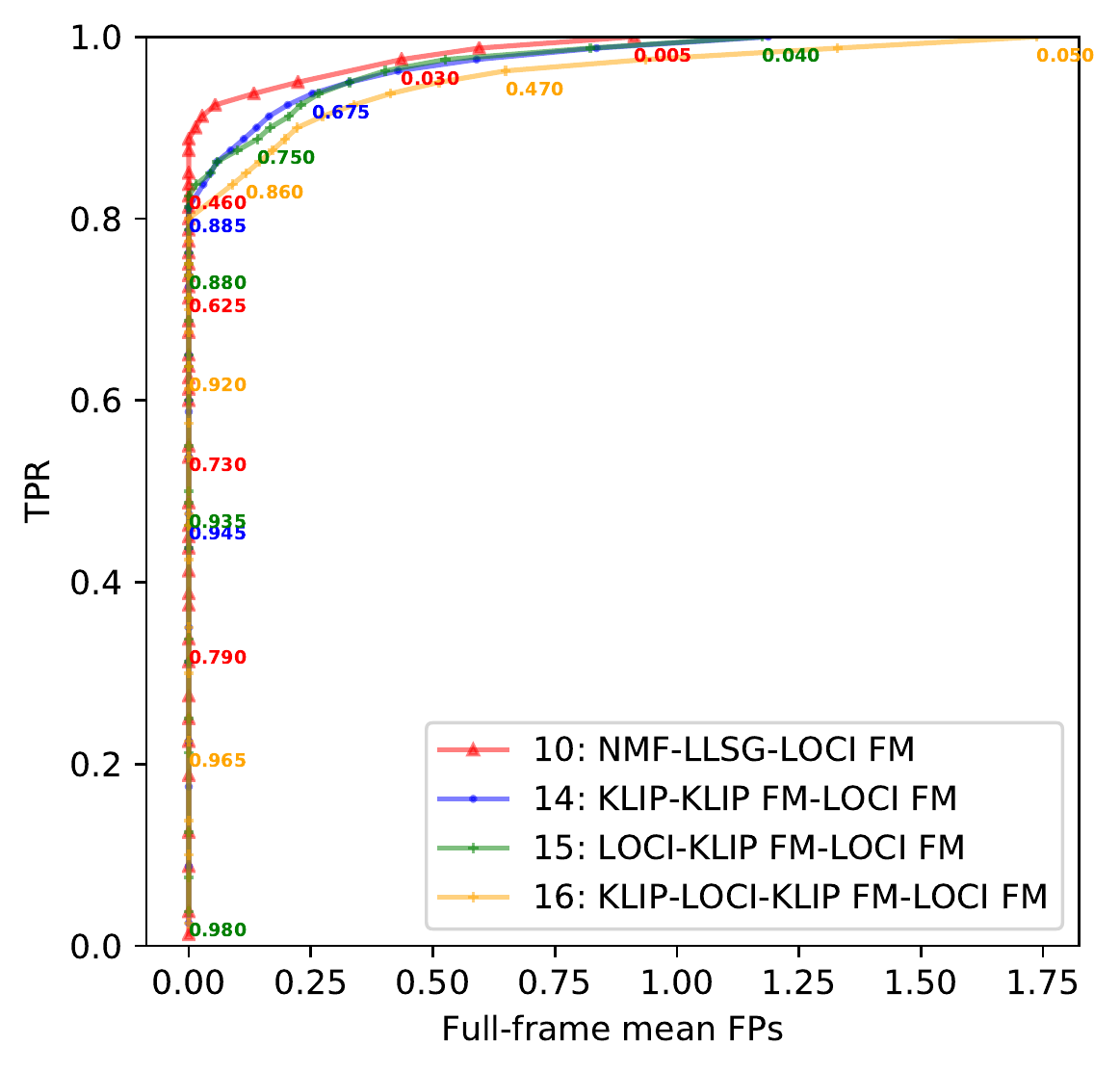}}\\
    \subfloat[NACO at $8\lambda/D$]{\includegraphics[width=160pt]{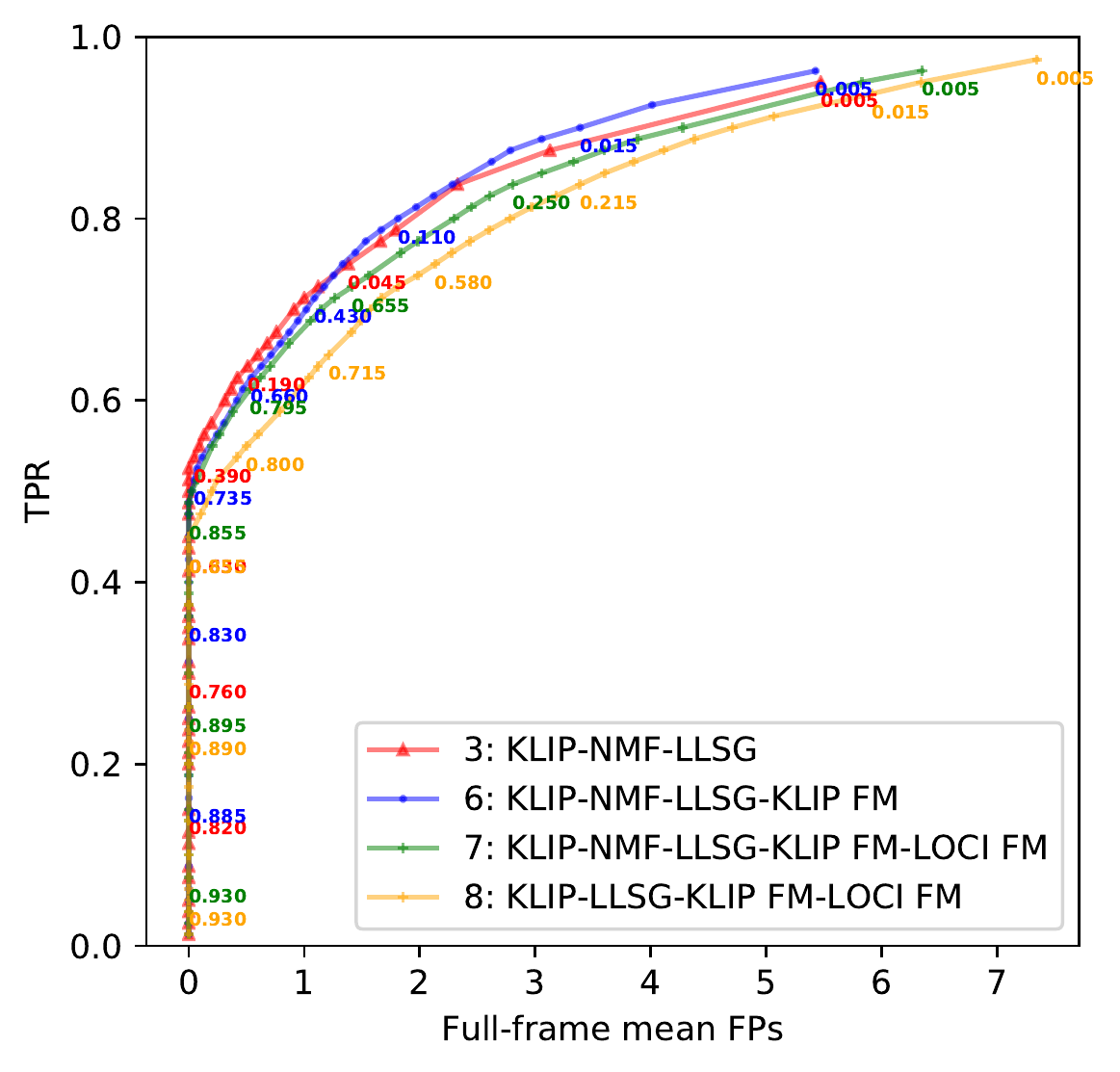}}
  \subfloat[SPHERE at $8\lambda/D$]{\includegraphics[width=160pt]{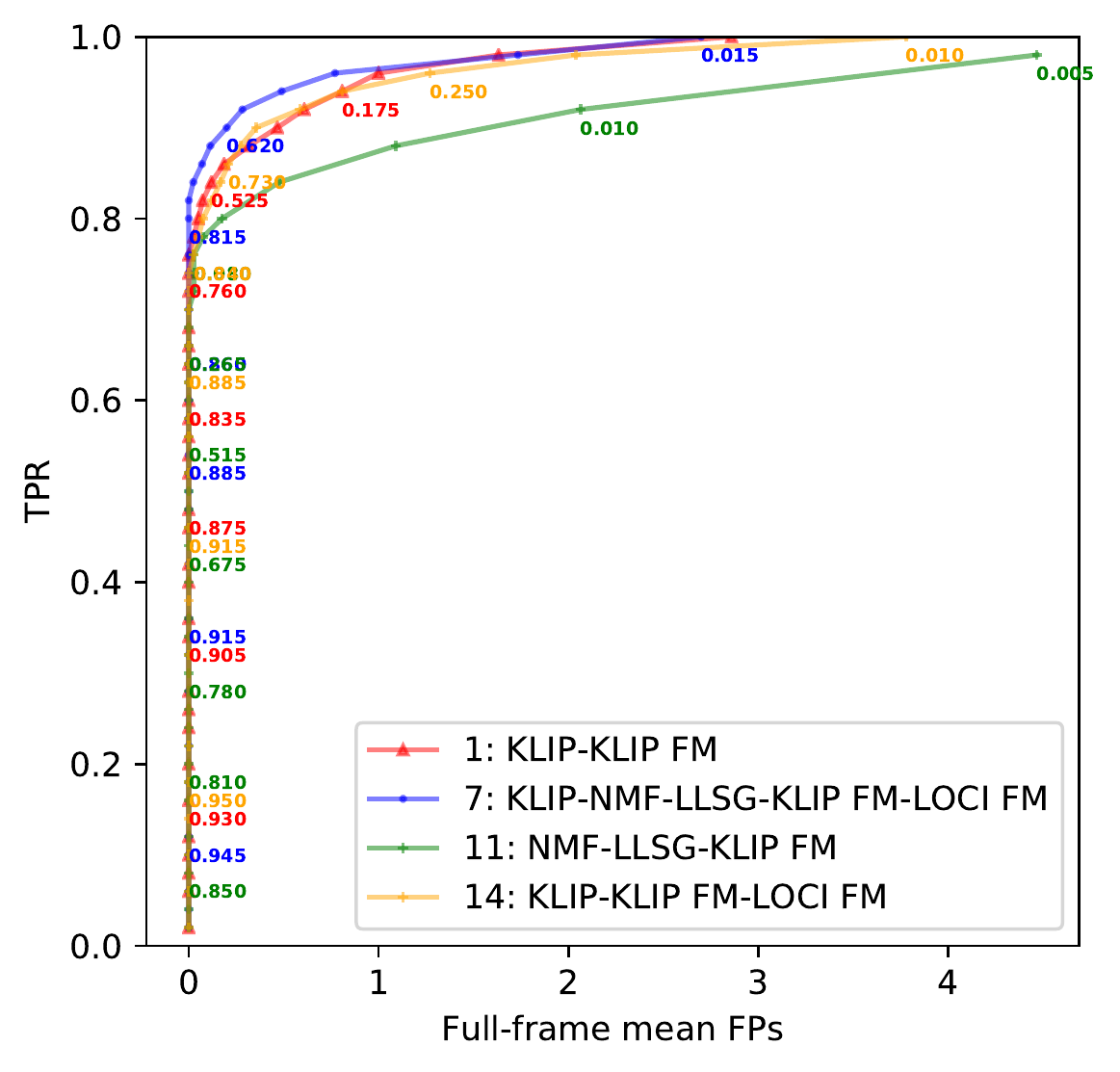}}
  \subfloat[LMIRCam at $8\lambda/D$]{\includegraphics[width=160pt]{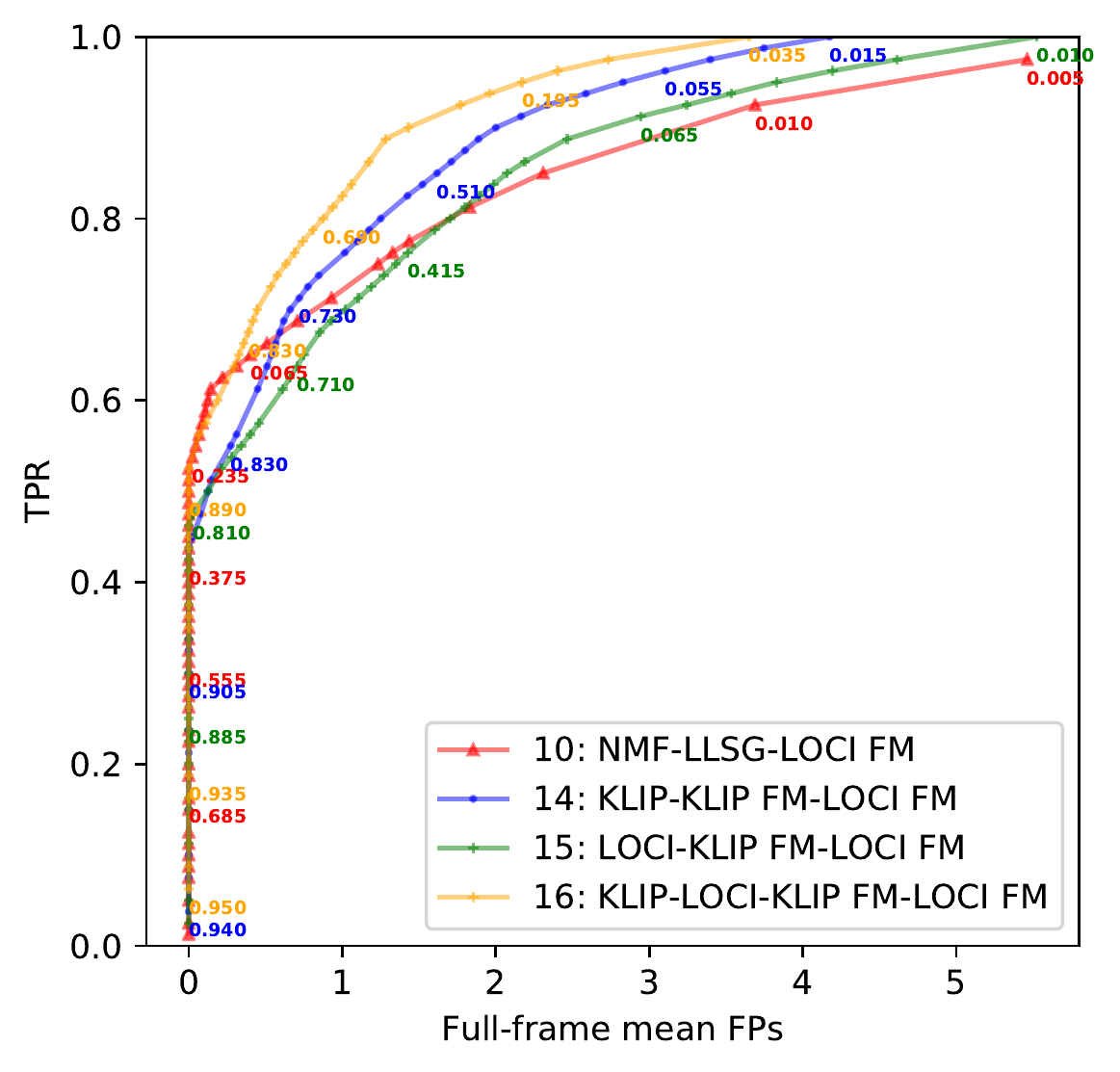}}
  
  \caption{\label{combi} ROC curves for the NACO, SPHERE and LMIRCam data sets, with the four best combinations of PSF subtraction techniques used to generate the RSM map algorithm.}
\end{figure*}

The ROC curve computation follows the same procedure as in the previous section with a region of one arcsecond considered for all three data sets and the curves being computed for the two same angular separations. The main characteristics of the ROC curves for the 16 selected combinations of PSF subtraction techniques may be found in Appendix D. The two parameters we used to select the best combinations, are the maximum TPR reached without any FP, and the average number of FPs inside the entire frame at TPR=1\footnote{The average number of FPs at TPR=1 is estimated by taking the highest threshold corresponding to a TPR of 1, or if a TPR of 1 cannot be reached, the smallest probability threshold we considered in our study, i.e., 0.5\%.}. The first parameter is the most important one, as it gives clues about the highest contrast the algorithm can reach without any false detection. The second gives a measure of the number of bright background structures that have not been properly treated by both the PSF subtraction techniques and the RSM algorithm.

The results presented in appendix D\footnote{The ROC curves summary for the 16 selected combinations is presented in appendix D via a table and a set of figures presenting the same results. Fig.~D.1 provides a graphical comparison between the combinations via bar charts while Tab.~D.1 provides the detailed results.} show large differences in terms of performance between the considered combinations highlighting the importance of the PSF subtraction techniques selection. Based on the two metrics introduced in the previous paragraph, we selected the four best combinations for each data set. The ROC curves for the two considered angular separations are presented in Fig.~\ref{combi}. When comparing these curves with the ones in Figs.~\ref{FMKLIP} and \ref{FMLOCI}, we see that the improvement of the RSM map performance occurs mainly at larger separations when considering multiple PSF subtraction techniques. The ROC curves are indeed very close to the ones obtained with the KLIP-FM RSM and LOCI-FM RSM for the $2\lambda/D$ radial distance, while the gap is much wider for the $8\lambda/D$. Apart from the combinations 7 and 14 (see Fig.~D.1), which are selected for multiple data sets, the other combinations are specific to each data set. This seems to demonstrate that the selection of an optimal combination should be done at least on an instrument-specific basis. The definition of a single optimal combination for the entire set of annuli seems also difficult, as we often observe that higher performance at short separations goes hand in hand with lower performance at large separations. A last element to consider for the selection of the optimal combination is the threshold value for which the first false positive is observed, which should be as small as possible since large values imply the presence of bright structures in the probability map. We conclude from Fig.~\ref{combi} that in a global sense, the best combinations are the combinations 3, 11, and 10 for  the NACO, SPHERE, and LMIRCam data sets, respectively. Looking at these three combinations, we see that they share a common structure, being composed of the LLSG and NMF PSF subtraction techniques, in addition to a LOCI- or KLIP-based PSF subtraction technique. The performance of this particular combination is probably due to the differences between these PSF subtraction techniques in terms of residuals noise profile. These differences should help to better average out the speckle noise via the RSM algorithm. This structure therefore appears to be an interesting starting point when studying a new data set. We further characterise these three combinations in Sect.~\ref{CC} by estimating their contrast curve. 

The results presented in this section demonstrate the dependence of the optimal set of PSF subtraction techniques on the instrument providing the ADI sequence, but also on the angular separation, although a common underlying structure could be seen. A larger set of ADI sequences would be needed to determine whether a single optimal set of PSF subtraction techniques may be identified for a given instrument or even multiple instruments, which could be very helpful when dealing with large surveys.

\section{Forward-backward model}

In this section, we discuss an additional improvement of the original RSM map by considering a forward-backward approach for the estimation of the probability, $\xi_{1,i_a}$. The current approach relies solely on past observations to construct the cube of probabilities while the entire cube of residuals is available for the estimation, that is, of both past and future observations. We propose therefore to replace the current forward approach by a forward-backward approach, which considers both past and future observations. This method computes two separate sets of probabilities, the forward probabilities as done in the original RSM framework:
 \begin{eqnarray}
\xi^f_{1,i_a}=\sum^{1}_{q=0} \frac{\eta_{1,i_a} p_{q,1} \; \xi^f_{q,i_a-1} }{\sum^{1}_{q=0} \sum^1_{s=0} \eta_{s,i_a} p_{q,s} \; \xi^f_{q,i_a-1}} ,
\end{eqnarray}
but also the backward probabilities, which rely on the probability estimated at index $i_a+1$ instead of index $i_a-1$ to compute the probability at current index $i_a$ as:
\begin{eqnarray}
\xi^b_{1,i_a}=\sum^{1}_{q=0} \frac{\eta_{1,i_a} p_{q,1} \; \xi^b_{q,i_a+1} }{\sum^{1}_{q=0} \sum^1_{s=0} \eta_{s,i_a} p_{q,s} \; \xi^b_{q,i_a+1}}.
\end{eqnarray}

Once both sets have been estimated, the final probabilities are obtained by multiplying the two sequences of probabilities. A normalisation factor is applied, making sure that the total probability equals 1 for every index $i_a$. The final probabilities are therefore given by:
\begin{eqnarray}
\xi_{1,i_a}=\frac{\xi^f_{1,i_a} \xi^b_{1,i_a}}{\sum^1_{s=0} \xi^f_{s,i_a} \xi^b_{s,i_a}} 
\end{eqnarray}

\begin{figure}[h!]
  \centering
  \includegraphics[width=240pt]{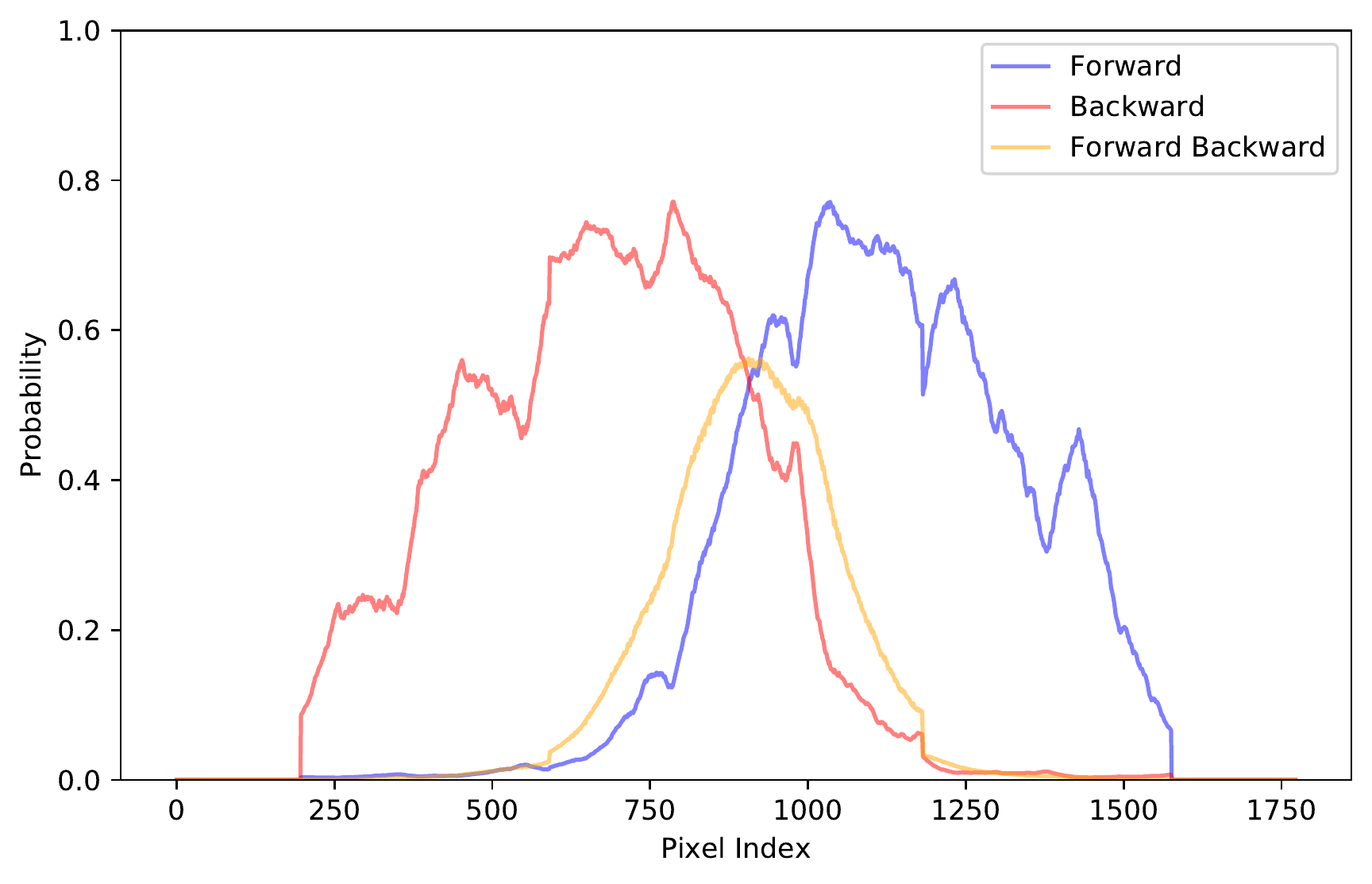}

  \caption{\label{fberi} Evolution of the probabilities for the forward, backward, and forward-backward approaches using KLIP, around the location of a planetary candidate injected in the 51 Eridani data set (radial distance of  $4\lambda/D$ with a contrast of 3.76 $\times10^{-5}$) .}
\end{figure}

Because the RSM features a short-term memory, the probability of being in the planetary regime builds up when we get closer to the planetary signal but with a small latency. As can be seen from Fig.~\ref{fberi}, this latency leads to a shift of the main peak towards the future for the forward approach and towards the past for the backward approach. When relying on the forward-backward approach these shifts cancel out and the main peak is centred on the true position of the planetary signal. The forward-backward approach should therefore allow to reach a higher precision in terms of exoplanet astrometry. 

In order to investigate the ability of both approaches to derive an accurate astrometric measurement for the detected planetary signal, we propose performing a series of simulations based on the SPHERE data set. We study the evolution of the astrometric precision for a range of contrasts, considering again a radial distance of $2 \lambda/D$ and $8 \lambda/D$. As done in the previous sections, we base our simulation on synthetic data sets, on the basis of which we apply a KLIP RSM map using the forward and forward-backward version of the RSM algorithm. The negative fake companion (NEGFC) method \citep{Lagrange10,Marois10,Wertz17} is also applied on the synthetic data sets, allowing for a comparison with a technique dedicated to astrometry\footnote{We relied on the function provided by the VIP package \citep{Gonzalez17} for the computation of the position via the NEGFC using a simplex (Nelder-Mead) optimisation.}. For each radial distance, we inject fake companions at 16 different position angles. The set of considered contrasts are computed based on the KLIP RSM contrast curve, estimated using the approach proposed in the next section (see Fig.~\ref{ccklip}). We define two sets of contrasts ranging from one to six times the sensitivity limit at the considered radial distance in Fig.~\ref{ccklip}, with a step size of 0.5. In the case of the RSM map, the computation of the position is done by fitting a two-dimensional Gaussian to the detected planetary signal. The astrometric error bars for the three considered methods are computed as the root mean squared (rms) position error between the obtained position and the injected fake companion true position, averaged over the two axes. The rms is estimated over the 16 fake companions injected at each radial distance, for every contrast.

\begin{figure}[t]
  \centering
  \subfloat[$2 \lambda/D$]{\includegraphics[width=200pt]{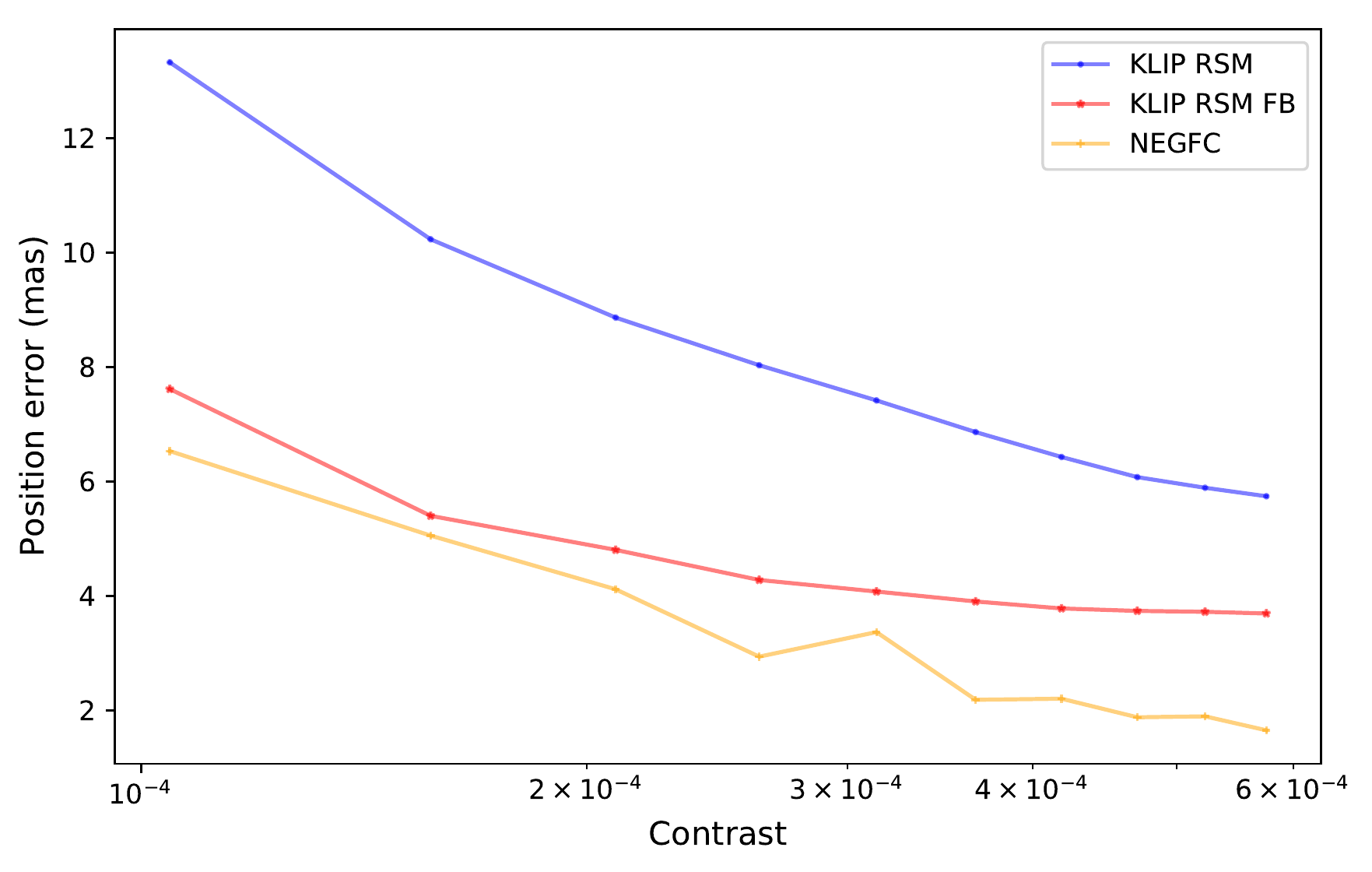}}\\
  \subfloat[$8 \lambda/D$]{\includegraphics[width=200pt]{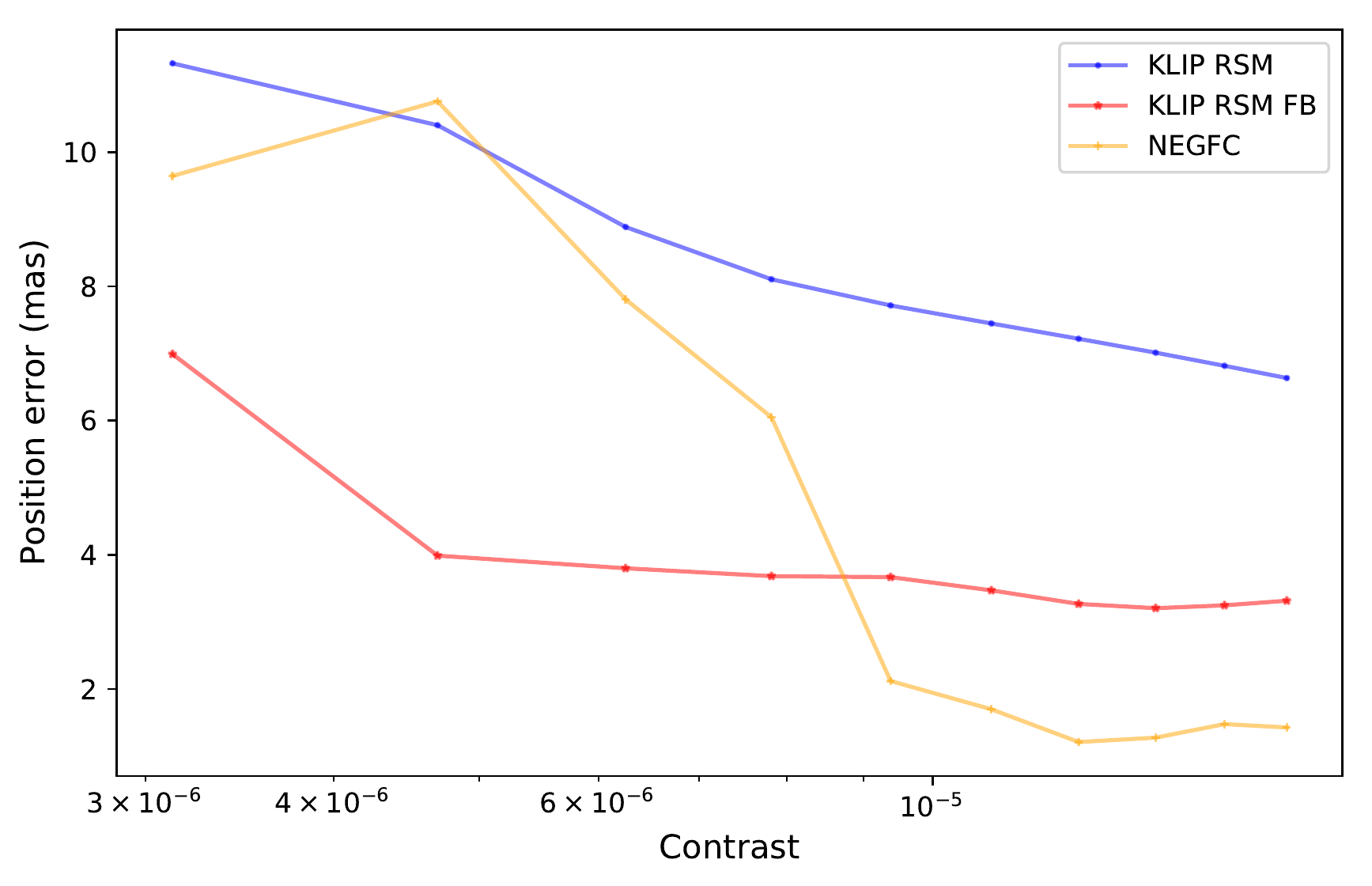}}

  \caption{\label{poserror} RMS position errors averaged over the two axes expressed in mas for the KLIP RSM map using the forward (blue) and forward-backward (red) versions of the RSM approach and for the NEGFC approach (orange). The two graphs show the dependence of the averaged RMS position error on the contrast for respectively a radial distance of $2\lambda/D$ and $8\lambda/D$.}
\end{figure}

The results from Fig.~\ref{poserror} demonstrate clearly the ability of the forward-backward approach to decrease the position error compared to the original forward approach. As can be seen from Fig.~\ref{poserror}, the RSM forward-backward approach performs better than the NEGFC approach at large radial distances and for high contrasts. However, for lower contrast, the RSM forward-backward approach reaches a noise floor around 4 mas, higher than the noise floor of the NEGFC approach, which is between 1.5 and 2 mas. This higher noise floor may be explained partly by the profile of the planetary signal in the RSM map. As can be seen from Eq.~\ref{proba}, the RSM approach response to a planetary signal is non linear and dependent on neighbouring pixels, leading potentially to asymmetries in the azimuthal direction, even in the forward-backward case. The algorithm architecture also leads  to non-linearities along the radial axis because of the annulus-wise probabilities computation. Finally, as can be seen from Fig.~\ref{fberi}, the forward-backward approach reduces the amplitude of the planetary signal within the probability map. All these elements affect the Gaussian fit and therefore the astrometric precision that the RSM algorithm can reach. Nevertheless, as demonstrated by the results from Fig.~\ref{poserror}, the RSM forward-backward approach can reach a higher astrometric precision, especially at large radial distances and high contrasts. This is due to the better ability of the RSM algorithm to detect faint companions. It is also worth noting that the computation time is much lower when using the RSM map than with the NEGFC approach or the more advanced Markov Chain Monte Carlo version of NEGFC approach. The RSM forward-backward approach provides therefore a good first estimate, especially for high contrast targets, which can then be used, for lower contrasts, as an initial position for more advanced astrometry techniques.

\begin{figure}[h!]
  \centering
  \subfloat[Original RSM map]{\includegraphics[trim =60 20 0 0,clip ,width=127pt]{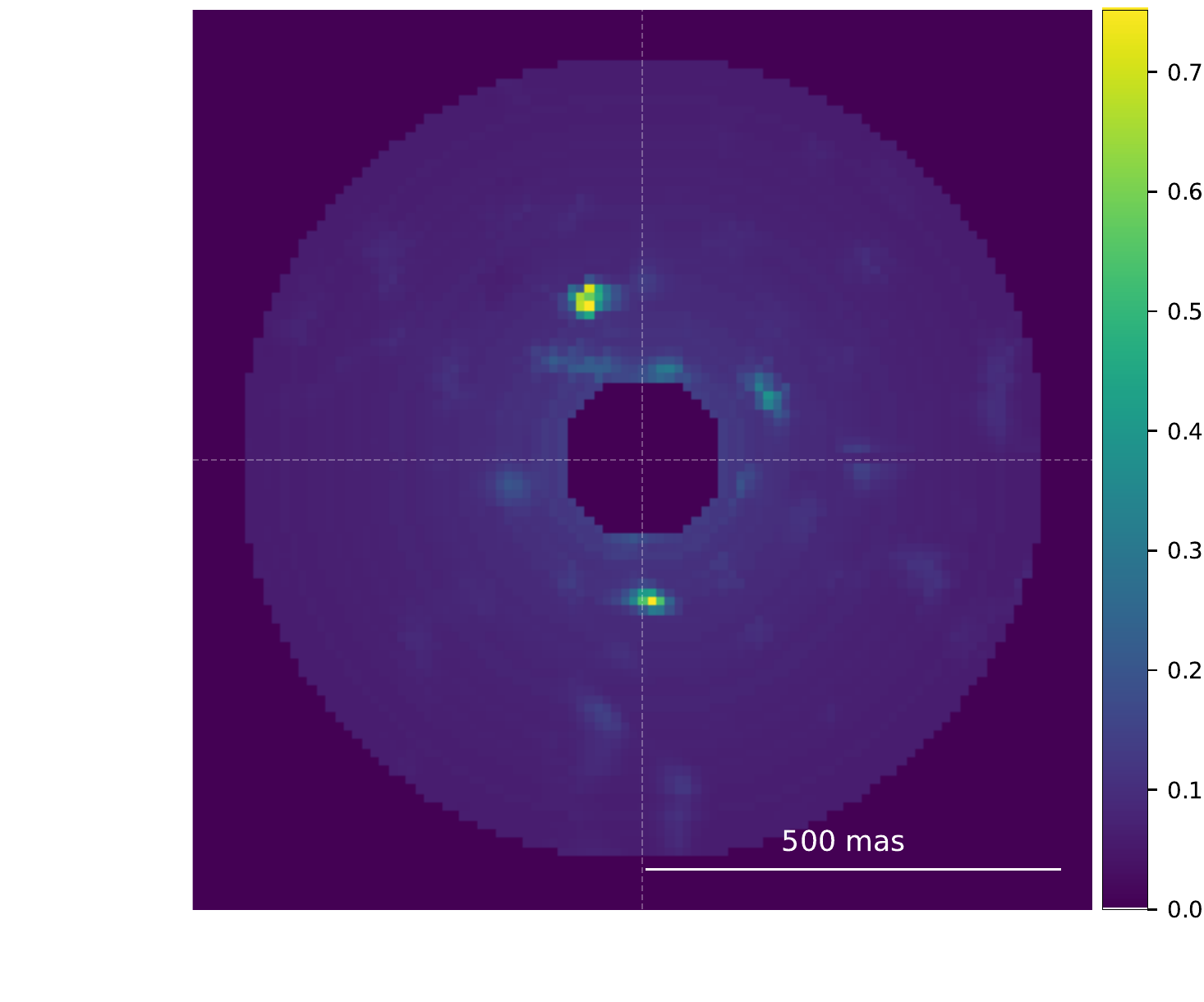}}
  \subfloat[Forward-Backward RSM map]{\includegraphics[trim =60 20 0 0,clip ,width=130pt]{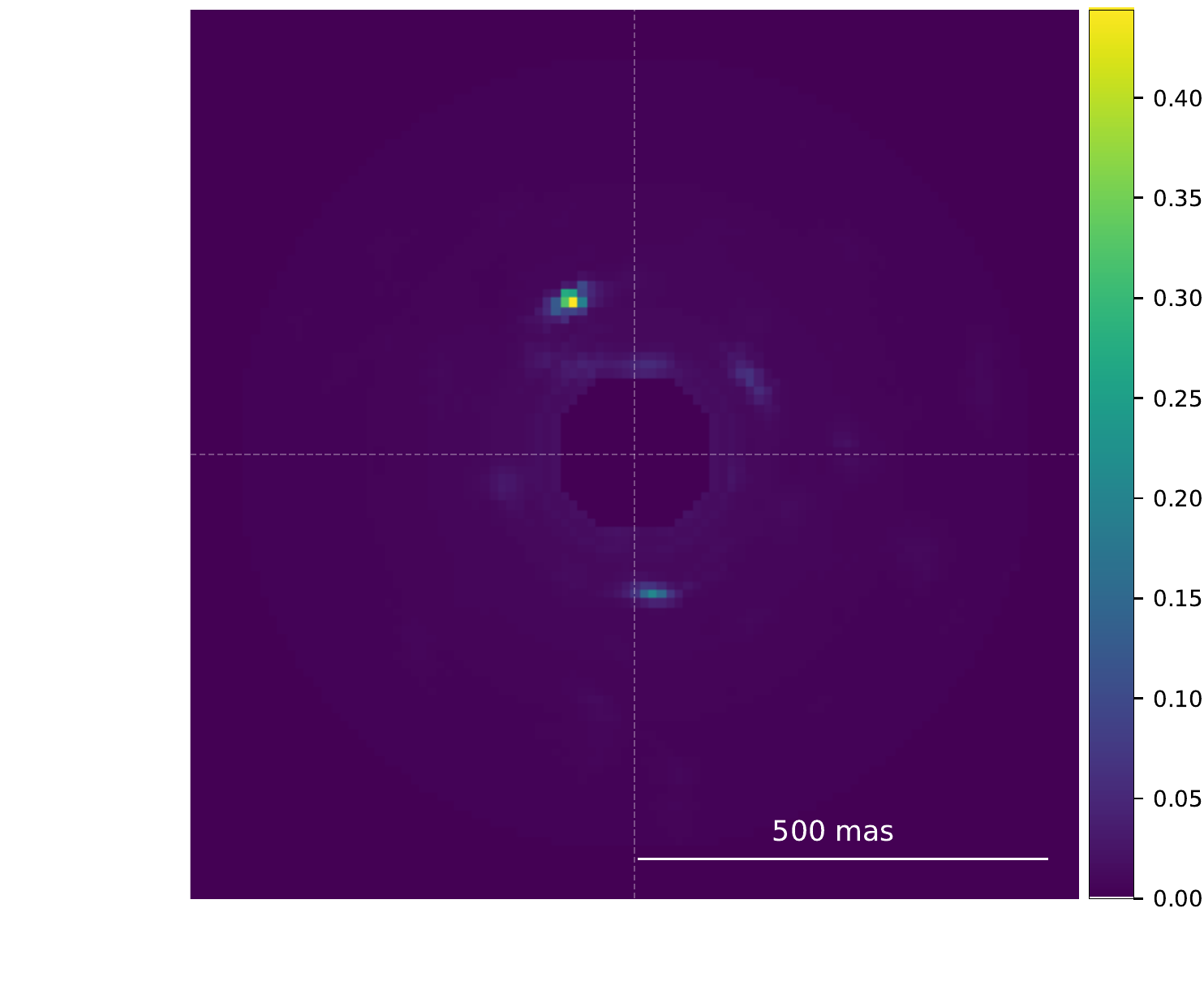}}

  \caption{\label{bgdnoisefb} RSM map generated with forward and forward-backward approach for the LMIRCam data set using KLIP with 18 principal components, a FOV rotation expressed in terms of FWHM of 0.3 and a Gaussian distribution. A square root-based scale has been used to increase the background speckle noise brightness.}
\end{figure}

Another advantage of the forward-backward approach lies in its ability to reduce the background speckle noise and smooth the probability curve, the noise being treated differently by the forward and backward components. Looking at Fig.~\ref{bgdnoisefb}, we see that the level of the residual speckle noise has reduced drastically for the LMIRCam data set, the brightest speckle probability decreasing by around 40 percent. However this noise reduction comes along with a reduced brightness of the planetary signal. This reduced brightness is also illustrated in Fig.~\ref{fberi}, where the peak value obtained with the forward-backward approach is lower than the one obtained with the two other approaches, leading to a reduced detection threshold. This reduction of the planet signal strength does not impact the performance of the forward-backward approach in terms of ROC curve, with similar results for both the forward and forward-backward approaches. The forward-backward approach outperforms even sightly the forward approach at small separation (see Appendix B for a comparison between the original RSM approach and the forward-backward version in the case of KLIP). 

\section{Contrast curve}
\label{CC}

This last section is devoted to the estimation of contrast curves based on our RSM framework. When relying on probability map for exoplanet detection, we cannot use the traditional procedure to compute the contrast curve. This procedure selects the contrast corresponding to a TPR of 0.5 for a given probability of observing a FP \citep{Jensen_Clem_2017}. A 5$\sigma$ threshold is usually chosen, which corresponds to a $3 \times 10^{-7}$ false alarm probability under a Gaussian noise hypothesis. In the case of the RSM map, the probability provided as an output by the algorithm is a non linear function of the underlying likelihoods, past observations, and the transfer matrix (see Eq.~\ref{proba}), which precludes us from defining a similar probability threshold for the FP. We  therefore rely on a procedure similar to the one we used for the ROC curve computation. Considering a given flux, we inject fake companions at different position angles for a given radial distance. The resulting synthetic data sets allow for the computation of a threshold-dependent TPR. The next step is to select a convenient threshold. As it is not possible to reach a 5$\sigma$ confidence empirically, we select the threshold that leads to the first detection of a FP within the entire frame \citep[as done in][but considering the entire frame instead of the selected annulus]{Jensen_Clem_2017}. This threshold definition provides a direct link with the background noise level while avoiding the usual shortcomings of standard contrast curves, namely, the noise Gaussianity assumption and the definition of the region used to empirically estimate the first two moments of the noise.

\begin{figure}[h!]
  \centering
  \subfloat[]{\includegraphics[width=128pt]{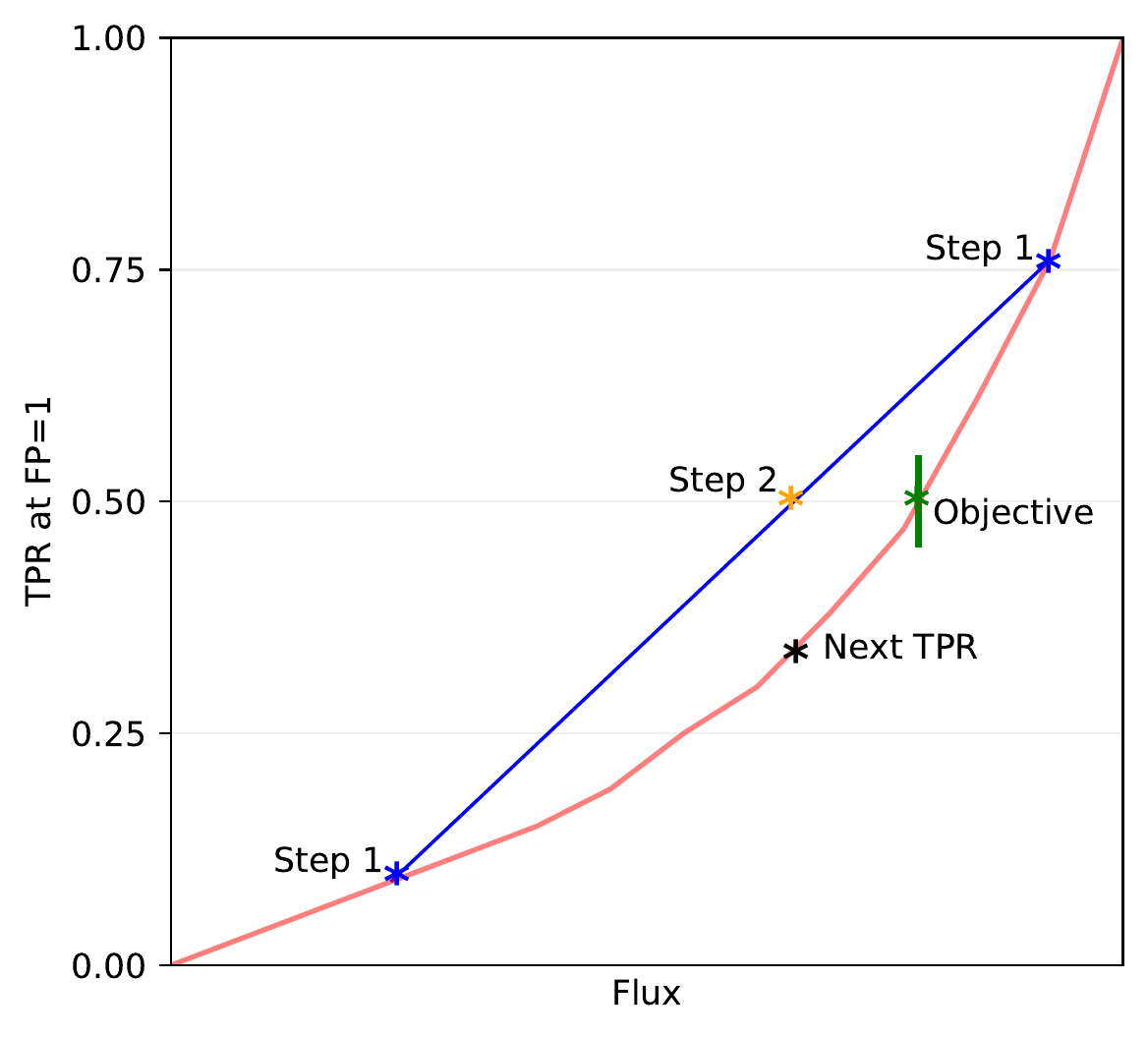}}
  \subfloat[]{\includegraphics[width=128pt]{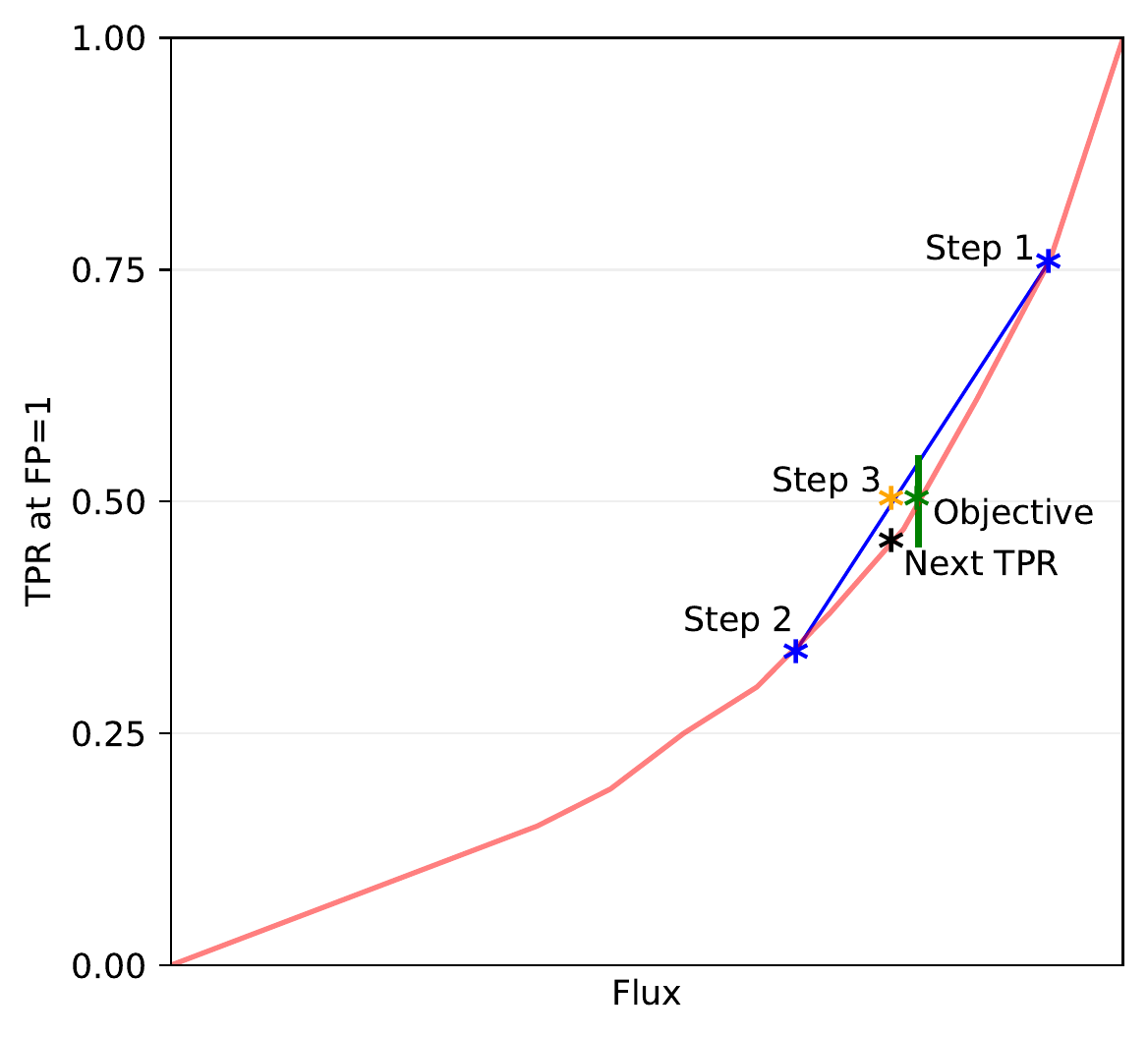}}

  \caption{\label{ccproc} Procedure for the estimation of the contrast curve, the orange star corresponds to the linear interpolation, the black star to the corresponding TPR, the blue stars to the two previous estimations used for the linear interpolation and the green star to the flux corresponding to a TPR of 0.5. The red curve makes the link between the flux and the true positive rate in the case of a single false positive while the green line shows the tolerance's interval of $\left[ 0.45,0.55\right]$ .}
\end{figure}

We repeat the steps presented in the previous paragraphs on sets of fake companions injected with different flux values until a TPR of 0.5 is found when detecting the first FP. We use an iterative procedure based on linear interpolations to minimise the number of flux values to be considered before reaching the TPR of 0.5. The procedure starts with the estimation of the TPR for a selected pair of upper and lower flux values (step 1 in Fig.~\ref{ccproc}, respectively below and above the TPR of 0.5 preferably). A linear interpolation then allows the determination of a best guess for the next flux value to be estimated (step 2). Once the TPR is estimated for this new flux value, a new linear interpolation is performed with this last flux value as upper bound (resp. lower bound) if the TPR is above 0.5 (resp. below 0.5) keeping the previous lower flux value (resp. upper flux value) (step 3). The procedure is repeated until a TPR $\in \left[ 0.45, 0.55\right] $ is found using a tolerance interval with a size of $0.1$. These steps are summarised in Fig.~\ref{ccproc}.

\begin{figure}[h!]
  \centering
  \subfloat[NACO]{\includegraphics[width=240pt]{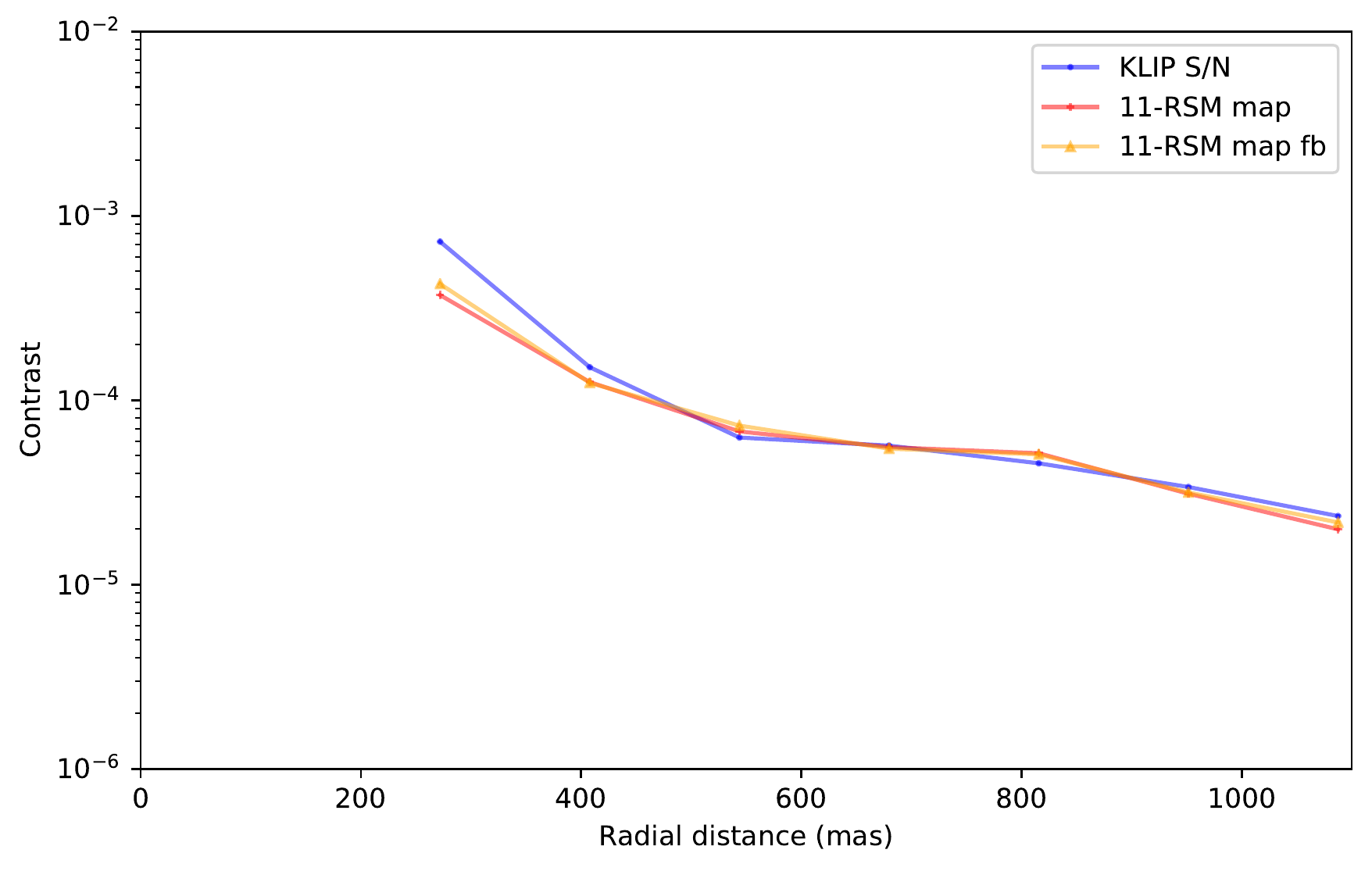}}\\
  \subfloat[SPHERE]{\includegraphics[width=240pt]{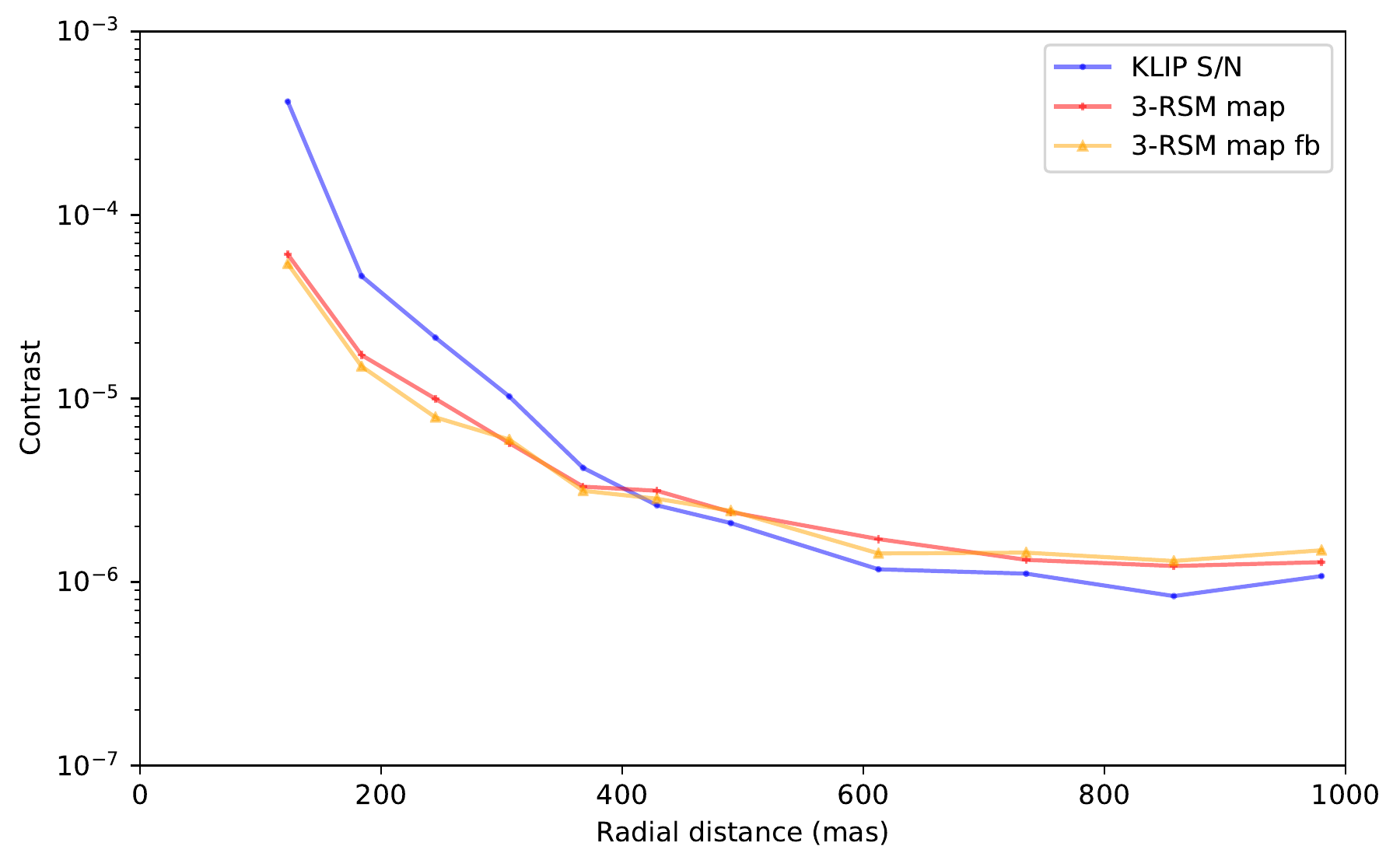}}\\
  \subfloat[LMIRCam]{\includegraphics[width=240pt]{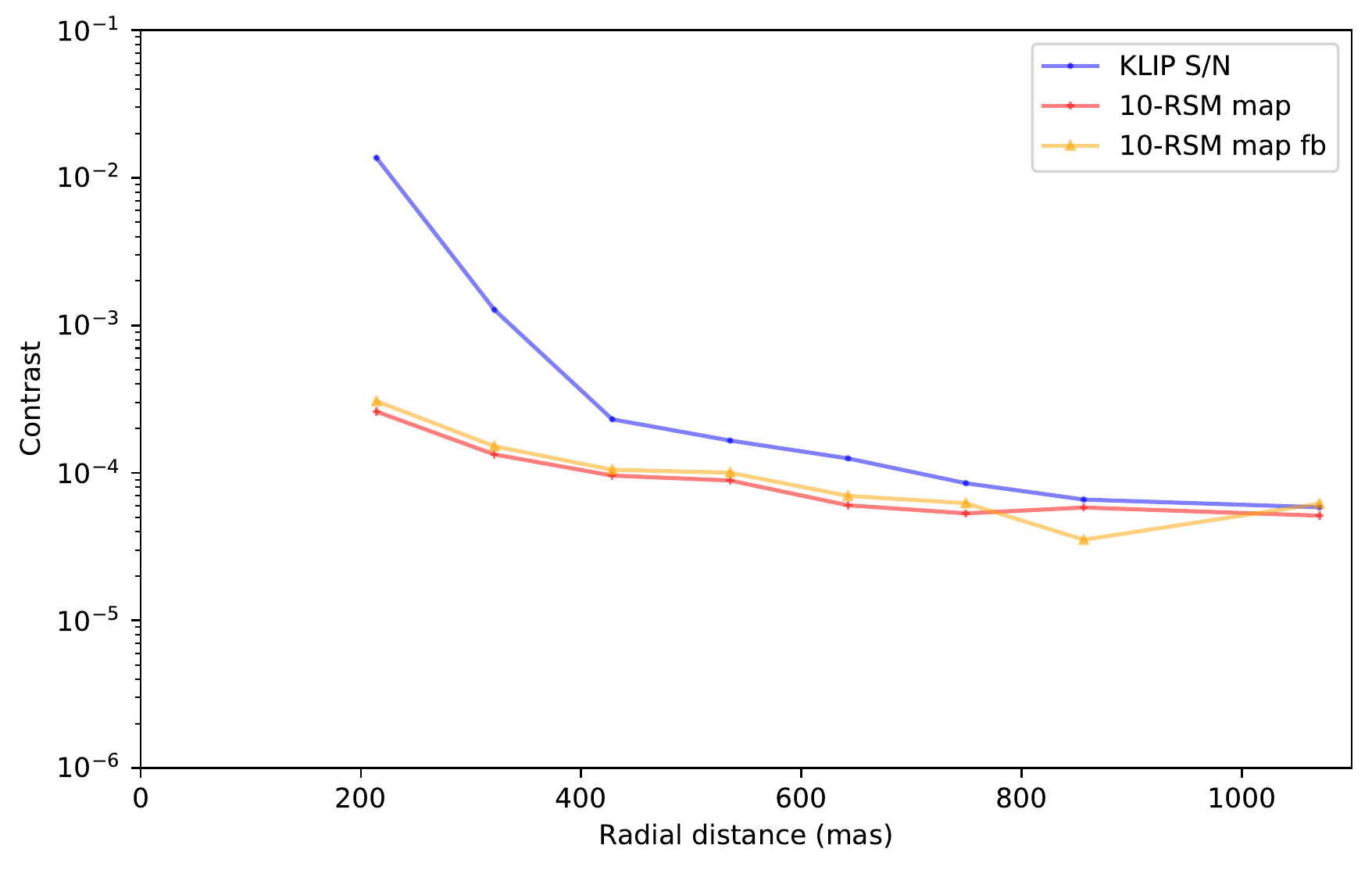}}
  
  \caption{\label{ccesti} Contrast curves for the three data sets using the optimal combination of PSF subtraction techniques obtained in Sect.~\ref{optimix} (resp.\ 3, 11, and 10) with the region $\left[ 2\lambda /D, 16\lambda /D \right]$ considered for the SPHERE data set and the region $\left[ 2\lambda /D, 8\lambda /D \right]$ considered for the other two data sets to get the contrast in the first arcsecond.}
\end{figure}

We estimated contrast curves for the three data sets, considering the original RSM map using the optimal combination found in Sect.~\ref{CC}, the forward-backward version of the algorithm, and the simple KLIP S/N map. We computed the contrast from a radial distance of $2\lambda/D$ up to $8\lambda/D$ using a step-size of one $\lambda/D$ and then up to $16\lambda/D$ for the SPHERE data set with a step-size of two $\lambda/D$. The resulting contrast curves may be found in Fig.~\ref{ccesti}. These contrast curves confirm the findings made with the ROC curves, the RSM map reaching in most cases a significantly higher contrast for small angular separations and providing comparable contrast at larger separations. The gap between the contrast curves of the KLIP S/N map and the RSM map is significantly larger for the LMIRCam data set, highlighting the interest of the RSM approach for this instrument. However, the lower number of frames and the smaller angular rotation for this particular data set may explain these results. A larger number of LMIRCam data sets is therefore needed to confirm these findings. We also find that the RSM map and the forward-backward version have a similar performance for all angular separations. Considering its higher precision and its lower background noise level, the forward-backward version seems to be a promising alternative to the standard RSM map.

\section{Conclusion}

In order to improve the RSM map sensitivity to faint exoplanets, we implemented several improvements at different levels. We first considered a forward-model version of the RSM map for two different PSF subtraction techniques, KLIP and LOCI. The computation of forward-modelled PSFs allowed to take into account PSF distortions due to the speckle field subtraction. We demonstrated the interest of the forward model at small angular separations through the estimation of ROC curves. These ROC curves highlight the radial dependence of the optimal PSF crop size, with larger crop sizes leading to better results for small angular separations while the reverse is true for smaller crop sizes. This could be explained by the fact that larger PSF crop sizes better account for the self-subtraction patterns, whose intensity decreases with radial distance as the relative movement of astrophysical signals increases with the distance to the host star.

In a second step, we investigated the question of the optimal selection of the PSF subtraction techniques to be included in the estimation of the RSM map. Relying again on ROC curves, we demonstrated the importance of this selection with sometimes large differences between the performance of the considered combinations. The optimal combinations seems to depend on the instrument, but also on the angular separation. 

We proposed then an improvement directly related to the way probabilities are estimated within the RSM algorithm by replacing the original forward approach by a forward-backward approach. This approach relies on both past and future observations to compute a final probability. This allows us to deal more efficiently with background noise as speckles are not treated in the same way when relying on a forward and backward approach. Another advantage of the forward-backward version of the RSM map is its ability to extract more precisely the planetary astrometry.

We finally implemented a new framework to compute contrast curves in the case of probability maps, which cannot be treated as S/N maps. We kept the TPR of 0.5 while removing the Gaussian based noise threshold definition, replacing it with a threshold based on the detection of the first FP. Using a procedure based on successive linear interpolations, we were able to compute the contrast curve for the original RSM map and the forward-backward version using the optimal combination obtained previously and compare their performance with the one of a simple KLIP S/N map. The results highlighted the ability of the RSM map to detect fainter companions at small angular separations (below 400 mas) and the overall similar results for larger separations. The contrast curves also confirmed the interest of the forward-backward approach as it provides similar contrast curves while reducing the speckle background noise and giving a higher precision in terms of astrometry. 

As the selection of the optimal combination of PSF subtraction techniques to be included in the RSM map as well as the parametrisation of these PSF subtraction techniques are relatively time-consuming, an interesting future development of the RSM map would be to create an automated parametrisation procedure that depends on angular separation.

\begin{acknowledgements}
This work was supported by the Fonds de la Recherche Scientifique - FNRS under Grant n$^{\circ}$ F.4504.18 and by the European Research Council (ERC) under the European Union's Horizon 2020 research and innovation program (grant agreement n$^{\circ}$ 819155). We thank our colleagues A.-L. Maire for sharing the SPHERE data set and Arianna Musso-Barcucci for sharing the LMIRCam data set. We also thank Gilles Louppe for his precious advice.
\end{acknowledgements}

\bibliographystyle{aa}
\bibliography{FMRSM.bib}

\begin{thebibliography}{26}
\expandafter\ifx\csname natexlab\endcsname\relax\def\natexlab#1{#1}\fi

\bibitem[{Absil {et~al.}(2013)Absil, Milli, Mawet, Lagrange, Girard, Chauvin,
  Boccaletti, Delacroix, \& Surdej}]{Absil13}
Absil, O., Milli, J., Mawet, D., {et~al.} 2013, A\&A, 559, L12

\bibitem[{Amara \& Quanz(2012)}]{Amara12}
Amara, A. \& Quanz, S.~P. 2012, Monthly Notices of the Royal Astronomical
  Society, 427, 948

\bibitem[{Bloemhof {et~al.}(2001)Bloemhof, Dekany, Troy, \&
  Oppenheimer}]{Bloemhof_2001}
Bloemhof, E.~E., Dekany, R.~G., Troy, M., \& Oppenheimer, B.~R. 2001, The
  Astrophysical Journal, 558, L71

\bibitem[{Bottom {et~al.}(2017)Bottom, Ruane, \& Mawet}]{Bottom17}
Bottom, M., Ruane, G., \& Mawet, D. 2017, Research Notes of the {AAS}, 1, 30

\bibitem[{Cantalloube {et~al.}(2015)Cantalloube, Mouillet, Mugnier, Milli,
  Absil, Gomez~Gonzalez, Chauvin, Beuzit, \& Cornia}]{Cantalloube15}
Cantalloube, F., Mouillet, D., Mugnier, L.~M., {et~al.} 2015, A\&A, 582, A89

\bibitem[{Dahlqvist {et~al.}(2020)Dahlqvist, Cantalloube, \&
  Absil}]{Dahlqvist20}
Dahlqvist, C.-H., Cantalloube, F., \& Absil, O. 2020, A\&A, 633, A95

\bibitem[{Delorme {et~al.}(2017)Delorme, Meunier, Albert, Lagadec, Coroller,
  Galicher, Mouillet, Boccaletti, Mesa, Meunier, Beuzit, Lagrange, Chauvin,
  Sapone, Langlois, Maire, Montargès, Gratton, Vigan, \&
  Fenouillet}]{Delorme17}
Delorme, P., Meunier, N., Albert, D., {et~al.} 2017

\bibitem[{Gomez~Gonzalez {et~al.}(2018)Gomez~Gonzalez, Absil, \&
  Van~Droogenbroeck}]{Gonzalez18}
Gomez~Gonzalez, C.~A., Absil, O., \& Van~Droogenbroeck, M. 2018, A\&A, 613, A71

\bibitem[{{Gomez Gonzalez} {et~al.}(2017){Gomez Gonzalez}, {Wertz}, {Absil},
  {Christiaens}, {Defr{\`e}re}, {Mawet}, {Milli}, {Absil}, {Van Droogenbroeck},
  {Cantalloube}, {Hinz}, {Skemer}, {Karlsson}, \& {Surdej}}]{Gonzalez17}
{Gomez Gonzalez}, C.~A., {Wertz}, O., {Absil}, O., {et~al.} 2017, \aj, 154, 7

\bibitem[{{Gomez Gonzalez, C. A.} {et~al.}(2016){Gomez Gonzalez, C. A.},
  {Absil, O.}, {Absil, P.-A.}, {Van Droogenbroeck, M.}, {Mawet, D.}, \&
  {Surdej, J.}}]{Gonzalez16}
{Gomez Gonzalez, C. A.}, {Absil, O.}, {Absil, P.-A.}, {et~al.} 2016, A\&A, 589,
  A54

\bibitem[{Jensen-Clem {et~al.}(2017)Jensen-Clem, Mawet, Gonzalez, Absil,
  Belikov, Currie, Kenworthy, Marois, Mazoyer, Ruane, Tanner, \&
  Cantalloube}]{Jensen_Clem_2017}
Jensen-Clem, R., Mawet, D., Gonzalez, C. A.~G., {et~al.} 2017, The Astronomical
  Journal, 155, 19

\bibitem[{Lafreniere {et~al.}(2007)Lafreniere, Marois, Doyon, Nadeau, \&
  Artigau}]{Lafreniere07}
Lafreniere, D., Marois, C., Doyon, R., Nadeau, D., \& Artigau, E. 2007, ApJ,
  660

\bibitem[{Lagrange {et~al.}(2010)Lagrange, Bonnefoy, Chauvin, Apai, Ehrenreich,
  Boccaletti, Gratadour, Rouan, Mouillet, Lacour, \& Kasper}]{Lagrange10}
Lagrange, A.-M., Bonnefoy, M., Chauvin, G., {et~al.} 2010, Science (New York,
  N.Y.), 329, 57

\bibitem[{Maire {et~al.}(2019)Maire, {Rodet, L.}, {Cantalloube, F.}, {Galicher,
  R.}, {Brandner, W.}, {Messina, S.}, {Lazzoni, C.}, {Mesa, D.}, {Melnick, D.},
  {Carson, J.}, {Samland, M.}, {Biller, B. A.}, {Boccaletti, A.}, {Wahhaj, Z.},
  {Beust, H.}, {Bonnefoy, M.}, {Chauvin, G.}, {Desidera, S.}, {Langlois, M.},
  {Henning, T.}, {Janson, M.}, {Olofsson, J.}, {Rouan, D.}, {M\'enard, F.},
  {Lagrange, A.-M.}, {Gratton, R.}, {Vigan, A.}, {Meyer, M. R.}, {Cheetham,
  A.}, {Beuzit, J.-L.}, {Dohlen, K.}, {Avenhaus, H.}, {Bonavita, M.}, {Claudi,
  R.}, {Cudel, M.}, {Daemgen, S.}, {D\'{}Orazi, V.}, {Fontanive, C.},
  {Hagelberg, J.}, {Le Coroller, H.}, {Perrot, C.}, {Rickman, E.}, {Schmidt,
  T.}, {Sissa, E.}, {Udry, S.}, {Zurlo, A.}, {Abe, L.}, {Orign\'e, A.}, {Rigal,
  F.}, {Rousset, G.}, {Roux, A.}, \& {Weber, L.}}]{Maire19}
Maire, A.-L., {Rodet, L.}, {Cantalloube, F.}, {et~al.} 2019, A\&A, 624, A118

\bibitem[{Marois {et~al.}(2000)Marois, Doyon, Racine, \& Nadeau}]{Marois00}
Marois, C., Doyon, R., Racine, R., \& Nadeau, D. 2000, Publications of the
  Astronomical Society of the Pacific, 112, 91

\bibitem[{Marois {et~al.}(2006)Marois, Lafreniere, Doyon, Macintosh, \&
  Nadeau}]{Marois06}
Marois, C., Lafreniere, D., Doyon, R., Macintosh, B., \& Nadeau, D. 2006, ApJ,
  641

\bibitem[{Marois {et~al.}(2010)Marois, Macintosh, \& Véran}]{Marois10}
Marois, C., Macintosh, B., \& Véran, J.-P. 2010, in Adaptive Optics Systems
  II, ed. B.~L. Ellerbroek, M.~Hart, N.~Hubin, \& P.~L. Wizinowich, Vol. 7736,
  International Society for Optics and Photonics (SPIE), 595 -- 606

\bibitem[{{Mawet} {et~al.}(2014){Mawet}, {Milli}, {Wahhaj}, {Pelat}, {Absil},
  {Delacroix}, {Boccaletti}, {Kasper}, {Kenworthy}, {Marois}, {Mennesson}, \&
  {Pueyo}}]{Mawet14}
{Mawet}, D., {Milli}, J., {Wahhaj}, Z., {et~al.} 2014, \apj, 792, 97

\bibitem[{Pairet {et~al.}(2019)Pairet, Cantalloube, Gomez~Gonzalez, Absil, \&
  Jacques}]{Pairet19}
Pairet, B., Cantalloube, F., Gomez~Gonzalez, C., Absil, O., \& Jacques, L.
  2019, Monthly Notices of the Royal Astronomical Society, 487, 2262

\bibitem[{Pueyo(2016)}]{Pueyo16}
Pueyo, L. 2016, The Astrophysical Journal, 824, 117

\bibitem[{Ren {et~al.}(2018)Ren, Pueyo, Zhu, Debes, \& Duchêne}]{Ren18}
Ren, B., Pueyo, L., Zhu, G.~B., Debes, J., \& Duchêne, G. 2018, ApJ, 852

\bibitem[{Ruffio {et~al.}(2017)Ruffio, Macintosh, Wang, \& Pueyo}]{Ruffio17}
Ruffio, J.-B., Macintosh, B., Wang, J.~J., \& Pueyo, L. 2017, ApJ, 842

\bibitem[{{Samland} {et~al.}(2017){Samland}, {Molli{\`e}re}, {Bonnefoy},
  {Maire}, {Cantalloube}, {Cheetham}, {Mesa}, {Gratton}, {Biller}, {Wahhaj},
  {Bouwman}, {Brandner}, {Melnick}, {Carson}, {Janson}, {Henning}, {Homeier},
  {Mordasini}, {Langlois}, {Quanz}, {van Boekel}, {Zurlo}, {Schlieder},
  {Avenhaus}, {Beuzit}, {Boccaletti}, {Bonavita}, {Chauvin}, {Claudi}, {Cudel},
  {Desidera}, {Feldt}, {Fusco}, {Galicher}, {Kopytova}, {Lagrange}, {Le
  Coroller}, {Martinez}, {Moeller-Nilsson}, {Mouillet}, {Mugnier}, {Perrot},
  {Sevin}, {Sissa}, {Vigan}, \& {Weber}}]{Samland17}
{Samland}, M., {Molli{\`e}re}, P., {Bonnefoy}, M., {et~al.} 2017, \aap, 603,
  A57

\bibitem[{Soummer {et~al.}(2012)Soummer, Pueyo, \& Larkin}]{Soummer12}
Soummer, R., Pueyo, L., \& Larkin, J. 2012, ApJ, 755

\bibitem[{Sparks \& Ford(2002)}]{Sparks02}
Sparks, W.~B. \& Ford, H.~C. 2002, The Astrophysical Journal, 578, 543

\bibitem[{{Wertz} {et~al.}(2017){Wertz}, {Absil}, {G{\'o}mez Gonz{\'a}lez},
  {Milli}, {Girard}, {Mawet}, \& {Pueyo}}]{Wertz17}
{Wertz}, O., {Absil}, O., {G{\'o}mez Gonz{\'a}lez}, C.~A., {et~al.} 2017, \aap,
  598, A83

\end{thebibliography}

\begin{appendix}

\section{Mathematical notations for RSM map, KLIP FM and LOCI FM}
                                
\begin{table*}[t]
                        \caption{Description of the mathematical notations for the variables used in the RSM detection map, KLIP FM, and LOCI FM computation.}
                        \label{variable}
\centering

                        \begin{tabular}{lll}
                        \hline
                        \hline
Symbol  & Dimension & Comments\\                        
 \hline \\
RSM map & &\\
\hline
$\bm{x}_{i_a}$ &$\theta \times \theta \times T L_a$ & Patch of residuals centred on pixel $i_a$\\
$F_{i_a}$& $T L_a$ & Realisation of a two-state Markov chain representing the state in which the system is for pixel $i_a$\\
$ \bm{m}$& $\theta \times \theta$ & Cropped planetary signal (off-axis PSF)\\
$\bm{\varepsilon_{s,i_a}}$& $2 \theta\times \theta \times T L_a$ & Error terms associated with the two regimes\\
$S_{i_a}$& $T L_a$ & State in which the system is for every pixel $i_a$\\
$\xi_{s,i_a}$& $2 \times T L_a$& Probability associated with state $s$ for every pixel $i_a$\\
$\eta_{s,i_a}$ & $2 \times T L_a$ & Likelihood of being in each state for every pixel $i_a$\\
$p_{q,s}$& $2 \times 2 $& Transition probabilities between the regimes\\
$\mu$& $1$ & Mean of the residuals contained in an annulus $a$, with width equal to $\theta$\\
$\sigma$& $1$ & Standard deviation of the residuals contained in an annulus $a$, with width equal to $\theta$\\
$\beta$& $1$ &  Parameter representing the intensity of the planetary signal in the cube of residual \\
$a$& $1$ & Annulus index\\
$L_a$& $1$ & Number of pixels included in the annulus $a$\\
$T$& $1$ & Number of frames in the cube of residuals\\
$i_a$&$1$& Index associated with every pixel from every frame in the annulus $a$ (ranges from $1$ to $T L_a$)\\
$\theta$& $1$ & Angular size of the considered planetary signal ( set to $1\lambda/D$)\\
\hline \\
KLIP FM & &\\
\hline
$\bm{i}$ &$N^{pix}$ & Vectorised science image/annulus before speckle subtraction \\
$ \bm{m}$& $N^{pix}$ & Vectorised planetary signal (off-axis PSF) inside the selected annulus\\
$ \bm{p}$& $N^{pix}$ & Vectorised forward-modelled planetary signal inside the selected annulus\\
$ \bm{x}$& $N^{pix}$ & Vectorised processed annulus after speckle subtraction\\
$ \bm{p}_{i_a}$/$ \bm{p}_{j}$& $\theta \times \theta$ & Derotated and cropped forward-modelled planetary signal for pixel $i_a$/frame $j$\\
$\bm{R}$& $N^R \times N^{pix}$&  Reference library matrix\\
$\bm{V}_K$ &$K \times N^R$ & Eigenvector matrix of the covariance matrix $\bm{R}\bm{R}^{\top}$with $\bm{V}_K=\left[ \bm{v}_1,\bm{v}_2,...,\bm{v}_k \right] $\\
$\bm{\Lambda}$&$N^R \times N^{pix}$& Diagonal matrix with the eigenvalues of the covariance matrix $\bm{R}\bm{R}^{\top}$ with $\bm{\Lambda}=diag(\mu_1,\mu_2,...,\mu_k)^{\top}$\\
$\bm{Z}_K$ &$K \times N^R$ & Karhunen-Loève image matrix \\
$\Delta \bm{Z}_K$ &$K \times N^R$ & Perturbation of the Karhunen-Loève image matrix\\
$\bm{M}$& $N^R \times N^{pix}$& Planet signal component in the reference library $\bm{R}$\\
${C}_{MR}$ & $N^R \times N^R$&  Covariance Matrix between $\bm{M}$ and $\bm{R}$\\
$N^{pix}$ &$1$&Number of pixels in the selected annulus of width equal to one FWHM\\
$N^{R}$ &$1$& Number of reference frames used for the speckle field computation\\
$K$ &$1$& Number of principal components used for the speckle field computation\\
\hline \\
LOCI FM & &\\
\hline
$c_k$  & $N^R$ & Factors of the linear combination used to model the speckle field\\
$\bm{o}_i$ &  $N^{pix}$ &  Selected annulus in the science image used for the estimation of the linear factors $c_k$\\
$\bm{o}_k$ &  $N^{pix}$ &  Selected annulus in frame k of the reference library used for the estimation of the linear factors $c_k$\\
$\bm{r}_k$ &  $N^{pix}$' &  Selected annulus in frame k of the reference library used for the computation of the speckle field\\
$\bm{m}_i$ &  $N^{pix}$' &  Selected annulus in frame corresponding to the science image in the planetary signal matrix\\
$\bm{m}_k$ &  $N^{pix}$' &  Selected annulus in frame k of the planetary signal\\
$N^{pix}$ &$1$& Number of pixels in the annulus with width equal to three FWHM used for the factors estimation \\
$N^{pix}$' &$1$& Number of pixels in the annulus with width equal to one FWHM used for speckle field computation\\
$N^{R}$ &$1$& Number of reference frames used for the speckle field computation\\
\hline
                        \end{tabular}
                                \end{table*}

\section{Estimation of the $\beta$ parameter and forward-backward model in the case of RSM KLIP}

\begin{figure*}[t]
  \centering
  \subfloat[NACO at $2\lambda/D$]{\includegraphics[width=160pt]{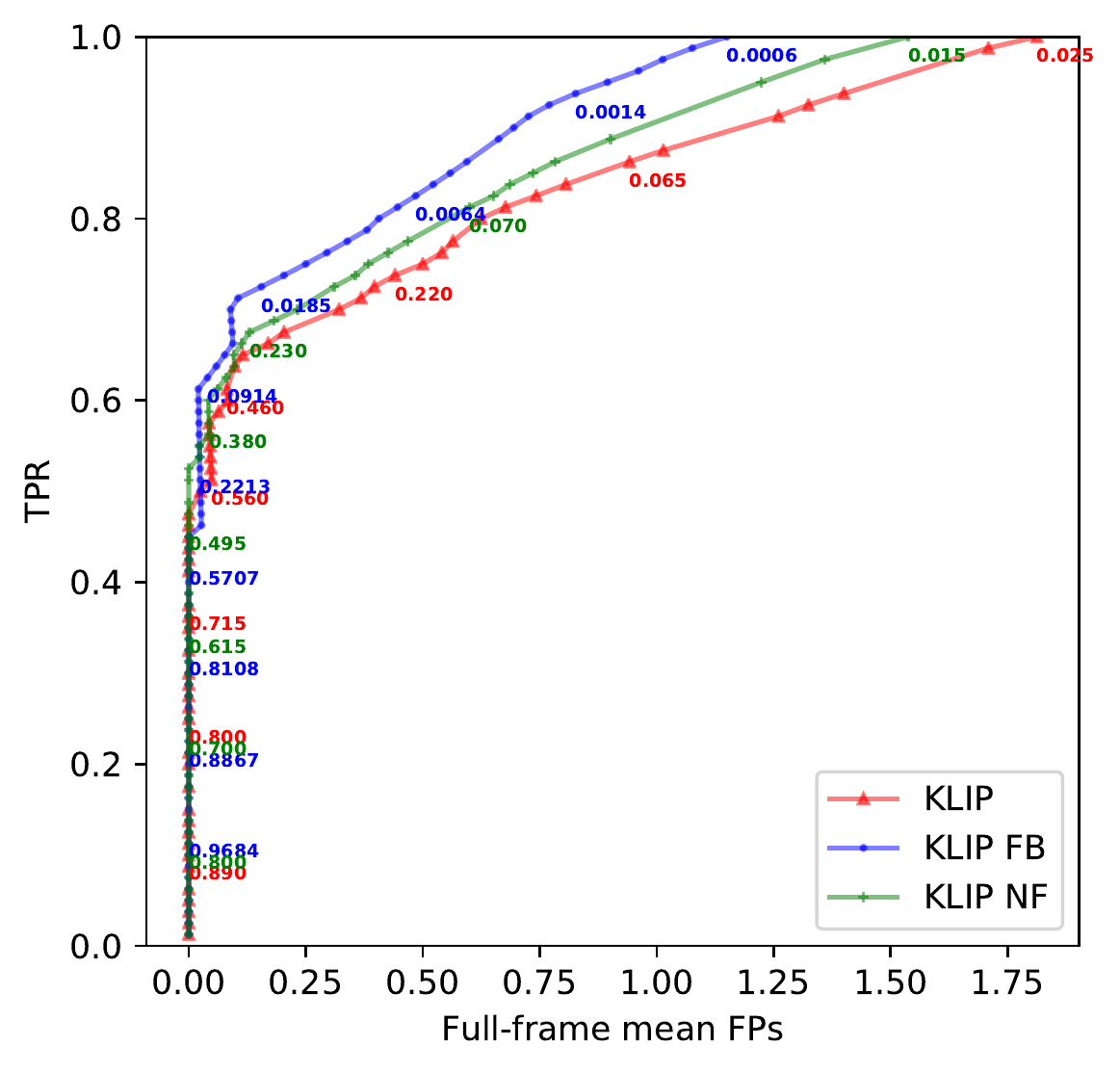}}
  \subfloat[SPHERE at $2\lambda/D$]{\includegraphics[width=160pt]{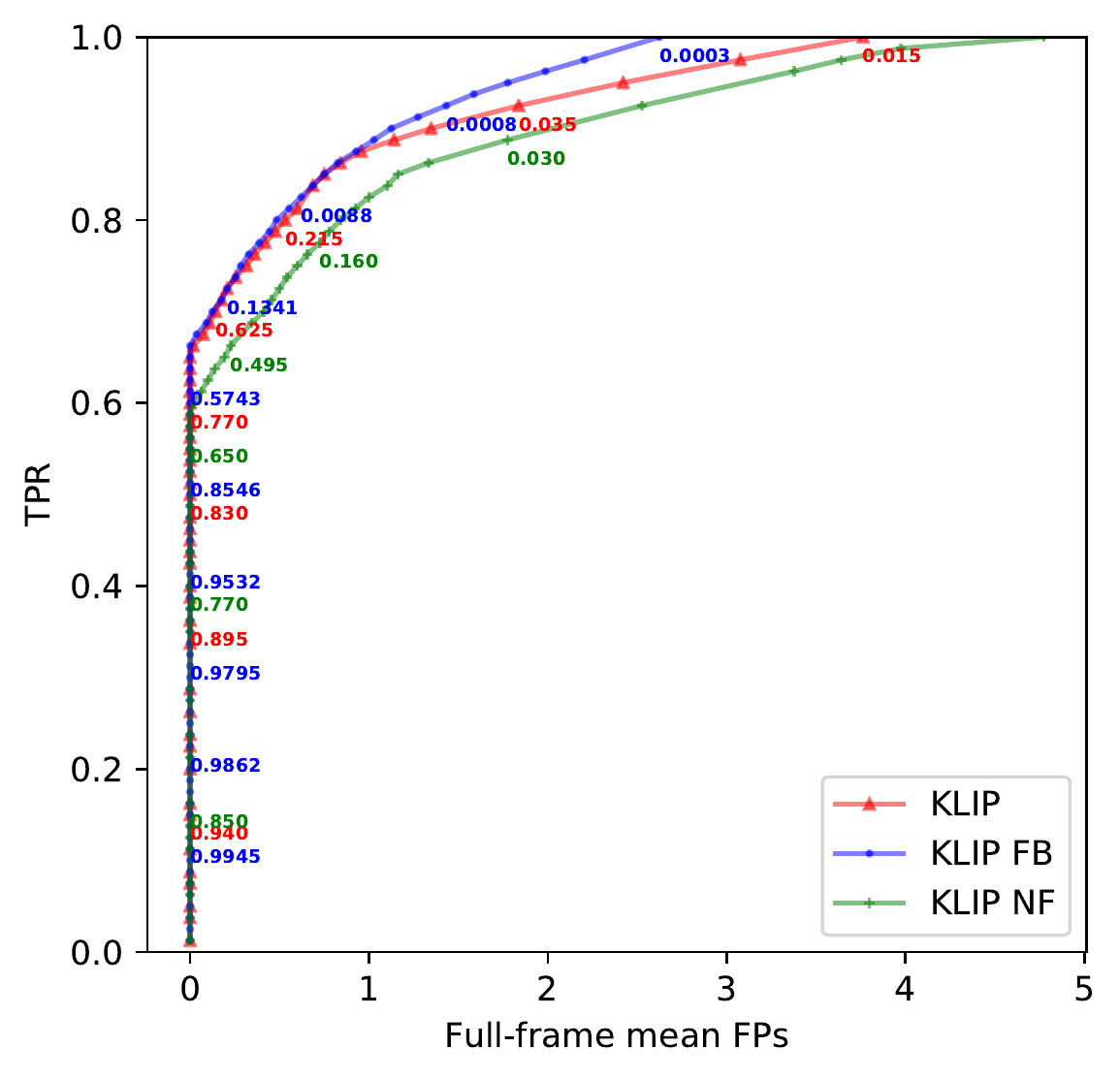}}
  \subfloat[LMIRCam at $2\lambda/D$]{\includegraphics[width=160pt]{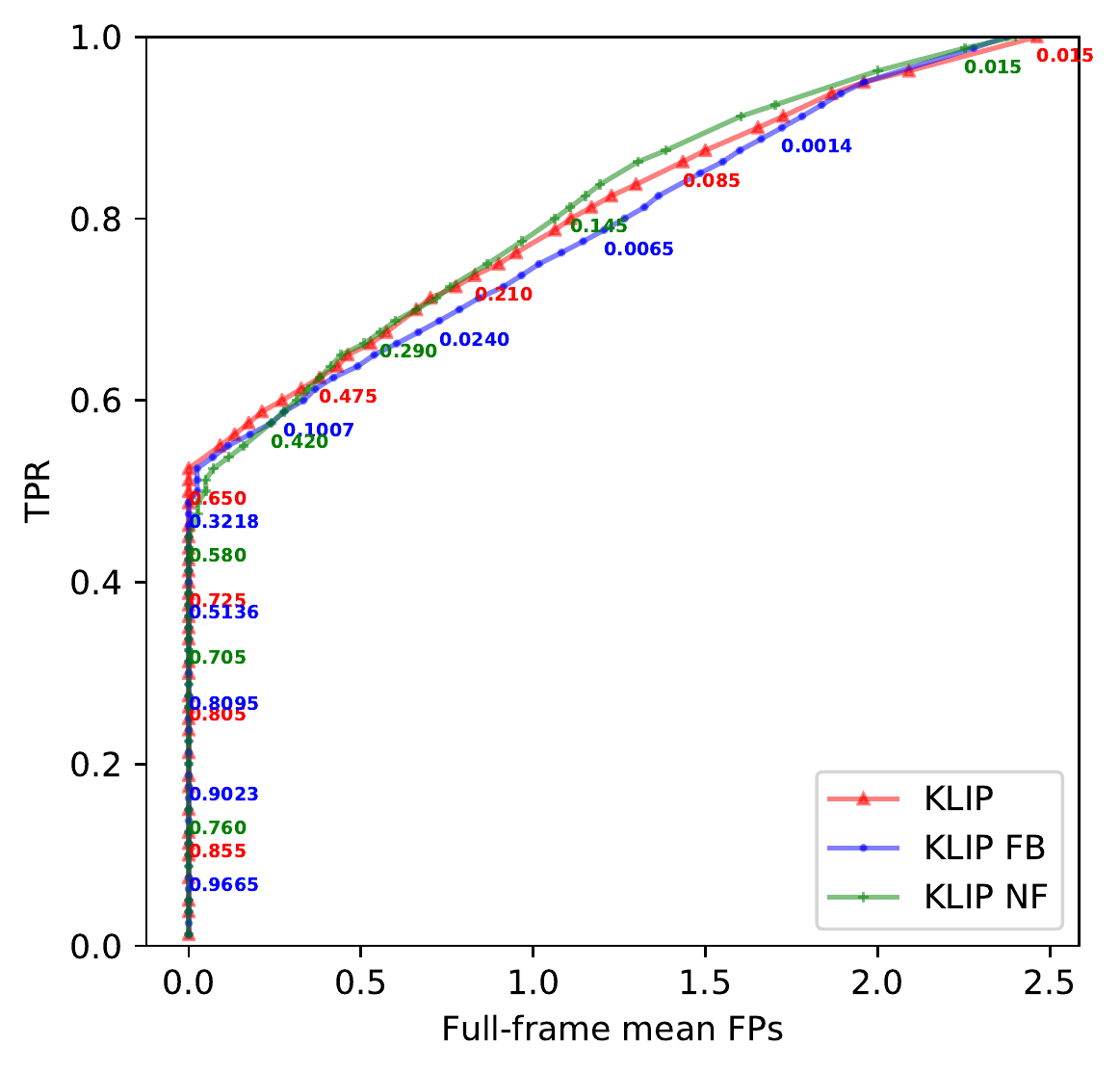}}\\
    \subfloat[NACO at $8\lambda/D$]{\includegraphics[width=160pt]{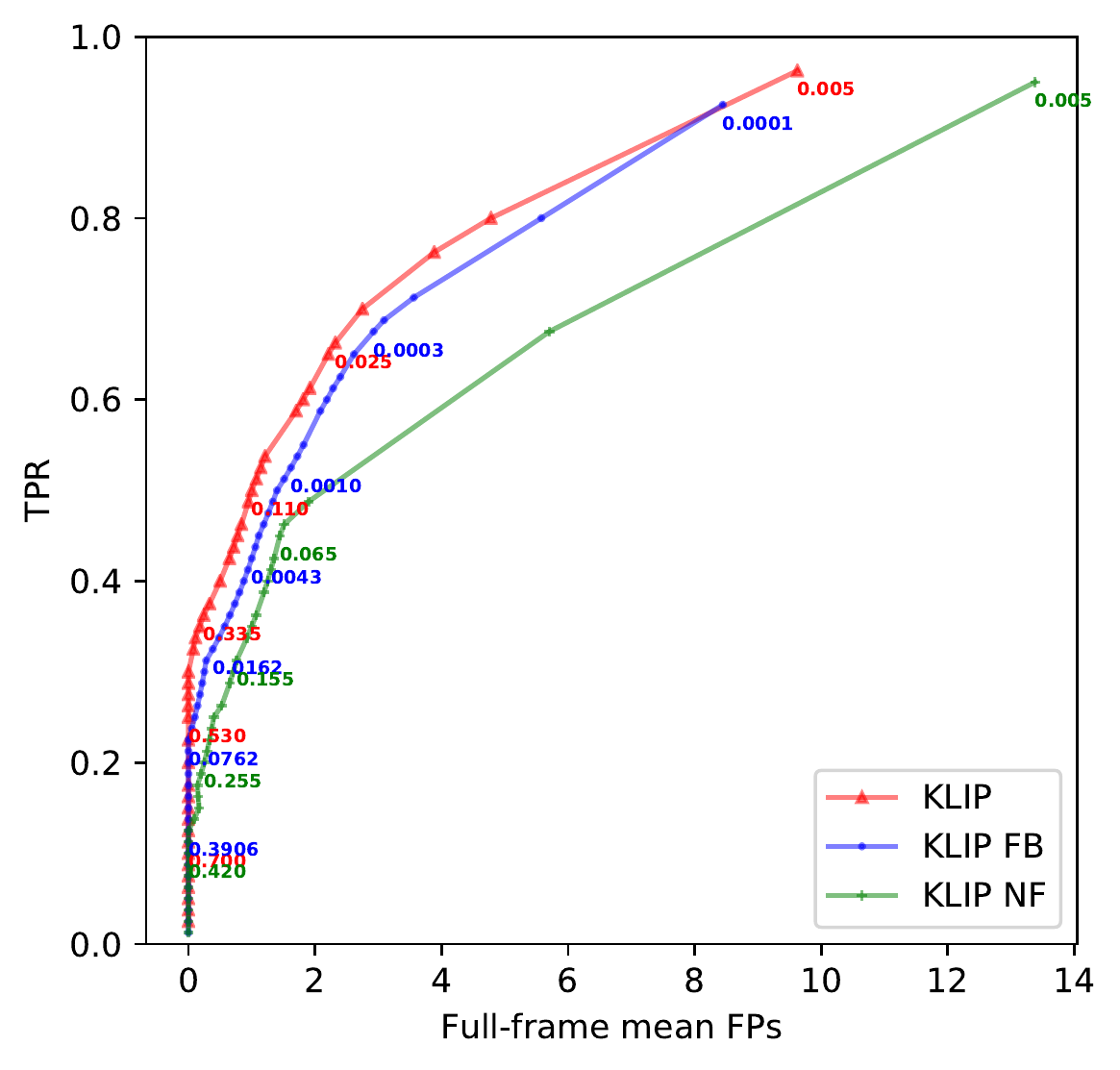}}
  \subfloat[SPHERE at $8\lambda/D$]{\includegraphics[width=160pt]{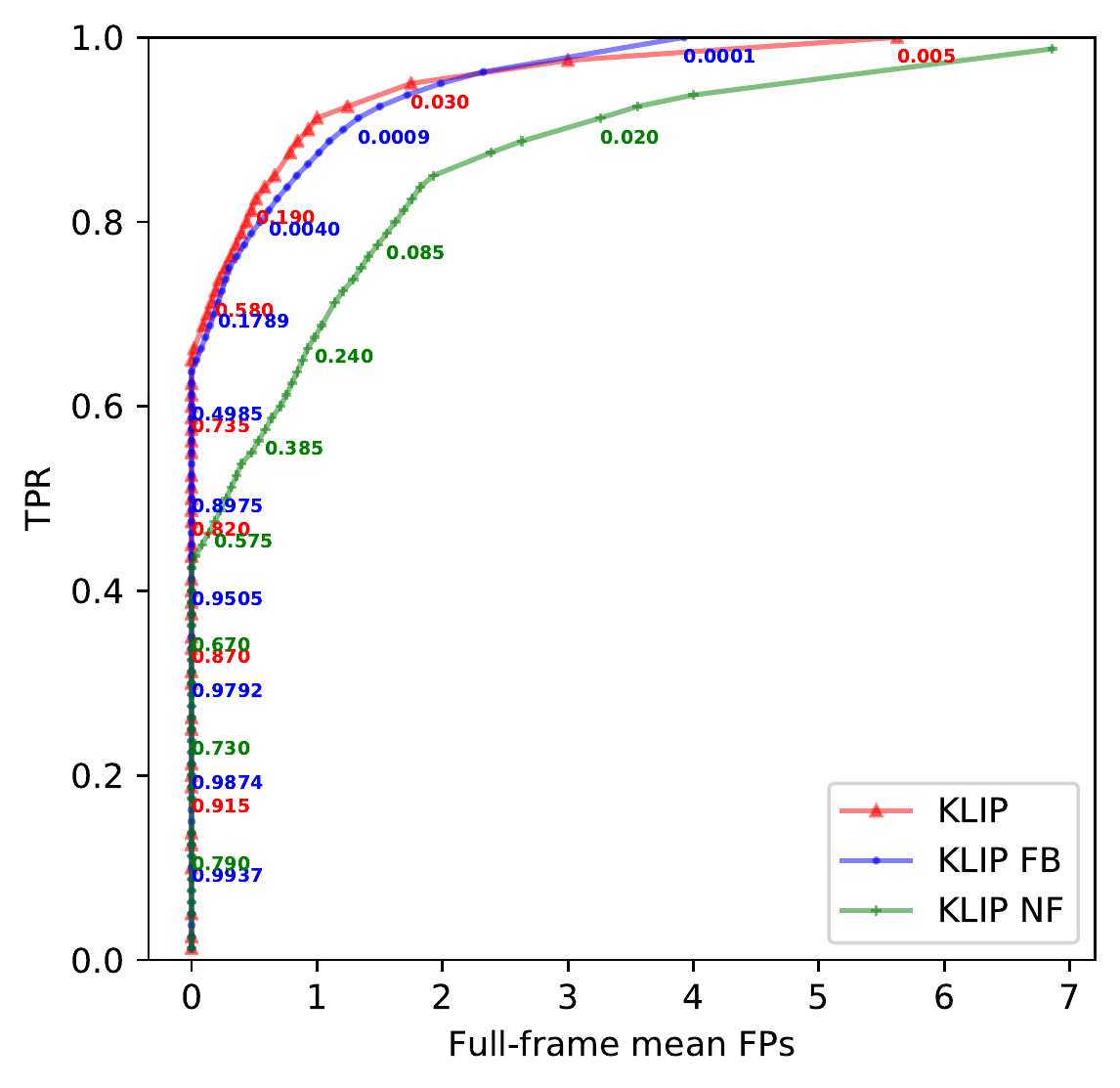}}
  \subfloat[LMIRCam at $8\lambda/D$]{\includegraphics[width=160pt]{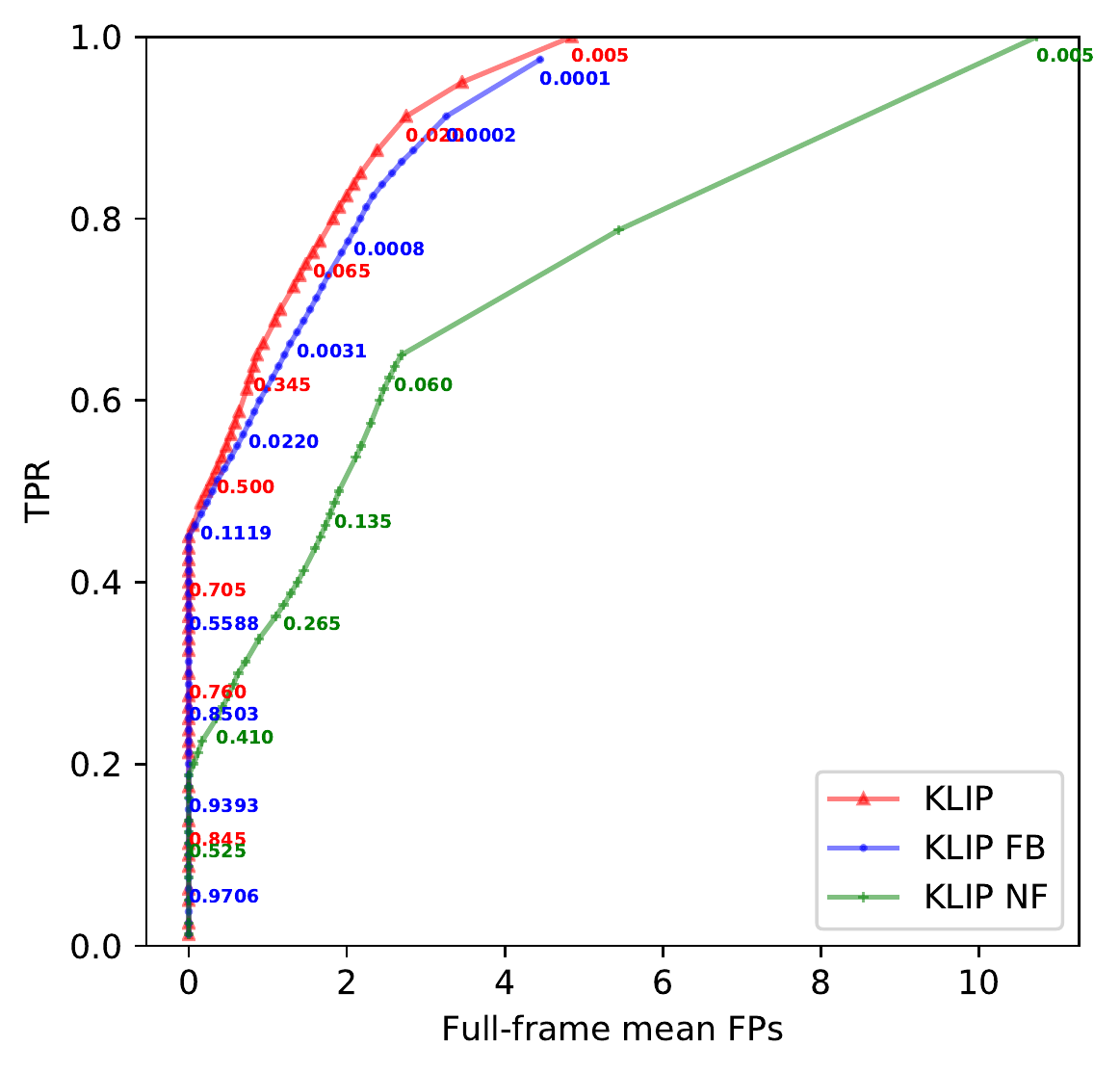}}
  
  \caption{\label{WSLOCI} ROC curves for the NACO, SPHERE, and LMIRCam data sets, with respectively the KLIP RSM map using the Gaussian maximum likelihood for the pre-optimisation of the flux parameter $\beta$ (red), the KLIP RSM map with no flux pre-optimisation (NF), which relies on the standard maximum likelihood used in the original RSM map for the estimation of flux parameter $\beta$ (blue), and the KLIP RSM map using the forward-backward approach for the probability estimation instead of the original forward approach (green).}
\end{figure*}

\section{Crop size comparison for FM LOCI RSM}

\begin{figure*}[t]
  \centering
  \subfloat[NACO at $2\lambda/D$]{\includegraphics[width=160pt]{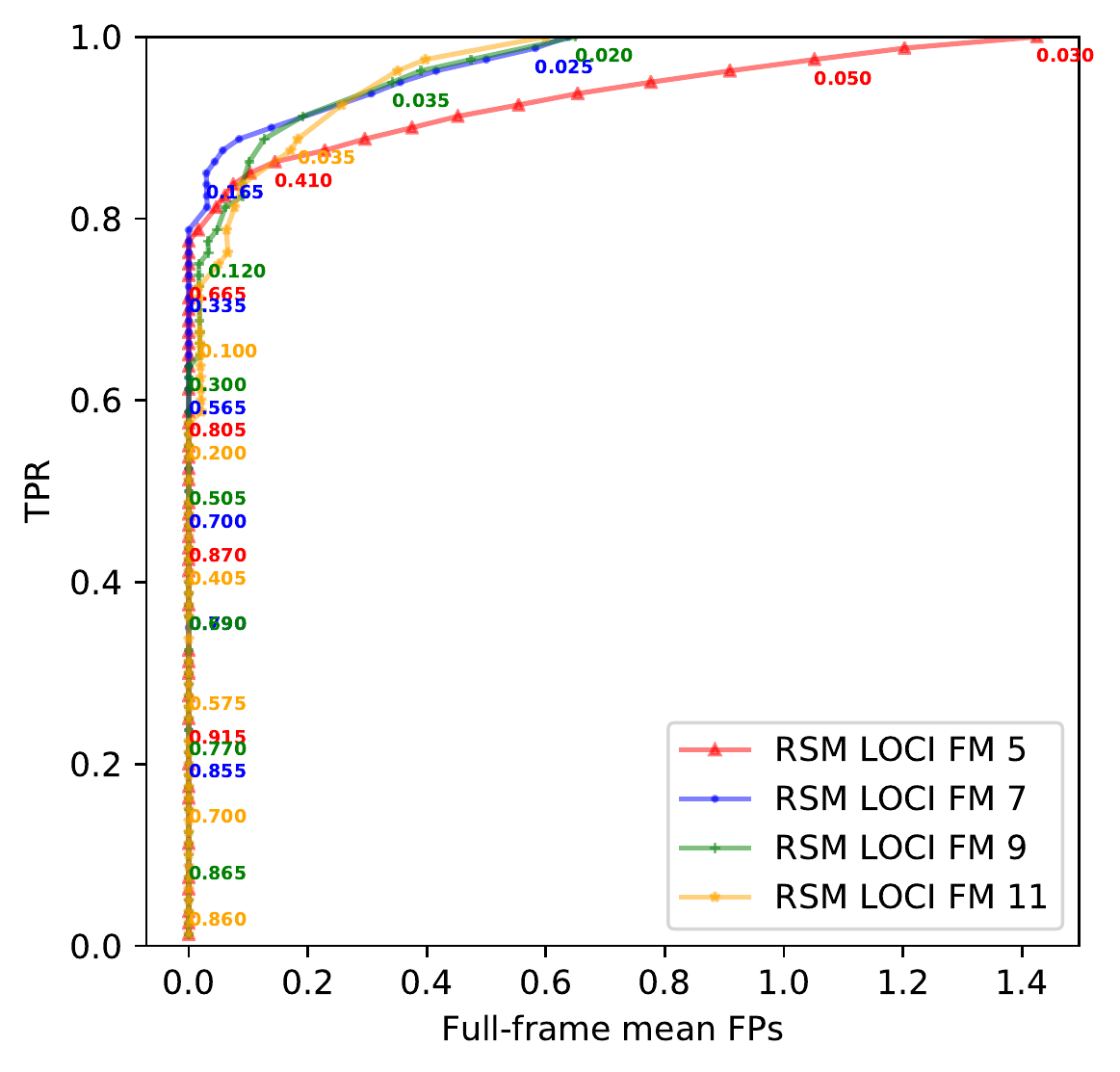}}
  \subfloat[SPHERE at $2\lambda/D$]{\includegraphics[width=160pt]{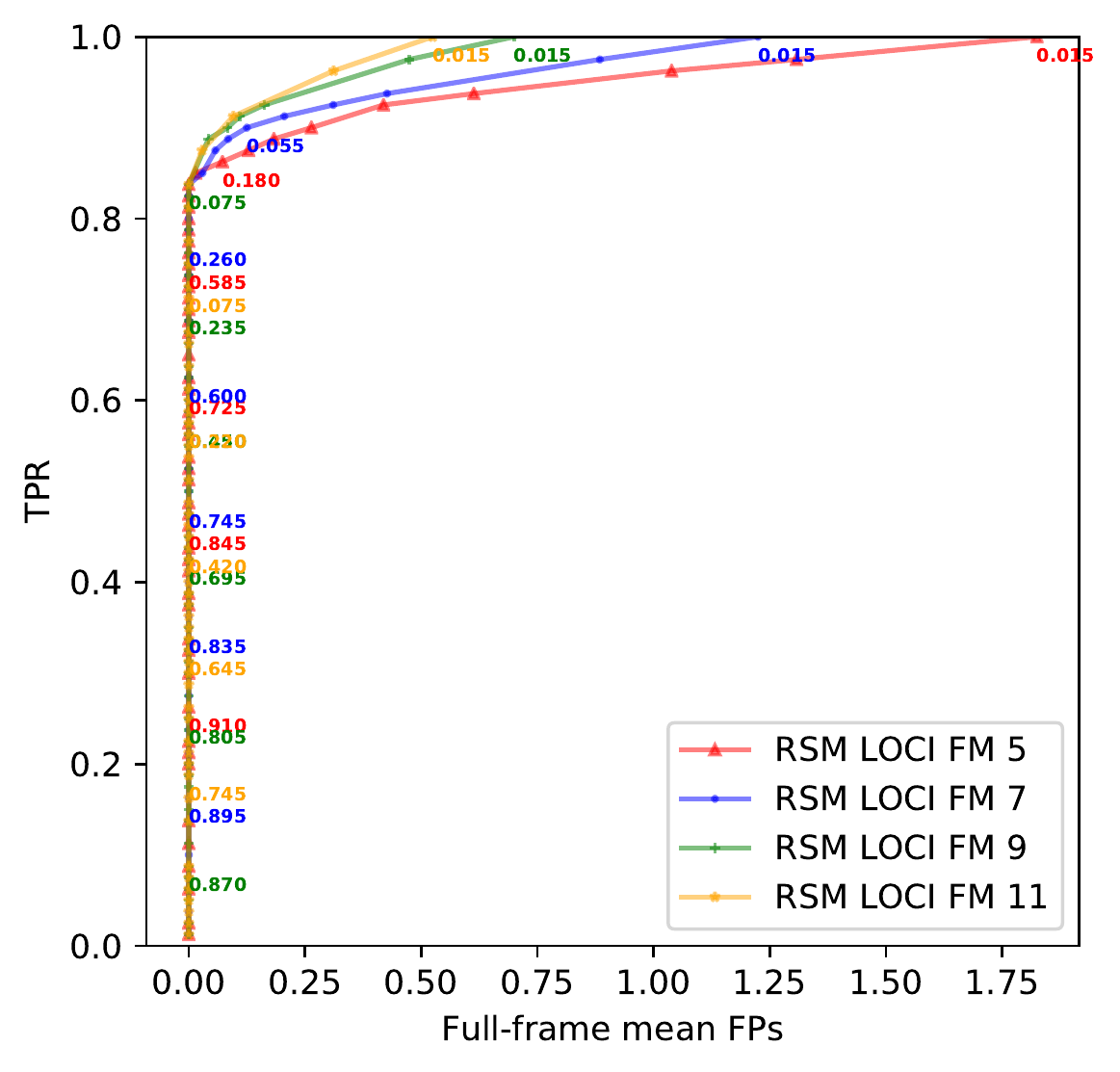}}
  \subfloat[LMIRCam at $2\lambda/D$]{\includegraphics[width=160pt]{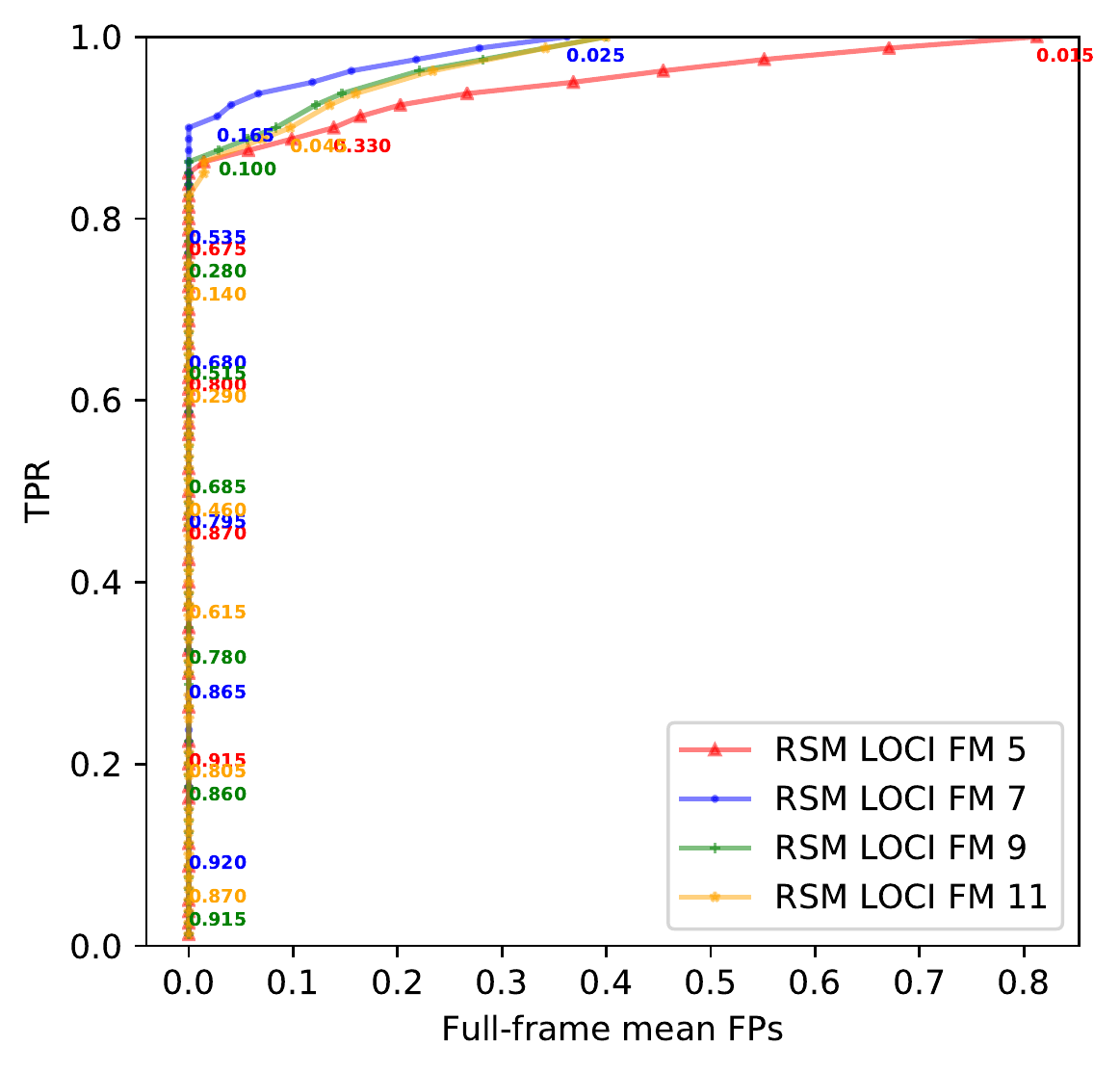}}\\
    \subfloat[NACO at $8\lambda/D$]{\includegraphics[width=160pt]{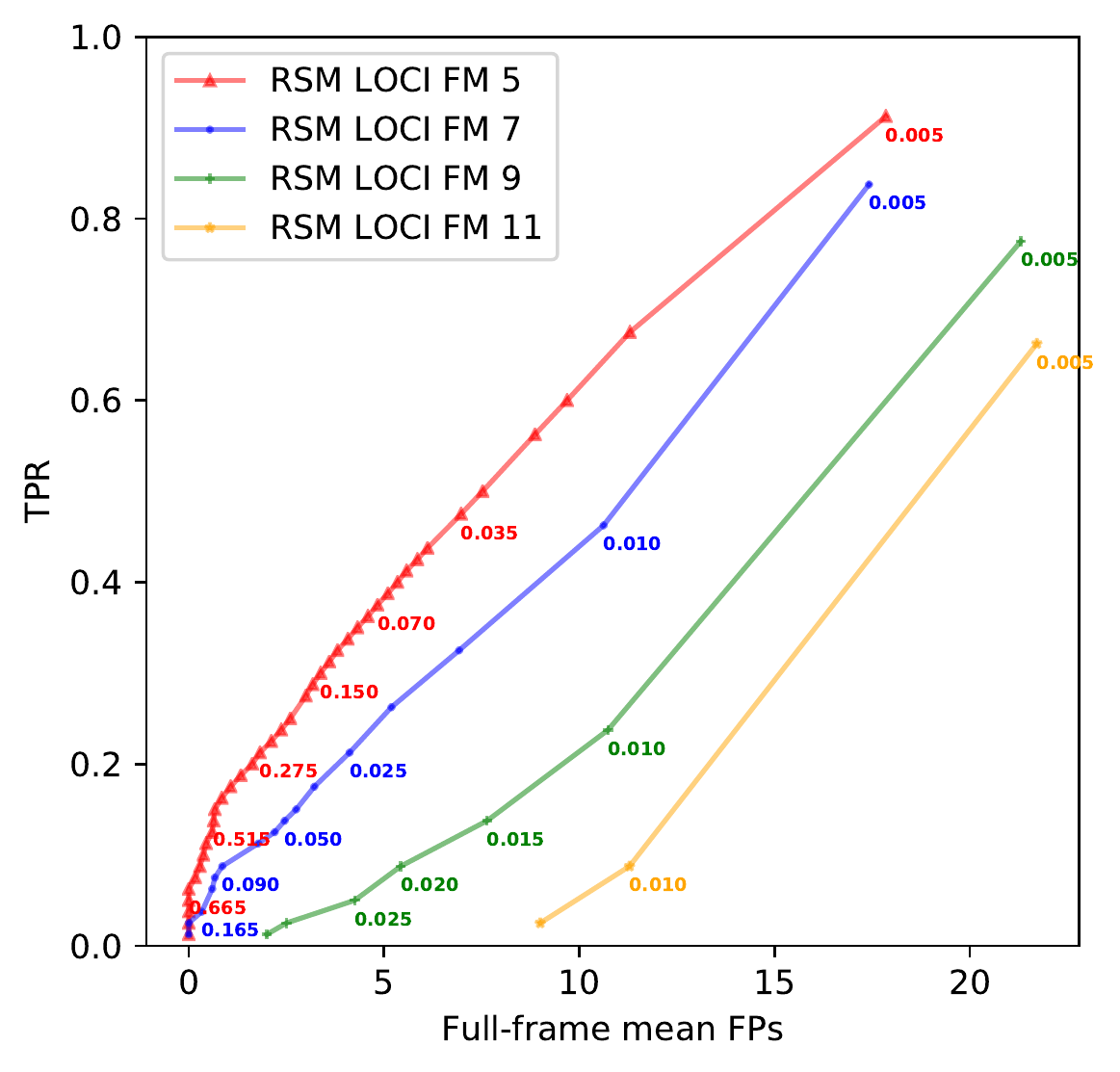}}
  \subfloat[SPHERE at $8\lambda/D$]{\includegraphics[width=160pt]{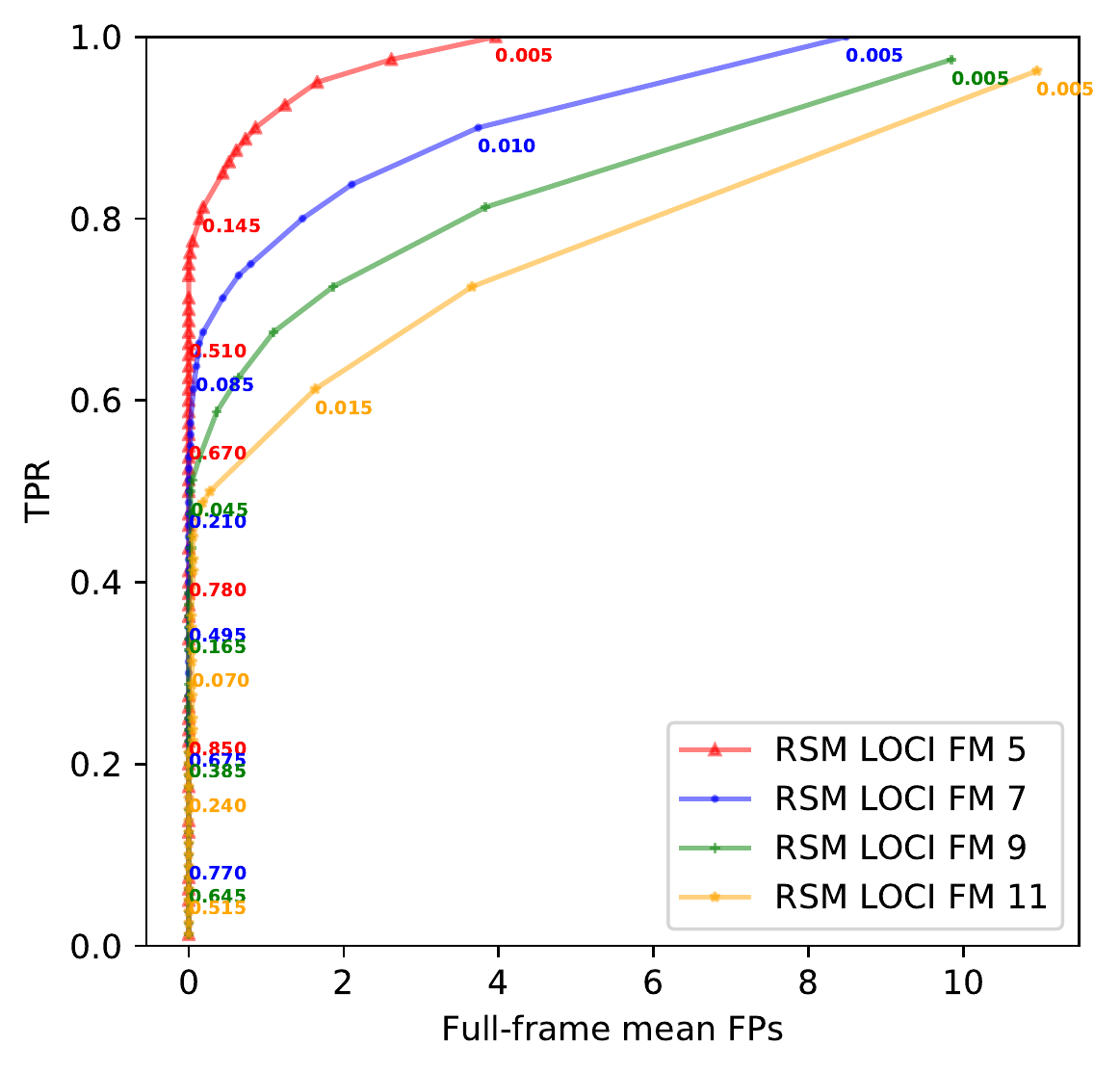}}
  \subfloat[LMIRCam at $8\lambda/D$]{\includegraphics[width=160pt]{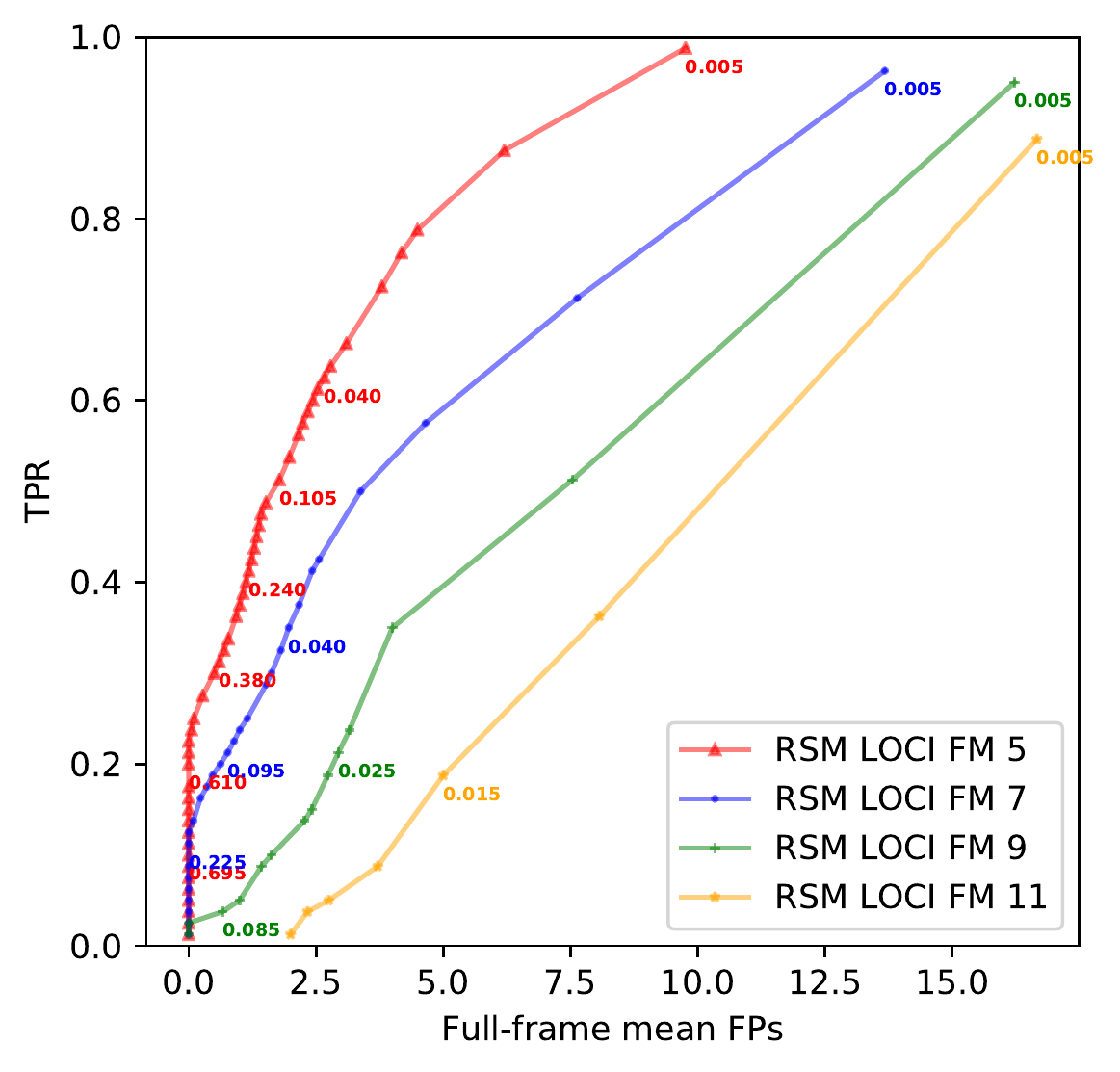}}
  
  \caption{\label{WSLOCI} ROC curves for the NACO, SPHERE, and LMIRCam data sets, with the LOCI-FM RSM map using respectively a crop size for the froward modelled PSF of 5 (red), 7 (blue), 9 (green), 11 (orange) pixels (FWHM$\approx 5$ pixels for all three data sets).}
\end{figure*}

\section{Performance comparison of PSF subtraction techniques combinations}

\begin{figure*}[t]
  \centering
  \subfloat[NACO ]{\includegraphics[width=370pt]{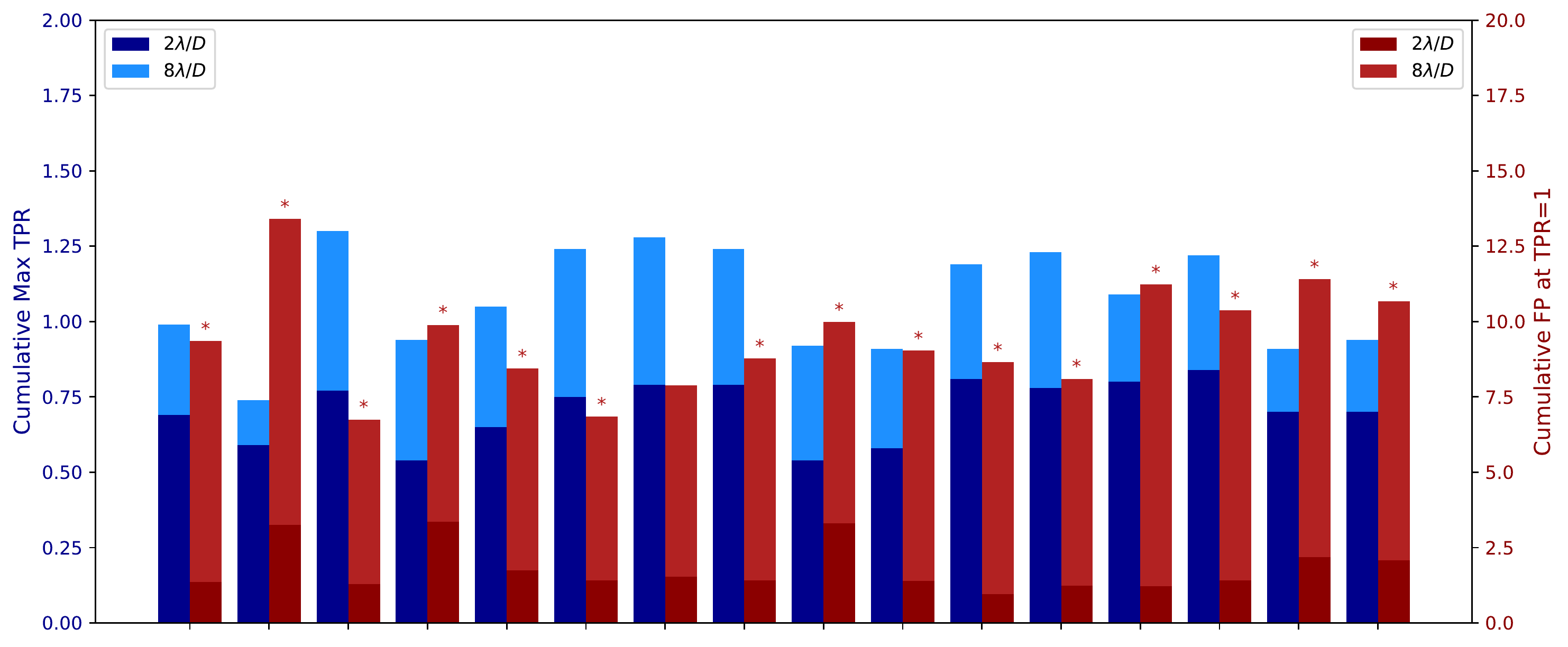}}\\
  \subfloat[SPHERE]{\includegraphics[width=370pt]{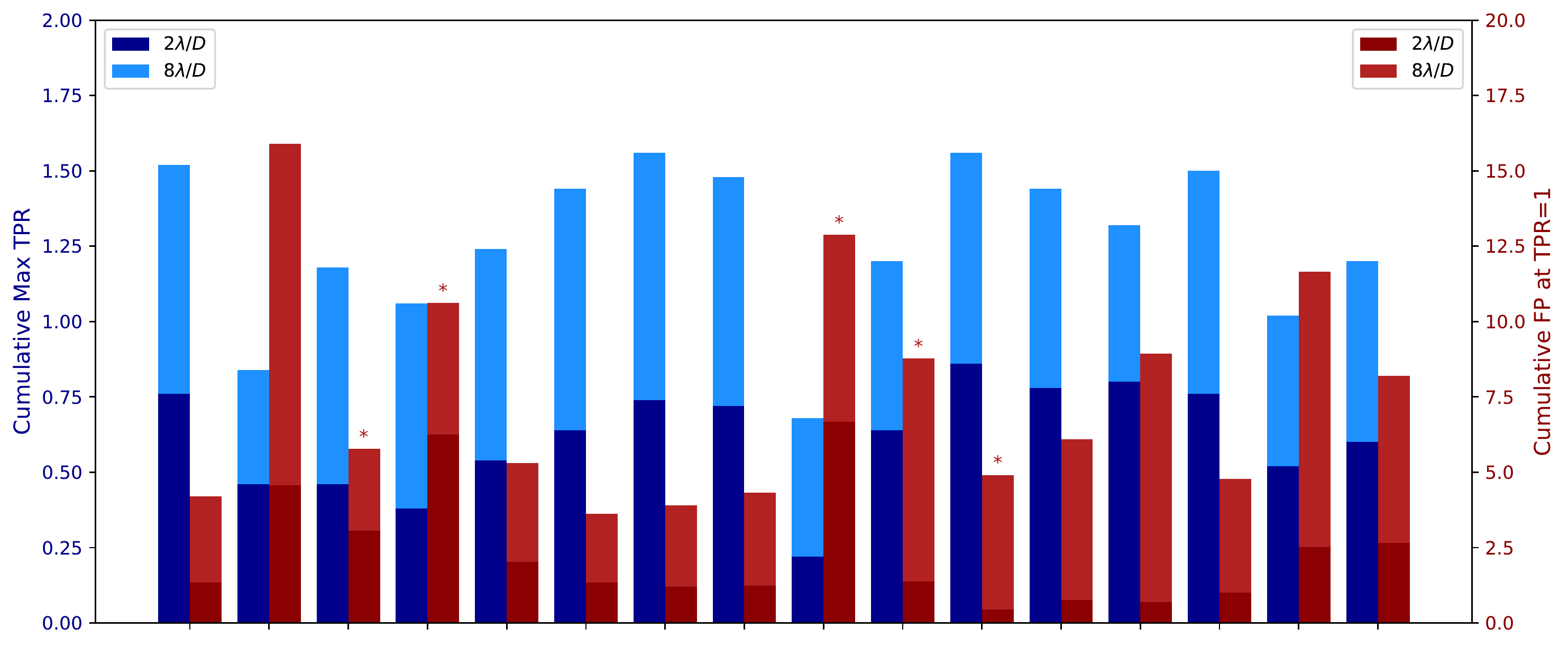}}\\
  \subfloat[LMIRCam]{\includegraphics[width=370pt]{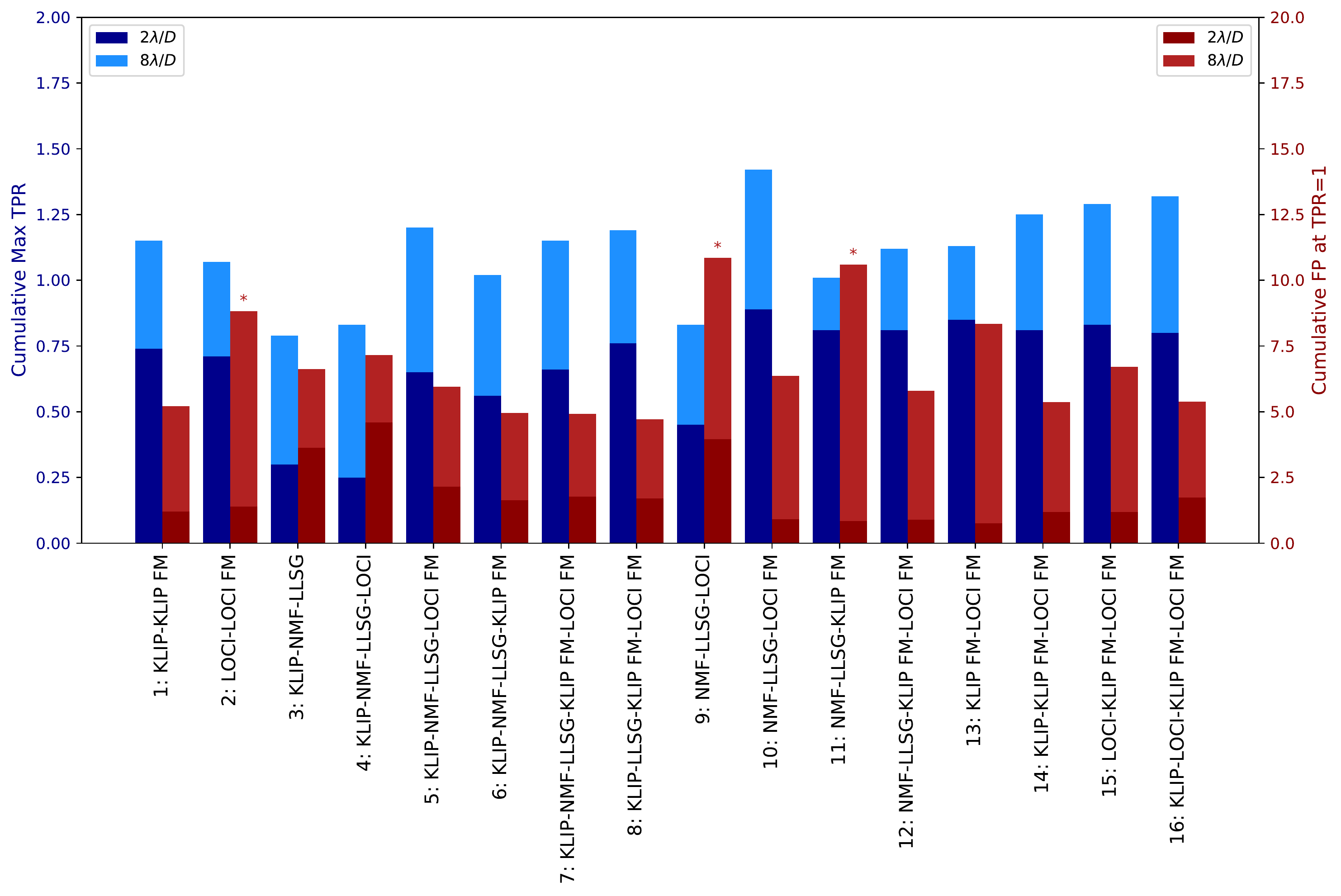}}\\

  \caption{\label{resmix} Cumulative maximum TPR reached without any FP (blue), and cumulative average number of FPs inside the entire frame at TPR=1 (red) for 16 different combinations of PSF subtraction techniques used to generate the RSM map and two radial distances. The dark-coloured bars are obtained at a radial distance of $2 \lambda/D$ while the light-coloured bars correspond to $8 \lambda/D$ for, respectively, the NACO (top), SPHERE (middle), and LMIRCam (bottom) data sets. The asterisks for some values of FP at TPR=1 indicate that a TPR of 1 has not been reached at a distance of $8 \lambda/D$ and that the smallest probability threshold was chosen instead (highest TPR). A high performance for a combination of PSF subtraction techniques corresponds to a tall blue histogram alongside a short red histogram.}
\end{figure*}

\begin{table*}
                        \caption{Maximum TPR reached without any FP and average number of FPs inside the entire frame at TPR=1 for 16 different combinations of PSF subtraction techniques used to generate the RSM map. The left values are obtained at a radial distance of $2\lambda/D$ while the right values correspond to $8\lambda/D$ for respectively the NACO, SPHERE and LMIRCam data sets. The asterisks for some values of FP at TPR=1 indicate that a TPR of 1 has not been reached and that the smallest probability threshold has been taken (highest TPR).}
                        \label{resmix}
\centering

                        \begin{tabular}{lllllll}
                        \hline
                        \hline
&\textbf{NACO}  & &\textbf{SPHERE}  & &\textbf{LMIRCam} \\                      
 Selected PSF subtraction techniques   & Max TPR & Max FP & Max TPR & Max FP & Max TPR & Max FP   \\      
 \hline
1: KLIP-KLIP FM&0.69/0.30&1.36/8.00*&0.76/0.76&1.34/2.86&0.74/0.41&1.20/4.01\\
2: LOCI-LOCI FM&0.59/0.15&3.26/10.15*&0.46/0.38&4.56/11.34&0.71/0.36&1.40/7.43*\\
3: KLIP-NMF-LLSG&0.66/0.53&1.28/5.47*&0.46/0.72&3.06/2.72*&0.30/0.49&3.63/2.99\\
4: KLIP-NMF-LLSG-LOCI&0.54/0.40&3.36/6.53*&0.38/0.68&6.26/4.36*&0.25/0.58&4.59/2.56\\
5: KLIP-NMF-LLSG-LOCI FM&0.65/0.40 &1.74/6.71*&0.54/0.70&2.02/3.28&0.65/0.55&2.15/3.80\\
6: KLIP-NMF-LLSG-KLIP FM&0.75/0.49&1.42/5.43*&0.64/0.80&1.34/2.28&0.56/0.46&1.64/3.31\\
7: KLIP-NMF-LLSG-KLIP FM-LOCI FM&0.79/0.49*&1.54/6.35&0.74/0.82&1.20/2.70&0.66/0.49&1.77/3.15\\      
8: KLIP-LLSG-KLIP FM-LOCI FM &0.79/0.45&1.42/7.35*&0.72/0.76&1.24/3.08&0.76/0.43&1.71/3.01\\ 
9: NMF-LLSG-LOCI  &0.54/0.38&3.31/6.68*&0.22/0.46&6.68/6.20*&0.45/0.38&3.96/6.89*\\        
10: NMF-LLSG-LOCI FM  &0.58/0.33&1.39/7.65*&0.64/0.56&1.38/7.40*&0.89/0.53&0.91/5.46\\        
11: NMF-LLSG-KLIP FM &0.81/0.38&0.95/7.70*&0.86/0.70&0.44/4.47*&0.81/0.20&0.84/9.76*\\          
12: NMF-LLSG-KLIP FM-LOCI FM&0.78/0.45 &1.24/6.86*&0.78/0.66&0.76/5.34&0.81/0.31&0.89/4.91\\       
13: KLIP FM-LOCI FM&0.80/0.29&1.22/10.01*&0.80/0.52&0.70/8.24&0.85/0.28&0.75/7.59\\        
14: KLIP-KLIP FM-LOCI FM&0.84/0.38&1.41/8.96*&0.76/0.74&1.00/3.78&0.81/0.44&1.19/4.18\\      
15: LOCI-KLIP FM-LOCI FM&0.70/0.21&2.18/9.23*&0.52/0.50&2.52/9.14&0.83/0.46&1.18/5.53\\
16: KLIP-LOCI-KLIP FM-LOCI FM&0.70/0.24&2.08/8.60*&0.60/0.60&2.66/5.54&0.80/0.52&1.74/3.65\\    
\hline
                        \end{tabular}
                                \end{table*}
                                
\section{Contrast curve for the SPHERE data set using KLIP RSM}
                                
\begin{figure*}[h!]
  \centering
  \subfloat{\includegraphics[width=320pt]{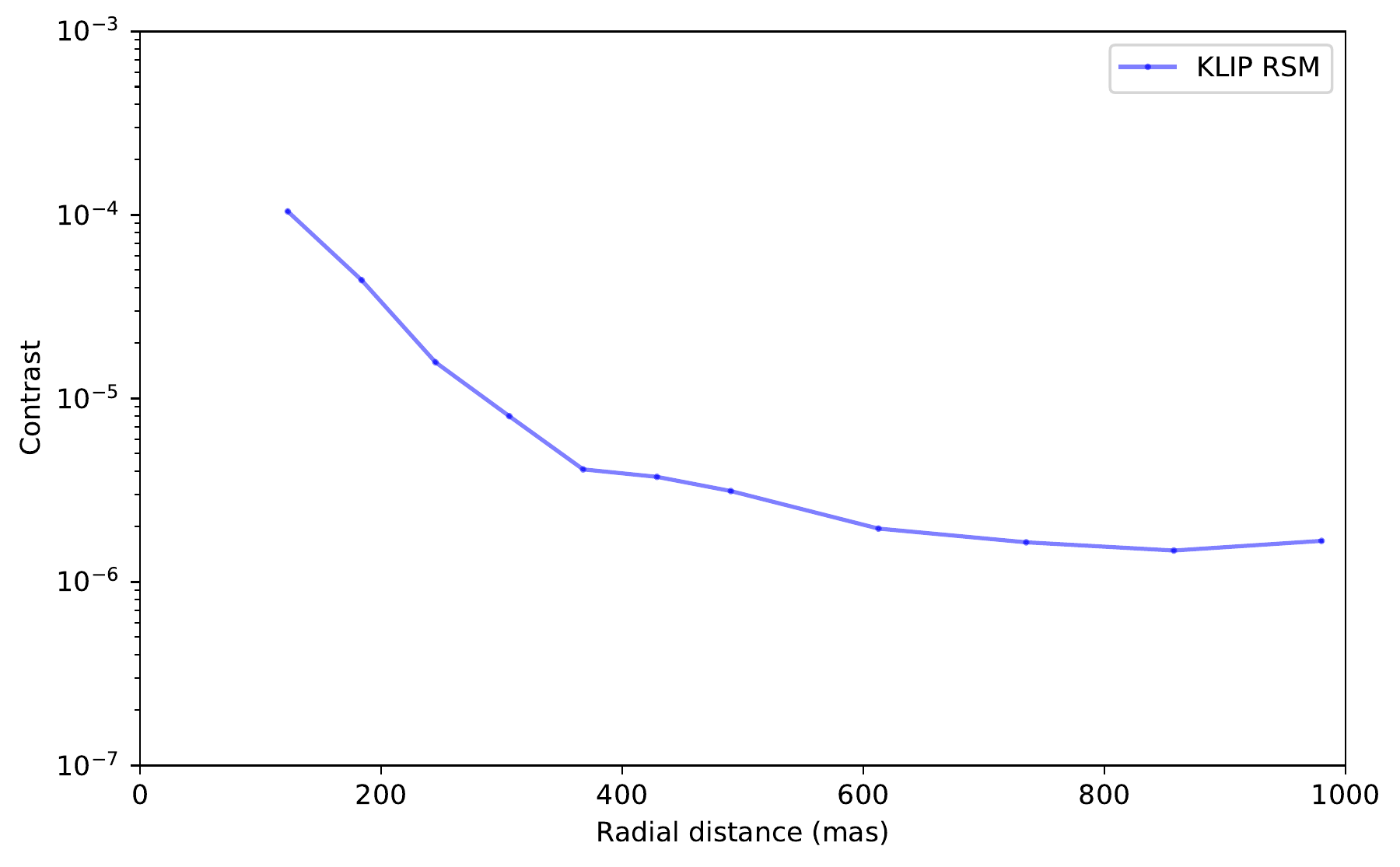}}\\
  
  \caption{\label{ccklip} Contrast curves for the SPHERE data set using KLIP RSM with 20 principal components and a FOV rotation expressed in terms of FWHM of 0.3. The region $\left[ 2\lambda /D, 16\lambda /D \right]$ has been considered to get the contrast in the first arcsecond.}
\end{figure*}

\end{appendix}

\end{document}